\documentclass[]{aa}  

\usepackage{graphicx}
\usepackage{caption,subcaption}
\usepackage{txfonts}
\usepackage{xcolor}
\usepackage[]{natbib}
\usepackage{lscape}
\usepackage[draft]{hyperref}
\usepackage{longtable}
\usepackage{amsmath}
\usepackage{breqn}
\usepackage{soul}
\usepackage{amssymb}
\usepackage{pifont}
\newcommand{\cmark}{\ding{51}}
\newcommand{\xmark}{\ding{55}}

\begin{document} 

   \title{Searching for molecular gas in/outflows in the nuclear regions of five Seyfert galaxies}

   \author{A. J. Domínguez-Fernández\inst{1,2}
           \fnmsep\thanks{\email{ajdfernandez@ucm.es}}
          \and
          A. Alonso-Herrero\inst{3}
          \and
          S. García-Burillo\inst{1}
          \and
          R. I. Davies\inst{4}
          \and
          A. Usero\inst{1}
          \and
          {A. Labiano}\inst{3}
          \and
          N. A. Levenson\inst{5}
          \and
          M. Pereira-Santaella\inst{6,7}
          \and
          M. Imanishi\inst{8,9}
          \and
          C. Ramos Almeida\inst{10,11}
          \and
          D. Rigopoulou\inst{6}
          }

   \institute{Observatorio Astronómico National (OAN), Alfonso XII 3, 28014 Madrid, Spain
         \and
             Departamento de Física de la Tierra y Astrofísica and IPARCOS-UCM (Instituto de Física de Partículas y del Cosmos de la UCM), Facultad de Ciencias Físicas, Universidad Complutense de Madrid, E-28040 Madrid, Spain
         \and             
             Centro de Astrobiología (CAB, CSIC-INTA), ESAC Campus, E-28692 Villanueva de la Cañada, Madrid, Spain
        \and
            Max Planck Institut für extraterrestrische Physik Postfach 1312, D-85741 Garching bei München, Germany
        \and
            Space Telescope Science Institute, Baltimore, MD, 21218, USA
        \and
            Department of Physics, University of Oxford, Keble Road, Oxford OX1 3RH, UK
        \and
            Centro de Astrobiología (CSIC-INTA), Ctra. de Ajalvir, Km 4, 28850, Torrejón de Ardoz, Madrid, Spain
        \and
            National Astronomical Observatory of Japan, National Institutes of Natural Sciences (NINS), 2-21-1 Osawa, Mitaka, Tokyo 181–8588, Japan
        \and 
            Department of Astronomical Science, SOKENDAI (The Graduate University of Advanced Studies), 2-21-1 Osawa, Mitaka, Tokyo 181-8588, Japan
        \and
            Instituto de Astrofísica de Canarias, Calle Vía Láctea, s/n, E-38205, La Laguna, Tenerife, Spain
        \and
            Departamento de Astrofísica, Universidad de La Laguna, E-38206, La Laguna, Tenerife, Spain
            }

   \date{Received Oct 21, 2019; accepted Mar 02, 2020}
 
  \abstract
  {AGN-driven outflows are believed to play an important role in regulating the growth of galaxies mostly via negative feedback. However, their effects on their hosts are far from clear, especially for low and moderate luminosity Seyferts. To investigate this issue, we have obtained cold molecular gas observations, traced by the CO(2-1) transition, using the NOEMA interferometer of five nearby (distances between 19 and 58 Mpc) Seyfert galaxies. The resolution of $\sim$0.3-0.8” ($\sim$30-100 pc) and field of view of NOEMA allowed us to study the CO(2-1) morphology and kinematics in the nuclear regions ($\sim$100 pc) and up to radial distances of $\sim$ 900 pc. We have detected  CO(2-1) emission in all five galaxies with disky or circumnuclear ring like morphologies. We derived cold molecular gas masses on nuclear ($\sim$ 100 pc) and circumnuclear ($\sim$ 650 pc) scales in the range from 10$^6$ to 10$^7$M$_{\odot}$ and from 10$^7$ to 10$^8$M$_{\odot}$, respectively. In all of our galaxies the bulk of this gas is rotating in the plane of the galaxy. However, non-circular motions are also present. In NGC 4253, NGC 4388 and NGC 7465, we can ascribe the streaming motions to the presence of a large-scale bar. In Mrk~1066 and NGC 4388, the non-circular motions in the nuclear regions are explained as outflowing material due to the interaction of the AGN wind with molecular gas in the galaxy disk.  We conclude that for an unambiguous and precise interpretation of the kinematics of the cold molecular gas we need a detailed knowledge of the host galaxy (i.e., presence of bars, interactions, etc) as well as of the ionized gas kinematics and the ionization cone geometry.}


   \maketitle

\section{Introduction}

The molecular gas is the main phase of the interstellar medium (ISM) in the central regions of galaxies \citep{schneider2007extragalactic} and is the material from which new stars are formed. In active galaxies, i.e., those harboring an accreting supermassive black hole (SMBH), it also plays a role in feeding the central engine. At the same time, the molecular gas in the galaxy disks may be affected by the AGN activity via feedback \citep[positive or negative, see ][and references therein]{fiore2017agn}.  Understanding the balance between feeding and feedback processes and how it changes through cosmic time is crucial to understand how galaxies and their SMBH coevolve. 

Seyfert galaxies and low-luminosity AGN have large amounts of molecular gas in their nuclear and circumnuclear regions with morphologies
resembling nuclear disks, rings, and mini-spirals  \citep[e.g.,][]{schinnerer2000distribution,schinnerer2000bars,garcia2003molecular,krips2007molecular,davies2014fueling,sani2012physical,combes2019alma,alonso2018resolving,alonso2019nuclear}. These morphologies might be related to the mechanisms transporting gas from kiloparsec scales all the way down to the SMBH sphere of influence. On nuclear and circumnuclear scales, the molecular gas has been observed inflowing and outflowing depending on the galaxy \citep[][\citealt{gallimore2016high}]{garcia2012feeding,combes2013alma,combes2014alma,davies2014fueling,garcia2014molecular}. Still, it is  
unclear how the gas loses almost all its angular momentum on its way to the accretion disk on sub-pc scales \citep{lynden1974evolution,pringle1981accretion,garcia2005molecular,haan2009dynamical}. 

Molecular outflows on scales of tens to a few hundred of parsecs are predicted by some recent torus models in the context of the gas cycle of AGN, such as radiation-driven torus models \citep{wada2012radiation,honig2018dusty,williamson2018radiation,izumi2018circumnuclear}. Indeed, these nuclear molecular outflows taking place in the nuclear torus/disk and on circumnuclear scales have already been detected in a few nearby Seyfert galaxies \citep[][Garcia-Burillo et al. 2019]{combes2013alma, morganti2015radio,  gallimore2016high,alonso2018resolving,alonso2019nuclear}. On hundreds of parsecs and kiloparsec scales in Seyfert galaxies, at a given AGN luminosity  the outflow molecular phase shows maximum velocities below those of the ionized phase \citep{fiore2017agn}. This suggests that the AGN wind in this type of active galaxies has generally only a moderate impact on the ISM of their host galaxies. Indeed, when these molecular outflows can be spatially resolved the AGN wind is found generally to sweep rather than clear out the cold molecular gas completely \citep[][]{garcia2014molecular,morganti2015radio,alonso2018resolving,alonso2019nuclear}.

\begin{table*}
	\caption{General properties of the galaxies in our sample.}           
	\label{tab:properties}     
	\centering           
	\begin{tabular}{c c c c c c c c c c}  
		\hline\hline       
		Galaxy& Morphological & Environment & Ref. & AGN & Ref. & $D_{L}$ & Scale & $L_{X \ (2-10 \ keV)}$ & Ref. \\
		&Type & & &Type& & (Mpc) & (pc/")& (erg s$^{-1}$) & \\ 
		\hline                    
		Mrk 1066 & (R)SB0$^{+}$(s) & Minor merger & 1,2 & Sy 2 & 5 & 47.2 & 224 & $7.8 \times 10^{42}$ & 11 \\
		NGC 2273 & SB(r)a & Isolated & 3 & Sy 2 & 6 & 25.8 & 124 & $1.9 \times 10^{42}$ & 12\\
		NGC 4253 & (R')SB(s)a & Isolated & 3 & Sy 1.5 & 7,8 & 57.8 & 272 & $5.5 \times 10^{42}$ & 13\\
		NGC 4388 & SA(s)b & Virgo Cluster & 3 & Sy 1.9 & 9 & 19.4 & 93 & $3.2 \times 10^{42}$ & 13\\
		NGC 7465 & (R')SB(s)$0^{0}$ & NGC 7448 Group & 4 & Sy 2/L & 10 & 21.9 & 105 & $5.6 \times 10^{41}$ & 13\\
		\hline                  
	\end{tabular}
	
	\tablebib{Morphological types, distances and scales are from NED. The last two are computed in the cosmic microwave background frame. Environment: (1) \cite{gimeno2004catalog}, (2) \cite{smirnova2010seyfert}, (3) NED, (4) \cite{van1992study}. AGN types: (5) \cite{goodrich1983mrk}, (6) \cite{contini1998starbursts}, (7) \cite{osterbrock1985spectra}, (8) \cite{osterbrock1993spectroscopic}, (9) \cite{mason2015nuclear}, (10) \cite{ferruit2000hubble}. Spectral types are further discussed in the text. Intrinsic (absorption-corrected) X-ray luminosities in the 2-10 keV band: (11) \cite{marinucci2012link}, (12) \cite{awaki2009detection}, (13) \cite{ricci2017bat}.}

\end{table*}

\begin{table*}
	\caption{Main bars, [OIII]$\lambda$5007 emission and stellar disk inclination. } 
	
	\label{tab:Bars}     
	\centering           
	\begin{tabular}{c c c c c c c c c c c}   
		\hline\hline        
		Galaxy & PA$_{\rm bar}$ & R$_{bar}$ & Ref. & PA$_{\rm [OIII]}$ & Ref. & Open. angle & Ref. & i$_{\rm disk}$ & PA$_{\rm disk}$ & Ref. \\
		&  ($^{\circ}$) &  (") & & ($^{\circ}$) & &  ($^{\circ}$) & & ($^{\circ}$) & ($^{\circ}$) & \\
		\hline                    
		Mrk 1066 & 143 & 15.5 & 1 & 139 & 2 & 25 & 2 & 50 & 300 & 3\\
		NGC 2273 & 108 & 27 & 4 & 90 & 5 & 32 & 6 & 51 & 53 & 7\\
		NGC 4253 & 108 & $\sim$8 & 8 & 160 & 9 & 110 & 9 & 18 & 246 & 3\\
		NGC 4388 & 100 & 19 d & 10 & $\sim35$ (N)& 9 & $\sim50$ & 9 & 78 & 90 & 11\\
		& & & & 193 (S)& 9 & 92 & 9 & & &\\
		NGC 7465 & * &  * & * & 135 & 5 & 56 & 6 & $\sim60$ & $\sim$120-130 &12\\
		\hline                  
	\end{tabular}
    \tablefoot{Column (1) Galaxy Name,  Columns (2)-(3)-(4) PA of the major axis of the main bar, radius and reference, Columns (4)-(5) PA of the biconical outflow inferred from the [OIII] emission and reference, Columns (6)-(7) cone opening and reference, Columns (8)-(9)-(10) stellar galaxy disk PA, inclination and reference. The d in the radius of the bar of NGC 4388 denotes that the value is deprojected. *: For the PA of the major axis of the main bar of NGC 7465, its radius and the references, see the discussion in Section~\ref{subsec:ngc7465}.}
    
	\tablebib{(1)~\citet{afanasiev1998two};
(2) \citet{fischer2013determining}; (3) \citet{riffel2017gemini}; (4) \citet{moiseev2004structure};
(5) \citet{ferruit2000hubble}; (6) Our own measurements using the images in (5); (7) \citet{barbosa2006gemini};
(8) \citet{alonso1998near}; (9) \citet{mulchaey1996emissionii}; (10) \citet{veilleux1999kinematic}; (11) \citet{greene2014circumnuclear};
(12) \citet{merkulova2012study}.}

\end{table*}

In this context, we have started several programs to obtain sub-arcsecond resolution observations of the cold molecular gas in Seyfert galaxies. One of our goals is to study the gas cycle (inflows/outflows) in the nuclear (scales of tens of parsecs) and circumnuclear regions (scales of a few hundred parsecs) of active galaxies. We chose nearby galaxies (z$\lesssim$0.01) so we can achieve this physical resolution. We are using  both the NOrthern Extended Millimeter Array (NOEMA) in the northern hemisphere and the Atacama Large Millimeter Array (ALMA) in the southern hemisphere. We  selected our samples of Seyfert galaxies covering a range  AGN luminosities, nuclear obscurations and nuclear mid-infrared emission  properties including the detection of polycyclic aromatic carbon (PAH) emission \citep{honig2010dusty,alonso2014nuclear,alonso2016mid}. 
Another criterion to select Seyfert galaxies for our study is the availability of observations and the corresponding modeling of the ionized gas and stellar continuum emission, preferably taken with integral field units (IFU). These observations probe the AGN wind kinematics as well as the ionization cone geometry, and the host galaxy kinematics, respectively. As we shall see throughout this work, these observations are crucial to interpret the kinematics of the molecular gas.

In this paper we present NOEMA observations of the  ${}^{12}$CO(J= 2-1) transition at $\nu_{\rm rest}= 230.54$ GHz  of five northern Seyfert galaxies at distances between 19 and 58\,Mpc (assuming $\Lambda$CDM cosmology with $H_{0}= \ 73$ km s$^{-1}$ Mpc$^{-1}$, $\Omega_{\Lambda}= \ 0.73$ and $\Omega_{M}= \ 0.27$). In  Table~\ref{tab:properties}  we summarize main properties of the sample including the morphological types and the X-ray $2-10\,$keV intrinsic luminosities, i.e., corrected for absorption. Table~\ref{tab:Bars} lists the position angles (PA) of the main bar and the ionization cone as well as the PA and inclination of the host galaxy disk for all of our sources. With the NOEMA observations we  aim to study the properties and kinematics of the cold molecular gas in the nuclear (typically less than 100\,pc resolutions) and circumnuclear regions (typically the central $\sim 600$ pc).

This paper is organized as follows. In $\S$\ref{observations} we present our observations and describe the data reduction. In $\S$\ref{methodology} we briefly discuss the data analysis. In $\S$\ref{results} we present the results source by source and in $\S$\ref{molecularmass} the molecular gas masses. In Section $\S$\ref{discussion} we discuss the main results of the paper. Finally, we summarize our conclusions in $\S$\ref{conclusions}. 

\section{Observations and Data Reduction}\label{observations}

\subsection{NOEMA Observations}

We obtained NOEMA observations of the CO(2-1) transition ($\nu_{rest}$ = 230.54 GHz) and its underlying continuum ($\sim$1.3\,mm) between March 2015 and December 2016 in the A configuration except for one track of NGC 2273 that was observed with the B configuration. Due to the expansion of NOEMA, our observations were taken with a different number of antennas ranging from 6 to 8 (see Table~\ref{tab:obs}). Two galaxies (Mrk~1066 and NGC~2273) were observed with two different configurations
(see Table~\ref{tab:obs}) which were combined to produce the final datacubes. The bandpass, flux, phase and amplitude calibrations were done using standard calibrators of the NOEMA archive, observed at the beginning or/and at the end of the tracks. The full width at half maximum (FWHM) of the Gaussian primary beam of NOEMA at the CO(2-1) transition frequency for our local sources is $\sim22$", which can be considered as the diameter of the NOEMA field of view (FoV). The final data cubes were split into 360 channels, each of $\sim13$ km s$^{-1}$ width, to improve the signal-to-noise ratio.

We performed the data reduction using the \textit{Grenoble Image and Line Data Analysis Software} (GILDAS\footnote{https://www.iram.fr/IRAMFR/GILDAS}) with standard techniques. We used natural weight to clean all data cubes to obtain the best sensitivity. We employed the task {\tt UV$\_$AVERAGE} to produce the continuum images, which are the average of the line-free emission channels. Then, we subtracted the continuum from the original cubes to obtain the line-emission data cubes. In order to optimize the result of the deconvolution process, we used polygons that enclosed the emission above 3$\sigma$ during the cleaning of both the continuum images and line-emission data cubes. The synthesized beams sizes and their position angles are listed in Table~\ref{tab:obs} for each galaxy. The final resolutions are in the range 0.4\arcsec-0.7\arcsec. At the distances of our galaxies they correspond to physical sizes in the range 65-115$\,$pc.   

All maps but one (NGC~4388) presented here have not been corrected by primary beam attenuation. In the case of NGC~4388, the observed CO(2-1) emission is spread over the full NOEMA FoV. As the attenuation of the emission is expected to be greater as we move away from the phase center, we applied the primary beam correction correction to correct for this.

\subsection{Ancillary Data}

We downloaded {\it Hubble Space Telescope} (HST) fully-reduced images from the HLA and MAST archives in the WFPC2 F606W ($\sim$ broad-band $V$) and NICMOS F160W ($\sim$ broad-band $H$) filters to produce $V-H$ color maps. For NGC 4388, the best quality WFPC2 F606W image has a low signal-to-noise ratio to the northwest of the nucleus and thus we used the WFC3/UVIS F814W ($\sim$ broad-band $I$) image for the optical emission instead. As we shall see, these color maps trace the dust extinction in the galaxies and can be compared directly with the cold molecular gas emission.  

\begin{table*}
	\caption{Log of NOEMA observations and data reduction.}            
	\label{tab:obs}     
	\centering           
	\begin{tabular}{c c c c c c c c}   
		\hline\hline        
		Source& Integ. time & Conf. & Date & Beam & PA$_{\rm beam}$ & Cont. rms noise & Line rms noise\\
		& (h) & & (mm/yy) & ("$\times$ ")& ($^{\circ}$) & ($\mu$Jy beam$^{-1}$) & (mJy beam$^{-1}$) \\ 
		\hline                    
		Mrk1066 & 2.4 + 2.0 & 6A, 7A & 03/15, 02/16 & $0.44 \times 0.40$ & 40.3 & 41 & 1.7 \\
		NGC 2273 & 1.5 + 3.8 & 6A, 7B & 03/15, 03/16 & $0.72 \times 0.58$ & 32.3 & 46 & 1.4 \\
		NGC 4253 & 3.8 & 8A & 12/16 & $0.55 \times 0.30$ & 21.8 & 28 & 1.0\\
		NGC 4388 & 4.9 & 7A & 03/15 & $0.84 \times 0.36$ & 21.1 & 92 & 3.8\\
		NGC 7465 & 3.4 & 8A & 12/16 & $0.59 \times 0.30$ & 23.1 & 30 &1.2 \\
		\hline                  
	\end{tabular}

	\tablefoot{(1) Name of the source; (2) Total integration time; (3) Number of antennas and array configuration(s); (4) Date of observations; (5) Synthesized beam size; (6) Position angle of the synthesized beam; (7) Continuum root mean square noise; (8) Root mean square noise in a channel ($\Delta$v$\sim$ 13 km s$^{-1}$) of the continuum-free line cube.}
	
\end{table*}

\begin{table*}
	\caption{Continuum best-fit parameters.}            
	\label{tab:uvfit}     
	\centering           
	\begin{tabular}{c c c c c c c c c }   
		\hline\hline        
		Galaxy & Component & Function & RA (J2000) & Dec (J2000) & Major & Minor & PA & S$_{\nu}$  \\
		&&&&& (") & (") & ($^{\circ}$) & (mJy)\\
		\hline                    
		Mrk 1066 & AGN & EG & 02:59:58.59 & 36:49:13.82 & $0.9 \pm 0.1$ & $0.35 \pm 0.08$ & $120 \pm 6$ & $3.9 \pm 0.5$ \\
		NGC 2273 & AGN & EG & 06:50:08.64 & 60:50:45.01 & $1.3 \pm 0.2$ & $0.9 \pm 0.2$ & $0 \pm 20$ & $2.4 \pm 0.4$ \\
		& W1 & P & 06:50:08.46 & 60:50:44.88 & - &  - & - & $0.3 \pm 0.1$ \\
		& W2 & P & 06:50:08.33 & 60:50:44.77 & - & - & - & $0.2 \pm 0.1$ \\
		& N & P & 06:50:08.55 & 60:50:47.16 & - & - & - & $0.4 \pm 0.1$ \\
		NGC 4253 & AGN & CG & 12:18:26.52 & 29:48:46.57 & $0.37 \pm 0.04$ & $0.37 \pm 0.04$ & - & $1.9 \pm 0.2$ \\
		NGC 7465 & AGN & CG & 23:02:00.96 & 15:57:53.26 & 0.6 (fixed) & 0.6 (fixed) & - & $0.9 \pm 0.2$ \\
		\hline                  
	\end{tabular}
	\tablefoot{(1) Name of the galaxy; (2) 1.3 mm continuum component; (3) Function  used to fit the continuum: point source (P), circular Gaussian (CG), elliptical Gaussian (EG); (4) and (5) J2000 equatorial coordinates; (6), (7) and (8) Size and position angle of the component; (9) Total spatially-integrated flux.}

\end{table*} 

\section{Analysis of the NOEMA
Observations} \label{methodology}

\subsection{Fit of the continuum emission}

We first determined the location of the 1.3 mm continuum sources by fitting a point source model, and when needed, circular or elliptical Gaussians to account for any extended emission. We performed these fits in the uv-plane using the GILDAS task {\tt UV$\_$FIT}. When we fitted the extended emission,  we also derived the source deconvolved size. We identified the 1.3 mm continuum peak as the AGN position, except for NGC~4388. For this galaxy, whose continuum at 1.3 mm has not been detected, we used the position of 21\,cm continuum peak from \textit{Very Long Baseline Radio Interferometry} (VLBI) observations as the location of the nucleus \citep[][see further discussion in Section~\ref{subsec:ngc4388}]{giroletti2009faintest}. 

In Table~\ref{tab:uvfit} we summarize the results of the continuum fits, and in Section~\ref{discussion} we describe in detail the results for each galaxy.

\subsection{CO(2-1) moment 0 maps}

We first produced maps of the integrated CO(2-1) emission using the {\tt MOMENTS} task in GILDAS. To do that, we applied a clipping of 3$\sigma$ in each pixel of every velocity channel with significant line emission ($\gtrsim 3 \sigma$) of the continuum-free line cube and integrated over them. We calculated the rms noise in a channel ($\sigma_{ch}$) by running the task {\tt GO NOISE} over a wide range of channels ($\sim$ 50) without line emission and then we propagated it with the formula $\sigma_{0th \ moment}= \sqrt{N} \ \sigma_{ch} \ \delta v$, where N is the number of channels with line emission and $\delta v$ is the width of the channel ($\sim$ 13 km s$^{-1}$ in our case), to obtain the noise in the integrated intensity maps. 

\subsection{CO(2-1) moments 1 and 2 and kinematic modeling}

To model the CO(2-1) cold molecular gas kinematics and to look for non-circular motions we used ${}^{\rm 3D}$BAROLO \citep{teodoro20153d}. ${}^{\rm 3D}$BAROLO allows to fit a tilted-ring model to 3D (2 spatial $\times$ 1 spectral) data cubes. We followed an analogous procedure to that described by \cite{alonso2018resolving}, which we summarize here briefly. In the first run, we allowed 5 parameters to vary: systemic and rotation velocities, velocity dispersion as well as the disk inclination and position angle. We fixed the kinematic center to the coordinates measured from the peak of the 1.3\,mm continuum or that of the 21\,cm continuum in the case of NGC~4388, which are assumed to identify the AGN location. We also assumed a thin disk with a scale height of 100 pc \citep{iorio2016little}. We computed the fits by minimizing the absolute value of the subtraction of the observed velocity field from model and using uniform weighting. For the second run we fixed systemic velocity, inclination and position angle to the average values derived in the first run and rerun ${}^{\rm 3D}$BAROLO only allowing the rotation velocity and velocity dispersion to vary. This way we produced a simple rotating model which we then subtracted from the observed mean velocity field to look for deviations from circular motions. 

\begin{table}
	\caption{Average parameters of the ${}^{\rm 3D}$BAROLO rotating disk model.}             
	\label{tab:Baroloparameters}     
	\centering           
	\begin{tabular}{c c c c }   
		\hline\hline        
		Galaxy & v$_{\rm sys}$  & i  & PA  \\
		& (km s$^{-1}$) & ($^{\circ}$)& ($^{\circ}$) \\
		\hline                    
		Mrk 1066 & 3632 & 50 & 302 \\
		NGC 2273 & 1836 & 52 & 54 \\
		NGC 4253 & 3874 & 31 & 263 \\
		NGC 4388 & 2516 & 82 & 88 \\
		NGC 7465 & 1975 & 53 & 52 \\
		\hline                  
	\end{tabular}
\end{table}

The main outputs of ${}^{\rm 3D}$BAROLO used in this work are the observed mean-velocity field (1st-order moment), the observed velocity dispersion (2nd-order moment) and the velocity model maps. We decided to extract the first and second moments with ${}^{\rm 3D}$BAROLO for consistency, that is, to compare all kinematic maps (mainly the first moment and its model) extracted in the same way. Finally, we obtained the mean-velocity residuals maps by subtracting the models from the observed velocity fields.

In Table \ref{tab:Baroloparameters} we list the average values of the disk model parameters derived with ${}^{\rm 3D}$BAROLO. The i and PA derived from the CO(2-1) kinematics are in good agreement with those derived at other wavelengths in all galaxies (Table \ref{tab:properties}) but in two cases (NGC 4253 and NGC 7465), which are discussed in their corresponding sections (see Sections \ref{subsec:ngc4253} and \ref{subsec:ngc7465}).

The models produced this way rely on the assumption that the majority of the cold molecular gas resides in an infinitely-thin axisymmetric disk orbiting in pure circular orbits.  
To validate this assumption,  we fitted the velocities derived in each ring along the major axis with ${}^{\rm 3D}$BAROLO to a parametric rotational curve \cite[e.g.,][]{salak2016gas,barbosa2009gemini} given by the following expression:

\begin{dmath}
    v{_{obs}}=v{_{sys}}+ \sqrt{\frac{R^{2} G M}{(R^{2}+A^{2})^{3/2}}} sin(i)
    \label{eq.rot}
\end{dmath}

\noindent where v$_{obs}$ is the observed velocity, v$_{sys}$ is the systemic velocity, R is the radius of the ring considered in the plane of the sky, G is the gravitational constant, M is the mass enclosed inside the A scale length and i is the inclination of the galaxy. In this case, the only free parameters were M and A and we used the curve\_fit function, which is part of the SciPy library, to do the fit. The resulting rotational curves are depicted in the position-velocity diagrams along the major axis for each galaxy.

\subsection{Characterization of the non-circular motions} \label{characterization}

To make an appropriate interpretation of the CO(2-1) velocity residuals and additional kinematic components, it is essential to know the host galaxy PA and inclination, the geometry of ionization cones or at least their projected sizes and orientations as well as the orientations of the stellar bars.  

As discussed in the Introduction, our Seyfert galaxies have optical and/or NIR IFU observations from the literature and there are good estimates of the PA and inclination of the host galaxy disks (see Table~\ref{tab:Bars} for the values and references). For the AGN ionization cones we used the information available in the literature or our own estimations from [OIII]$\lambda 5007$ images or/and [OIII]$\lambda 5007$/H${\alpha}$+[NII] excitation maps (see Table~\ref{tab:Bars}). We note that the [OIII] line has a relatively high ionization potential (35.12 eV) and for that reason is suitable to trace the ionization bicone \citep[e.g.,][]{chen2019discovery}. The excitation maps help distinguish between  emission coming from the AGN wind and that from  star formation \citep[e.g.,][]{ferruit2000hubble}. For each galaxy, we plotted the axis and edges of the ionization cones as yellow dotted and green dashed lines, respectively.

Primary bars regulate the distribution and kinematics of cold gas that resides in the galaxy plane not only on kpc-scales, where the gas tends to follow the so-called x$_{1}$ orbits, but also on scales as small as hundreds of parsecs from the nucleus, where usually x$_{2}$ orbits dominate. We  list in Table~\ref{tab:Bars} the PA (plotted for each galaxy with a brown solid line in its corresponding figure), and radius of the primary bar (and the literature references). Since we are studying the gas motions in the galaxy plane, a key factor is to know whether the region of interest is inside or outside the corotation region of the bar \citep[e.g.,][]{garcia1994gas}. A rough prediction of the corotation radius\footnote{It should be noted that the term "corotation radius" is a generally accepted abuse of language. The corotation region has a certain radial extent \citep[e.g.,][]{combes2004galaxies}.} (R$_{CR}$) of the main bar can be obtained from the empirical ratio R$_{CR}$/R$_{bar}$$\sim1.2\pm0.2$ found by \cite{athanassoula1992existence}, which is in fairly good agreement with recent calculations \citep[e.g.,][]{aguerri2015bar}. Given the values of R$_{bar}$  for the bars in our sample (see Table \ref{tab:Bars}), this criterion places the regions proved by our NOEMA observations well inside the corotation region of each respective main bar. 
Some of the galaxies in our sample also show evidence of a nuclear/secondary bar. This kind of bars is believed to help gas not to get trapped in the Inner Linblad Resonance (ILR) of the main bar, favouring the inflow motions in the innermost regions of the galaxy \citep{Shlosman1989accretion}. It is less straightforward to obtain an estimation of its corotation radius as this requires high angular resolution data. Thus, we will discuss this issue in the corresponding section of the galaxies where it has been proposed the existence of such bars (Mrk 1066 and NGC 4253).

By definition, the mean velocity field (first moment) gives flux density-weighted velocities. Thus, the \textit{mean} velocity of each spaxel (or spectral pixel) is biased towards the brighter velocity components of its spectrum. Consequently, the modeled and residual velocities are also biased. Thus, analyzing only the mean-velocity residual map may lead to misinterpreting the motions of the gas. To overcome this issue, we extracted spectra from regions which show either peculiar velocity residuals or/and high velocity dispersion (always >13 km s$^{-1}$, that is, our spectral resolution) to search for additional kinematic components that cannot be explained by circular rotation. We expect that if the bulk motion of the gas is rotating in the plane of the galaxy, the brighter component of the extracted spectrum has a centroid close to the modeled velocity. Other velocity components reveal departures from that motion and are related to local gas flows.

We preferentially looked for non-circular motions along the derived kinematic minor axis of the galaxy or its surroundings because the rotational component of the galaxy is zero, and so any non-zero kinematics are due to radial motions. Thus, we produced position-velocity diagrams along the derived kinematic minor axis as an additional tool to reveal non-circular motions. We also produced position-velocity diagrams along the derived kinematic major axis to study the gas rotation. The position-velocity diagrams were constructed with ${}^{\rm 3D}$BAROLO using the default slit width given by the tool, that is, a single pixel, which ranges from $7.1 \times 10^{-2}$ to $7.6 \times 10^{-2}$ arcsec in the case of our galaxies. 

In the absence of non-circular motions, the position-velocity diagrams along the minor axis are expected to show emission around the systemic velocity with a typical width of a few tens of km s$^{-1}$ due to the combined effect of the intrinsic cloud-cloud velocity dispersion and the beam smearing \citep[e.g.,][]{garcia2014molecular}. However, the effect of the beam smearing is known to be more important in the central regions where the velocity gradient is steeper \citep[e.g.,][]{davies2011how,tacconi2013phibss,federrath2017sami}. For this reason, we only consider departures from circular rotation in these regions as truly signatures of non-circular motions if we can confirm them with extracted spectra or they are >100 km s$^{-1}$ and have a clear positive/negative pattern in opposite sides of the galaxy. The latter value has been chosen because is twice the expected dispersion velocity for a pure rotating disk \citep{garcia2014molecular}.

Finally, for a correct interpretation of the non-circular motions, we need to know where the cold molecular gas is. Streaming motions due to spiral arms or bars are naturally associated with the plane of the galaxy. This may not necessarily be the case for outflows. In the best-studied case of a molecular outflow in a Seyfert galaxy, NGC 1068, there is clear evidence of a three-dimensional morphology (a component in the plane of the galaxy plus a vertical component). However, in this Seyfert, the bulk of the outflowing molecular gas is in the galaxy disk \citep[see][for more details]{garcia2014molecular,garcia2019}. Thus, based on the observational evidence (see also the Introduction), it is reasonable to assume that the cold molecular outflows in the nuclear regions of low and moderate luminosity Seyferts take place in the plane of the galaxy.

\begin{figure*}[!ht]
	\centering
	\includegraphics[width=\hsize]{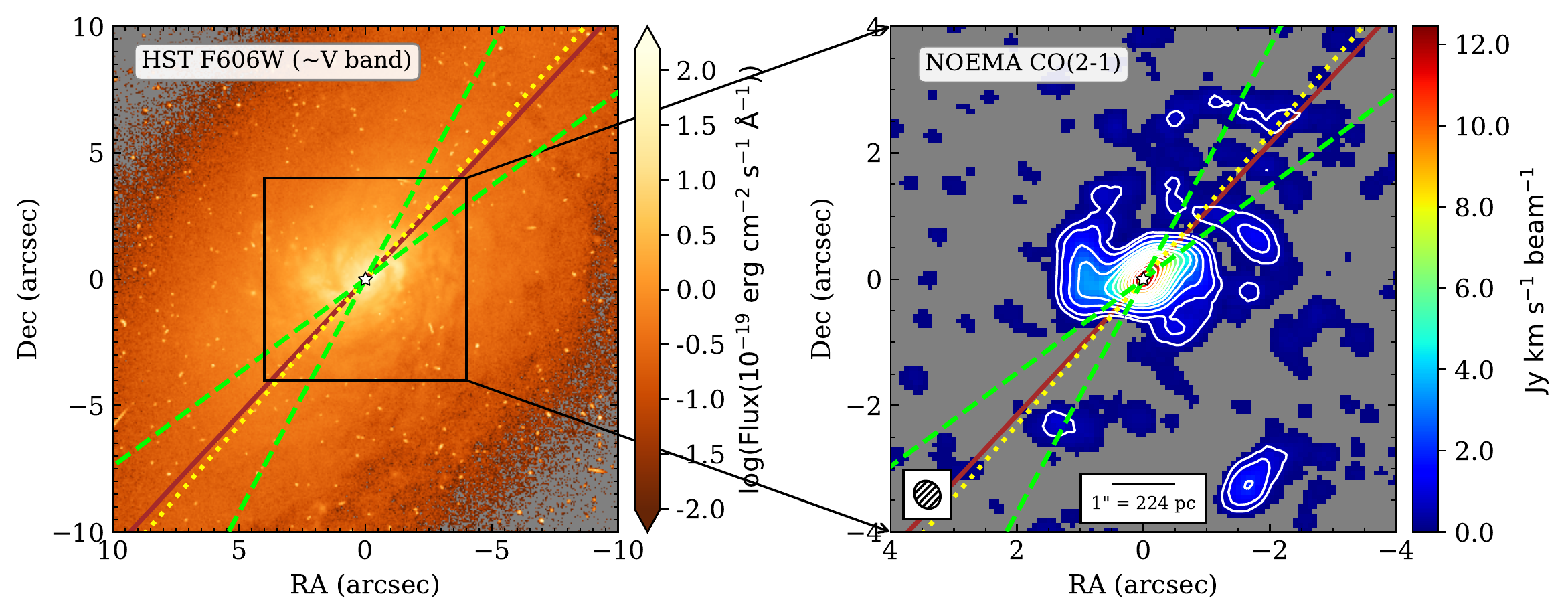}
	\caption{Mrk 1066. Left: HST/WFPC2 F606W ($\sim$ V band) image. Right: NOEMA CO(2-1) integrated intensity map (0th moment) with a 3-$\sigma$ clipping applied on each pixel for the channels with line emission in the continuum-free line cube (see $\S$\ref{methodology}). Masked pixels are shown in grey. The green dashed lines represent the edges of the ionization cone and the dashed yellow line the cone axis. The brown solid lines are the PA of the major axis of the main bars. The parameters and references for both the bicone and the main bar are listed in Table \ref{tab:Bars}. In both figures, north is up and east to the left. The hatched ellipse in the bottom left corner of the right panel is the beam size (see Table \ref{tab:obs}). The CO(2-1) integrated intensity contours are as follows: 0.5 ($\sim 4 \sigma$, $\sigma = 0.13$ Jy  km s$^{-1}$ beam$^{-1}$) and from 1 to 12 Jy km s$^{-1}$ beam$^{-1}$, in steps of 1.0 Jy km s$^{-1}$ beam$^{-1}$.}

	\label{fig:mrk1066_op_in}
\end{figure*}

\begin{figure*}[!ht]
	\centering
	\includegraphics[width=\hsize]{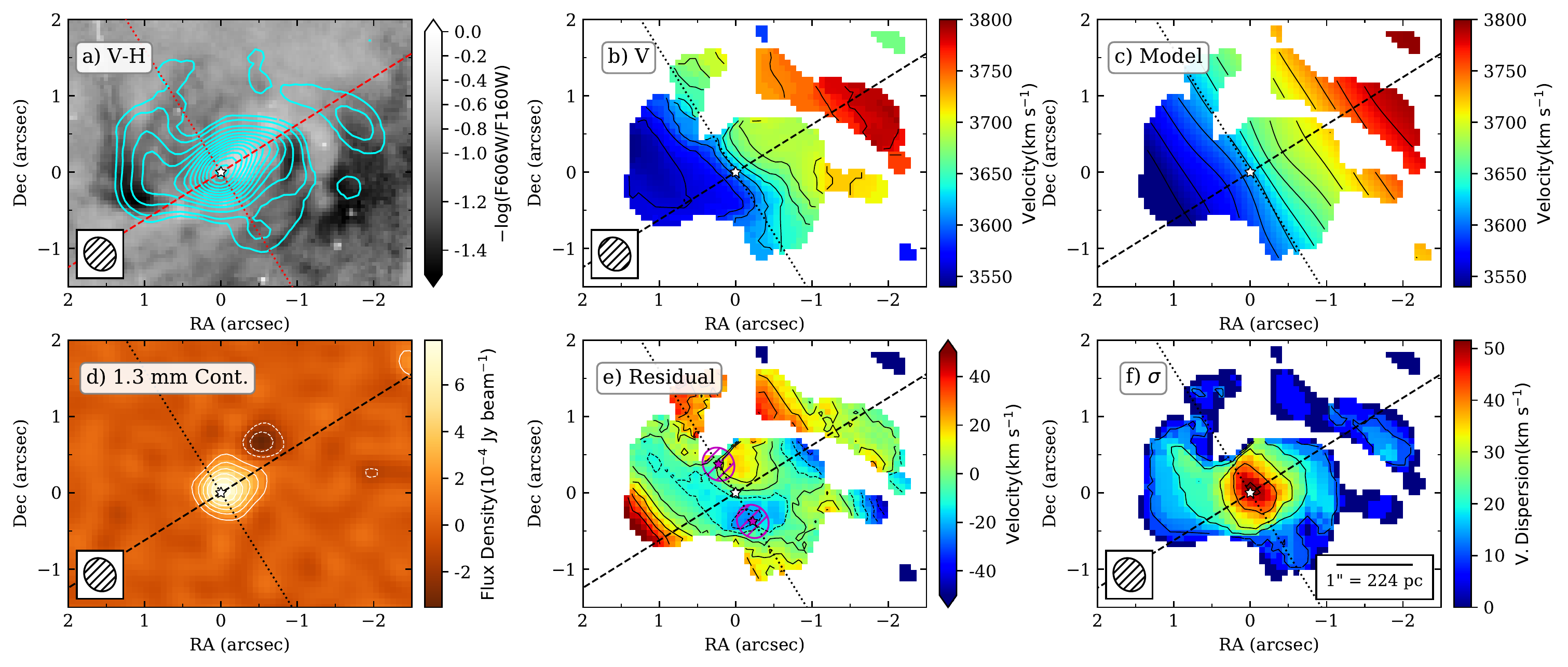}

	\caption{Mrk 1066.  Panel a: CO(2-1) integrated intensity contours (as in Fig. \ref{fig:mrk1066_op_in}) in blue overlaid on the $V-H$ color map. Light colors indicate low extinction and dark colors high extinction. Panel b: ${}^{\rm 3D}$BAROLO CO(2-1) observed velocity field. Panel c: ${}^{\rm 3D}$BAROLO velocity model. Velocity contours in b and c panels are from 3540 to 3800 km s$^{-1}$ in steps of 20 km s$^{-1}$. Panel d: the 1.3 mm continuum map with the contours at -6$\sigma$, -3$\sigma$ (white dashed line), 3$\sigma$, 6$\sigma$, 9$\sigma$, 12$\sigma$, 15$\sigma$ and 18$\sigma$ (white solid line) and  $\sigma$=41 $\mu$Jy beam$^{-1}$. Panel e: Residual velocity map. The contours are from $-50$ to 50 km s$^{-1}$ in steps of 10 km s$^{-1}$. The purple stars and ellipses mark those regions where we extracted  spectra (see Fig. \ref{fig:spectra_1066}). Panel f: ${}^{\rm 3D}$BAROLO CO(2-1) observed velocity dispersion  map. The  contours are in steps of 10 km s$^{-1}$, starting at 10 km s$^{-1}$. The ellipses in the bottom corners are the NOEMA synthesized beam (see Table \ref{tab:obs}) and red or black dashed (dotted) line is the major (minor) kinematic axis derived from the ${}^{\rm 3D}$BAROLO fit (see Table \ref{tab:Baroloparameters}). The white star marks the 1.3\,mm continuum peak.  }
	\label{fig:six_1066}
\end{figure*}

\begin{figure}

\centering

\begin{subfigure}[b]{0.45\textwidth}
    \centering
    \includegraphics[width=\textwidth]{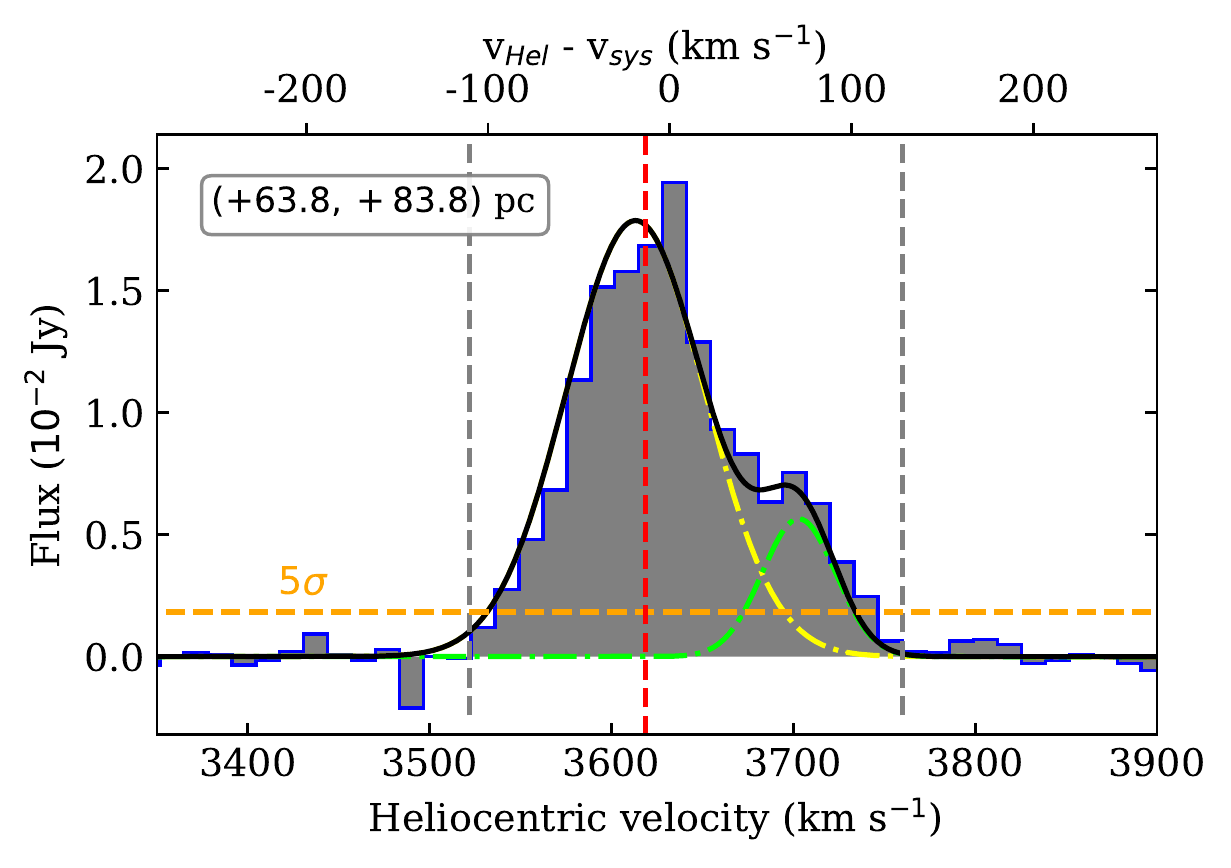}
\end{subfigure}
\begin{subfigure}[b]{0.45\textwidth}
    \centering
    \includegraphics[width=\textwidth]{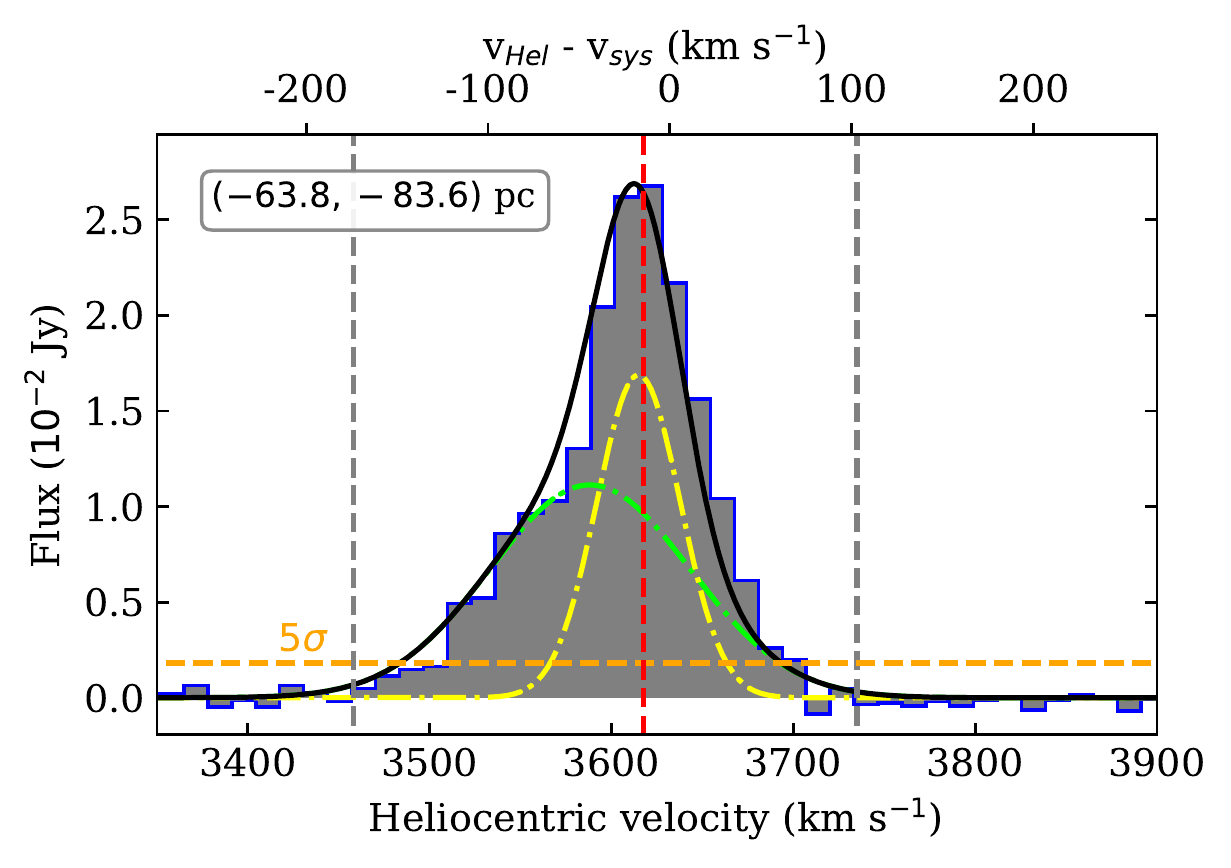}
\end{subfigure}

\caption{CO(2-1) spectra from the two nuclear regions of Mrk~1066 integrated over an area equal to the beam. The fitted Gaussians used to identify the different kinematic components are the color dash-dotted lines and their sum the black line. The horizontal orange dashed line is the 5$\sigma$ level, where $\sigma$ is the standard deviation in the channels outside of the spectral window (vertical grey dashed lines)}. We also show the modeled velocity (vertical red dashed line) for the rotating disk obtained with ${}^{\rm 3D}$BAROLO for the central pixel of the area (purple stars in Fig.~\ref{fig:six_1066}, panel e). Relative coordinates in pc from the AGN for the central pixel of the apertures are given in the top left boxes.

\label{fig:spectra_1066}
\end{figure}

\begin{figure}[ht!]
\centering
\begin{subfigure}[b]{0.46\textwidth}
    \centering
    \includegraphics[width=\textwidth]{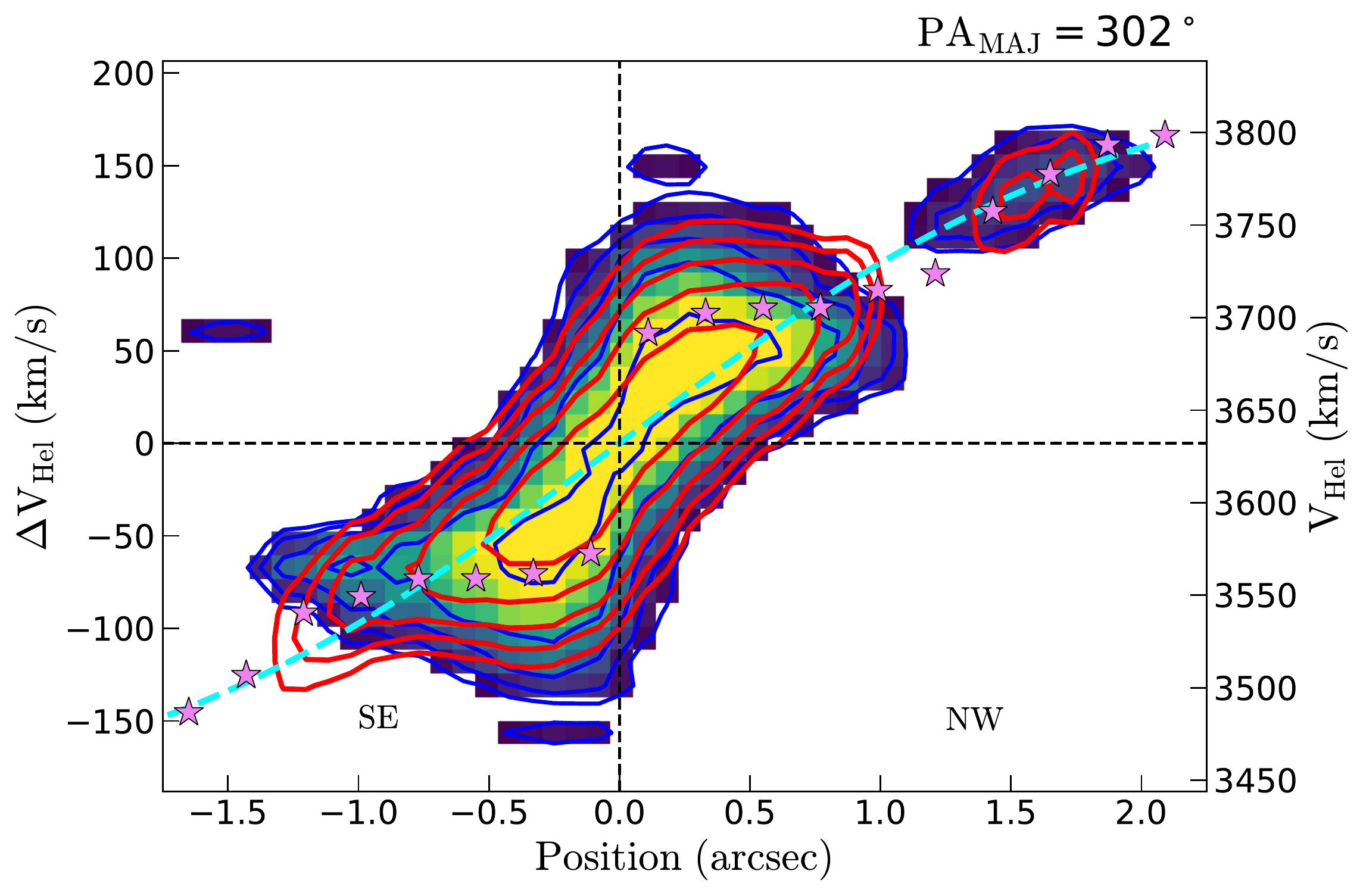}
\end{subfigure}\hfill
\begin{subfigure}[b]{0.46\textwidth}
    \centering
    \includegraphics[width=\textwidth]{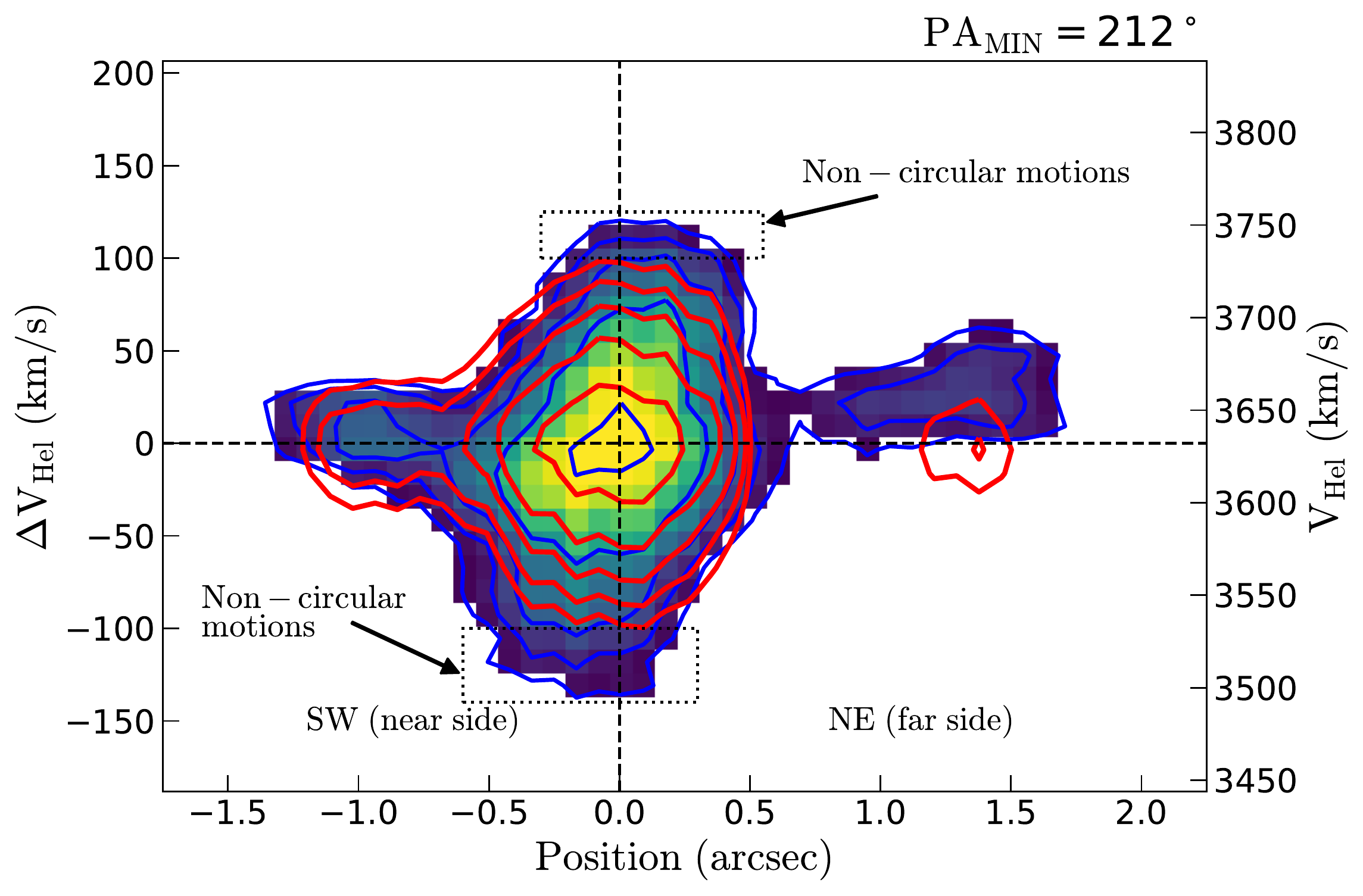}
\end{subfigure}\hfill

\caption{Position-Velocity diagrams taken along the kinematic major (PA$\sim$302$^{\circ}$) and minor (PA$\sim$212$^{\circ}$) axes of Mrk 1066 extracted with ${}^{\rm 3D}$BAROLO. The color images and the blue contours 1 ($3 \sigma$, see Table \ref{tab:obs}), 2, 4, 8 and 16 times 5.1 $\times 10^{-3}$ Jy beam$^{-1}$ represent the observed CO(2-1) line emission. Red contours depict the ${}^{\rm 3D}$BAROLO rotating disk model and their values are the same as the blue ones. The magenta stars represent the velocities in each ring of the model while the cyan dashed line is the best fit of these velocities to the parametric rotational curve given by Eq. \ref{eq.rot}. The black dotted lines (bottom panel) approximately enclose the emission which shows non-circular motions.} 
\label{fig:pv_1066}
\end{figure}

\section{Results for individual galaxies} \label{results}

\subsection{Mrk~1066 (UGC 02456)} \label{mrk1066}

This double barred galaxy harbors a Seyfert 2 nucleus \citep{goodrich1983mrk} and is classified as a Compton-thick candidate \citep{levenson2001seyfert,hernandez2015x}. The primary bar is oriented at a PA$\sim$ 143$^{\circ}$ (see Fig. \ref{fig:mrk1066_op_in}) and has a diameter of 31" ($\sim$7.0 kpc) \citep{mazzarella1993optical,afanasiev1998two}. In addition, it may have a nuclear bar oriented at PA=112$^{\circ}$ with a radius of 2.5" ($\sim$0.6 kpc) according to the works of \cite{afanasiev1998two} and \cite{riffel2017gemini}. Mrk~1066 shows signs of a recent minor merger \citep{gimeno2004catalog,smirnova2010seyfert}. \cite{bower1995radio} reported a jet-like structure extending $\sim$1".4 to the NW of the nucleus at PA=$315^{\circ}$ based on their HST [OIII]$\lambda$5007 image. The radio emission at 3.6, 6 and 21 cm shows a similar morphology \citep{ulvestad1989radio,nagar1999radio}.  

\subsubsection{Morphology}

The 1.3 mm continuum image (Fig. \ref{fig:six_1066}d) shows a bright source detected at a 18$\sigma$ level. There are also weak elongations along PA$\sim$300$^{\circ}$ and PA$\sim$350$^{\circ}$. However, it can be successfully fitted in the uv-plane with an elliptical Gaussian function of size 0".9 $\times$ 0".35 and PA=120$^{\circ}$ (see Table \ref{tab:uvfit}). The continuum source is thus oriented along approximately the major axis of the galaxy. The coordinates of the 1.3 mm continuum peak are coincident (within the errors) with those of the 3.6 and 20 cm continuum peaks \citep{nagar1999radio}.   

Most of the NOEMA cold molecular gas emission (right panel of Fig. \ref{fig:mrk1066_op_in}) in Mrk~1066 is concentrated within the nuclear region at $r<2"$ ($\la$450 pc) with a disk-like morphology, as observed at other wavelengths including the ionized gas \citep{riffel2011compact} and the $8.7\,\mu$m emission \citep{alonso2014nuclear,ramos2014mid}. There is also a spiral arm segment approximately 1" to the NW of the nucleus and more diffuse emission far away to the NW, and clouds to the SW and SE of the AGN within the inner 4" ($\sim$0.9\,kpc). The CO(2-1) emission peaks at the location of the AGN as traced with the 1.3 mm continuum peak. In the central region, the CO(2-1) emission shows a two-arm mini-spiral morphology starting from the edges of the disk-like structure which is also seen in  the $V-H$ color map \citep[see Fig.~\ref{fig:six_1066}, panel a, and also][]{martini2003circumnuclear}. The nuclear spiral morphology has also been  reported by \cite{riffel2011compact}. The concave sides of the arms show strong reddening, suggesting that the spiral trails. Finally, the more prominent dust lanes are in the SW indicating that this is the near side of the galaxy. 

The [OIII] emission is more prominent to the NW, suggesting that the SE side of the cone is obscured by the disk \citep{bower1995radio,fischer2013determining}.

\subsubsection{CO(2-1) Kinematics}

The CO(2-1) observed mean-velocity field of the inner region of Mrk 1066 derived with $^{\rm 3D}$BAROLO is shown in Fig. \ref{fig:six_1066}b. The general pattern of the isovelocities corresponds undoubtedly to rotation in the plane of the galaxy. This is similar to results for the warm molecular gas traced by the near-infrared (NIR) H$_2$ 2.12\,$\mu$m emission line \citep{riffel2011compact}. The central isovelocities present a characteristic S-shape, which is associated with oval structures. A similar feature is also present in the stellar velocity field \citep{riffel2011compact}. The  velocity dispersion map (Fig. \ref{fig:six_1066}f) presents the highest values (>30 km s$^{-1}$) close to the nucleus, and displays an oval shape elongated along PA$\sim$32$^{\circ}$. Starting from the central oval, there are two finger-like structures that approximately trace the spiral arms seen in the HST $V-H$ map. 

The $^{\rm 3D}$BAROLO velocity model of Mrk~1066 (panel c of Fig.~\ref{fig:six_1066},  parameters of the model are listed in Table \ref{tab:Baroloparameters})  fits reasonably well the observed velocity field. However, it does not reproduce the S-shape isovelocities and the higher velocities in the spiral arm segment. The CO(2-1) disk inclination and PA of the model are close to those obtained from the stellar component and warm molecular gas but not the systemic velocity, which is  $\sim50$ km s$^{-1}$ higher than that derived from the stellar component \citep{riffel2011compact,riffel2017gemini}. For comparison, the systemic velocity obtained for the HI component is 3610 km s$^{-1}$ \citep{mirabel1984neutral}, i.e., approximately in the middle of the molecular and stellar values. Furthermore, our value for the systemic velocity is close to that obtained for the strongest maser component reported by \cite{henkel2005new}.

As can be seen from Fig.~\ref{fig:six_1066} (panel e), the amplitude of the residuals is $\leq$40 km s$^{-1}$ ( [$(v_{obs}-v_{model})/sin(i)] \leq$ 52 km s$^{-1}$ deprojected). This small amplitude means that our model reproduces the bulk of the gas motions. Thus, we can conclude that the majority of the cold molecular gas traced by the CO(2-1) emission is rotating in the plane of the galaxy \citep[see also][]{riffel2011compact}. 
Apart from the deviations seen at the edge of the modeled FoV (likely artifacts due to the low SNR there), there are two regions $\sim$0".5 NE and SW from the AGN (marked with hatched circles in panel e of Fig.~\ref{fig:six_1066}) and along the kinematic minor axis with significant residuals. We extracted beam aperture CO(2-1) spectra and found that both show two velocity components
(see Fig.~\ref{fig:spectra_1066}). In both spectra, the centroid of one of the peaks is close to the velocity predicted by the ${}^{\rm 3D}$BAROLO model, suggesting that this component corresponds to the rotating gas in the disk of the galaxy. The other centroid is redshifted by $\sim 90\,{\rm km\,s}^{-1}$ in the spectrum extracted to the NE of the AGN and blueshifted by $\sim 30\,{\rm km\,s}^{-1}$ in the spectrum extracted to the SW. The presence of these non-circular motions can also be clearly seen in the position-velocity diagram taken along the kinematic minor axis, whereas 
the rotation pattern is well reproduced with the $^{\rm 3D}$BAROLO model along the kinematic major axis
(Fig.~\ref{fig:pv_1066}). 

\begin{figure*}[!ht]
	\centering
	\includegraphics[width=\hsize]{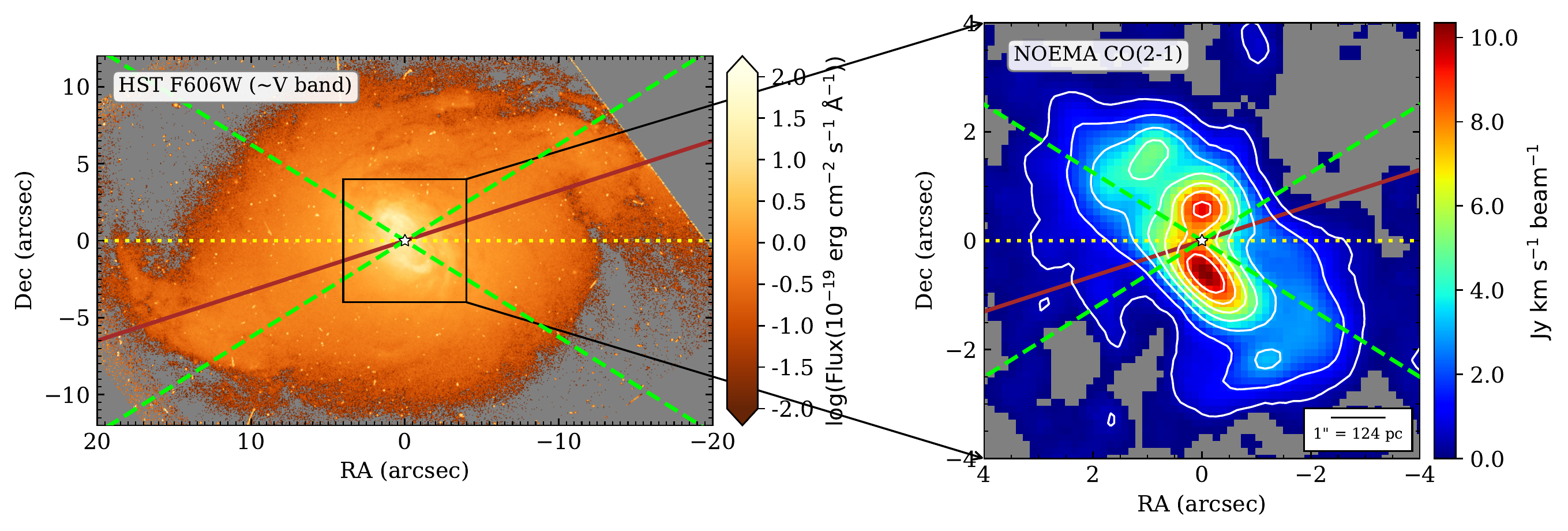}
	\caption{Same as Fig. \ref{fig:mrk1066_op_in} but for NGC~2273. CO(2-1) integrated-intensity contours: 0.5 ($5 \sigma$, $\sigma = 0.1$ Jy km s$^{-1}$ beam$^{-1}$) and from 1.5 to 9 Jy km s$^{-1}$ beam$^{-1}$, in steps of 1.5 Jy km s$^{-1}$ beam$^{-1}$. }
	\label{fig:ngc2273_op_in}
\end{figure*}

\begin{figure*}[!ht]
	\centering
	\includegraphics[width=\hsize]{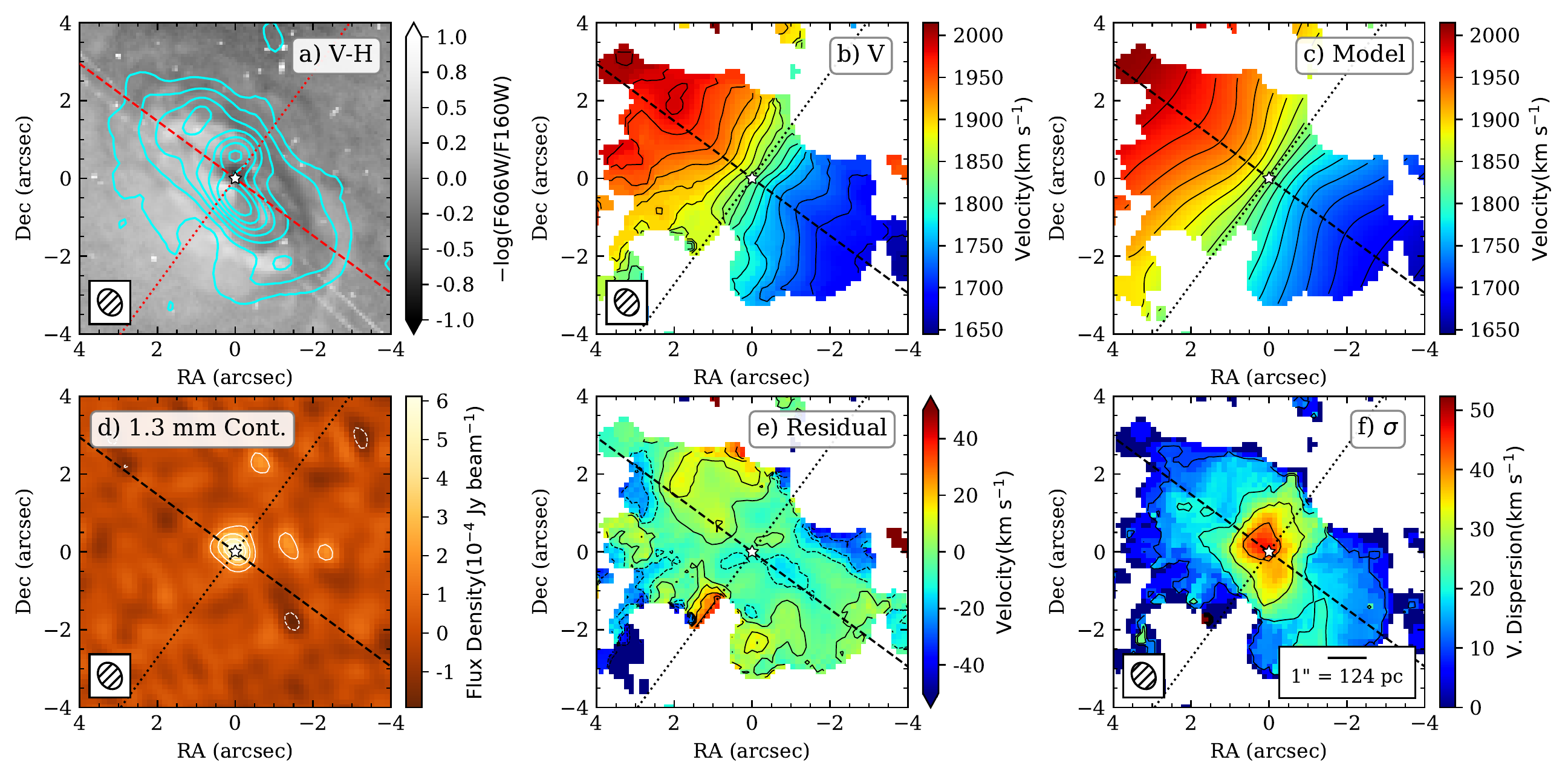}
	\caption{Same as Fig. \ref{fig:six_1066} but for NGC~2273. The velocity contours (b and c panels) are from 1645 to 2015 km s$^{-1}$ in steps of 20 km s$^{-1}$. The continuum contours are -3$\sigma$ (white dashed line), 3$\sigma$, 6$\sigma$ and 9$\sigma$ (solid lines) with $\sigma$=46 $\mu$Jy beam$^{-1}$. The residual velocity contours are from $-50$ to 50 km s$^{-1}$ in steps of 10 km s$^{-1}$. The velocity dispersion contours are in steps of 10 km s$^{-1}$ starting at 10 km s$^{-1}$.}
	\label{fig:six_2273}
\end{figure*}

\begin{figure}
\centering
\begin{subfigure}[b]{0.46\textwidth}
    \centering
    \includegraphics[width=\textwidth]{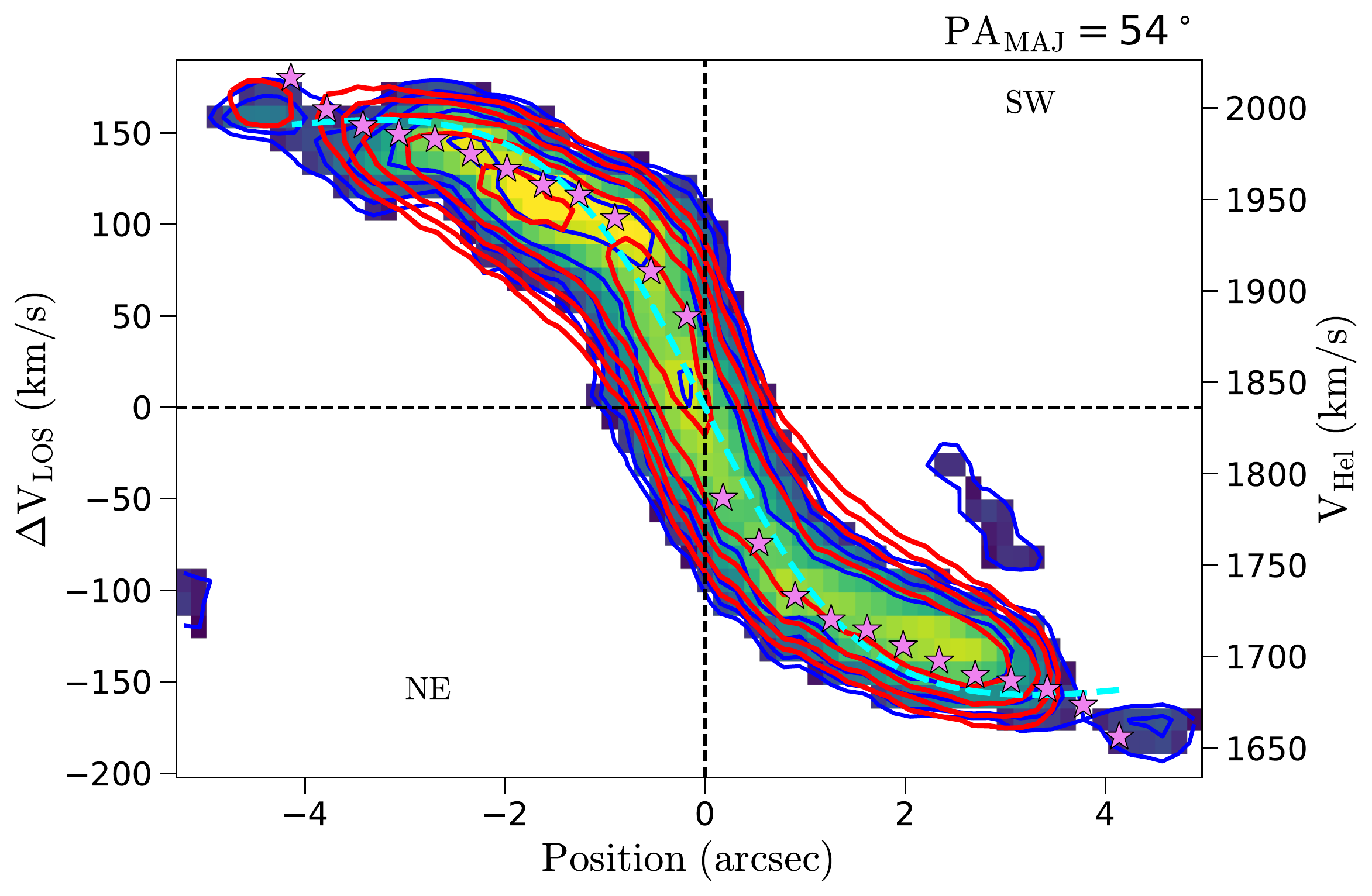}
\end{subfigure}\hfill
\begin{subfigure}[b]{0.46\textwidth}
    \centering
    \includegraphics[width=\textwidth]{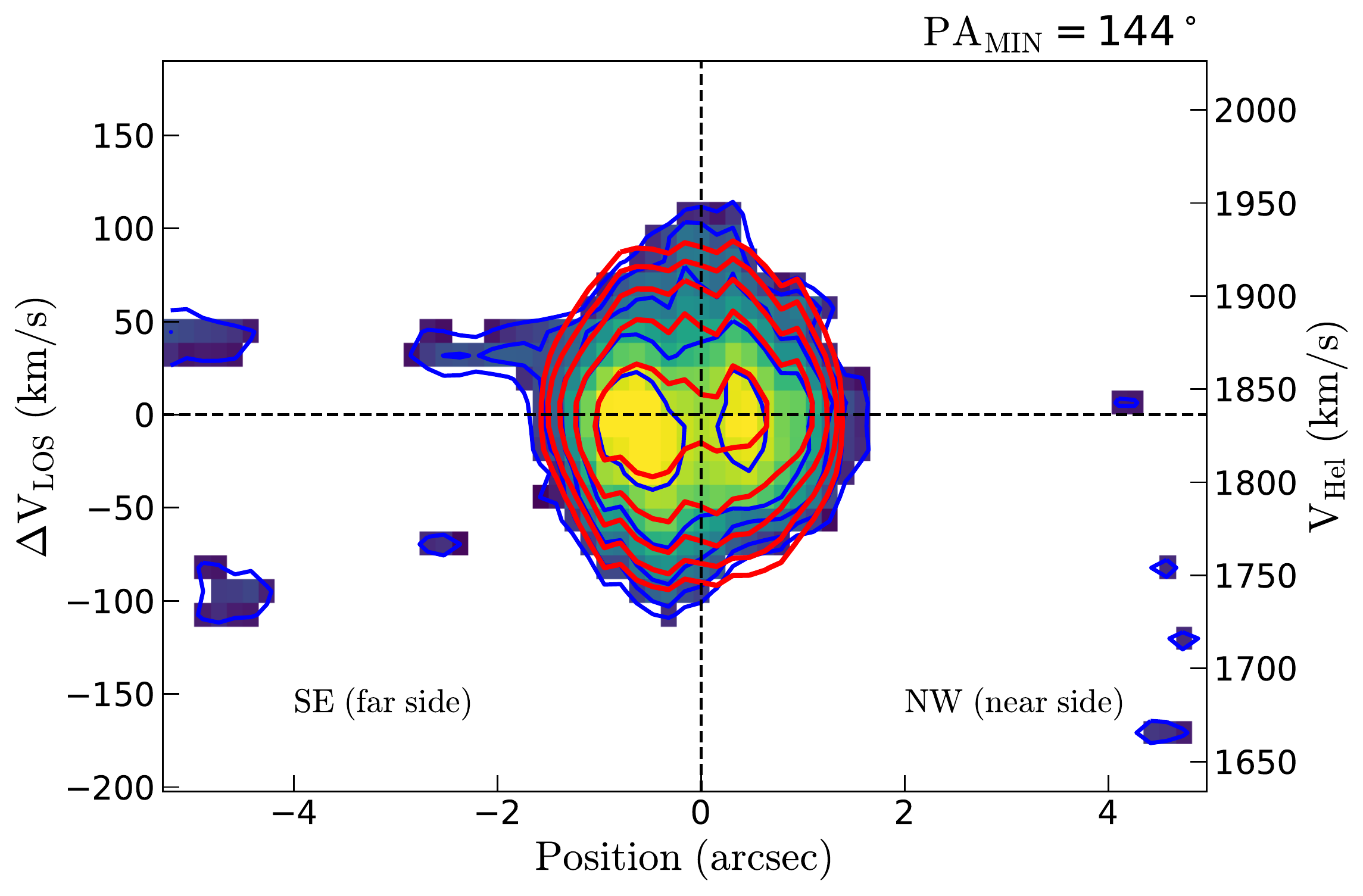}
\end{subfigure}\hfill

\caption{Position-Velocity diagrams taken along the kinematic major (PA$\sim$54$^{\circ}$) and minor (PA$\sim 144^{\circ}$) axes of NGC 2273 extracted with ${}^{\rm 3D}$BAROLO. Contours are 1 ($3 \sigma$, see Table \ref{tab:obs}), 2, 4, 8 and 16 times 4.2 $\times$ 10$^{-3}$ Jy beam$^{-1}$. Colors and lines as in Fig.~\ref{fig:pv_1066}.}
\label{fig:pv_2273}
\end{figure}

The amplitude of the non-circular velocity values of the cold molecular gas, observed along the minor axis, are similar to those measured in the nuclear/circumnuclear regions of other Seyfert galaxies \citep[see e.g.,][]{alonso2018resolving}. Assuming that the corotation radius of the nuclear bar is similar to its radius ($\sim$2".5), then  the non-circular motions are taking place within the corotation region of the nuclear bar. Then, if the non-circular motions are taking place in the disk of the galaxy, given its orientation, they are due to outflowing molecular gas.
 
In Fig.~\ref{fig:mrk1066_op_in} we overlaid the approximate orientation and projected size of the ionization bicone of Mrk~1066 modeled by \cite{fischer2013determining}. The bicone orientation (the angle between the plane of the sky and the end of the NW cone is i$_{NW \ cone}\sim$10$^{\circ}$) and opening angle ($\alpha=$15-25$^{\circ}$) would imply that the AGN wind does not intersect the galaxy disk (neither with their assumed inclination of the galaxy disk (i=54$^{\circ}$) nor with ours). Similarly, \cite{riffel2011compact} assumed that the bicone axis is close to the plane of the sky (i$_{NW \ cone}\gtrsim0^{\circ}$) and the projected opening angle is $\sim$ 20 to interpret their observations.  However, there is still room for a AGN wind-driven outflow if the base of the ionization cone crosses at least a small fraction of the galaxy disk. In this case, the AGN wind would be pushing the molecular gas radially outwards in the galaxy disk on scales of less than 100\,pc along the minor axis of the galaxy. This is similar to recent results found for the Seyfert galaxy NGC~3227 \citep{alonso2019nuclear}. 

\subsection{NGC 2273 (Mrk 620)} \label{ngc2273}

This Seyfert 2 galaxy \citep{contini1998starbursts} has a set of nested rings \citep{van1991study}. Its main bar has a radius of 27" and is oriented at PA=108$^{\circ}$ \citep[][see Fig. \ref{fig:ngc2273_op_in}, left panel]{moiseev2004structure}. The inner ring (r$\sim$2") is suspected to coincide with the Inner Linblad Resonance (ILR) of the main bar \citep{erwin2002double,erwin2003imaging}. At optical wavelengths, \cite{ferruit2000hubble} concluded that the regions $\sim 1".5$ to the north and south of the nucleus are likely star forming regions, based on the low values in their [OIII]$\lambda 5007$/H${\alpha}$+[NII] excitation map. NGC~2273 is classified as Compton-thick in X-rays  \citep[N$_{H}$ $\ge$ 10$^{24}$ cm$^{-2}$, ][]{comastri2004compton,awaki2009detection}. 

\subsubsection{Morphology}

The 1.3 mm continuum, shown in Fig. \ref{fig:six_2273}d, presents four distinct components. We associate the brightest with the AGN position.  This choice also maximizes the spatial correspondence between the CO(2-1), dust, and H${\alpha}$ morphologies (compare Fig.~\ref{fig:six_2273}a with the top panel of Fig.~7 of \citealt{ferruit2000hubble}). Moreover, the coordinates of 1.3\,mm continuum peak (Table~\ref{tab:uvfit}) are in good agreement with those derived from the 33 GHz (9.1 mm) continuum \citep{kamali2017radio}. The western components of the 1.3 mm continuum (termed W1 and W2 in Table \ref{tab:uvfit}) roughly agree with the extended radio component found in 3.6 and 20 cm observations \citep{nagar1999radio,mundell2009radio}. These components are aligned with the highly ionized gas which shows a jet-like structure extending $\sim2$" from the nucleus in the east direction. Finally, the north component (termed N in Table~\ref{tab:uvfit}) has no correspondence in either the 3.6 or 20 cm continuum images. 

\begin{figure*}[!ht]
	\centering
	\includegraphics[width=\hsize]{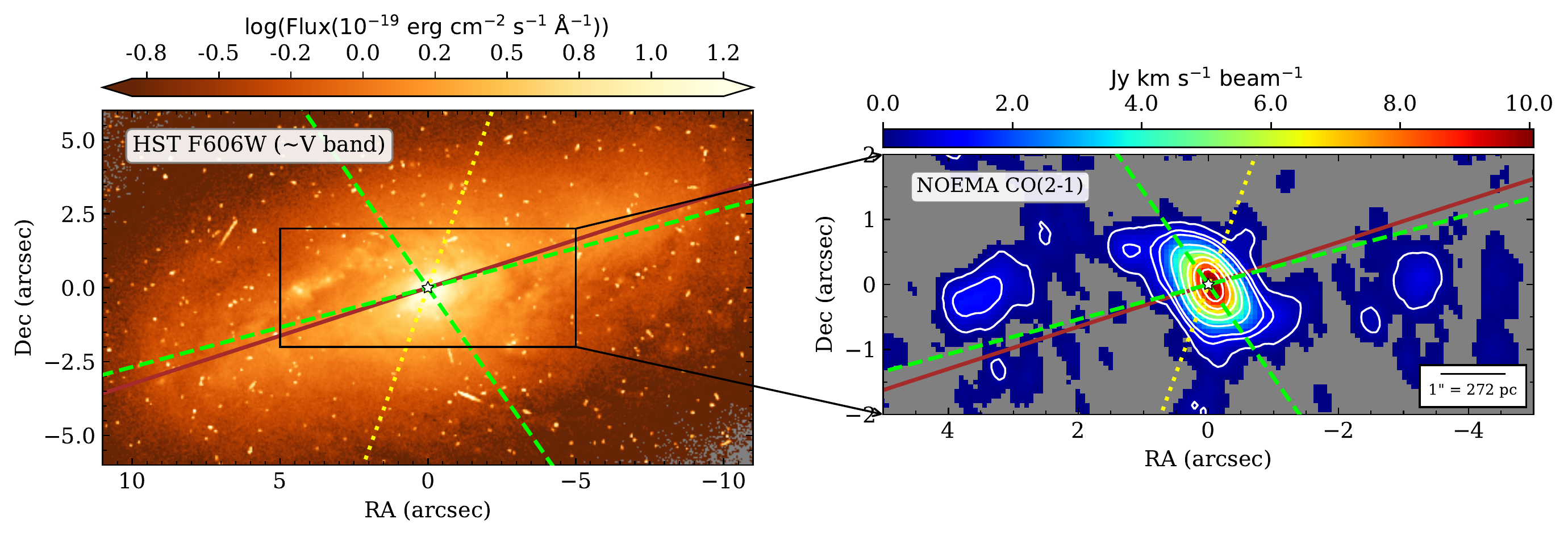}
	\caption{Same as Fig. \ref{fig:mrk1066_op_in} but for NGC~4253. CO(2-1) integrated-intensity contours: 0.3 ($5 \sigma$, $\sigma = 0.063$ Jy km s$^{-1}$ beam$^{-1}$), 1.0 and from 1.5 to 9 Jy km s$^{-1}$ beam$^{-1}$, in steps of 1.5 Jy km s$^{-1}$ beam$^{-1}$.}
	\label{fig:ngc4253_op_in}
\end{figure*}

\begin{figure*}[!ht]
	\centering
	\includegraphics[width=0.47\hsize]{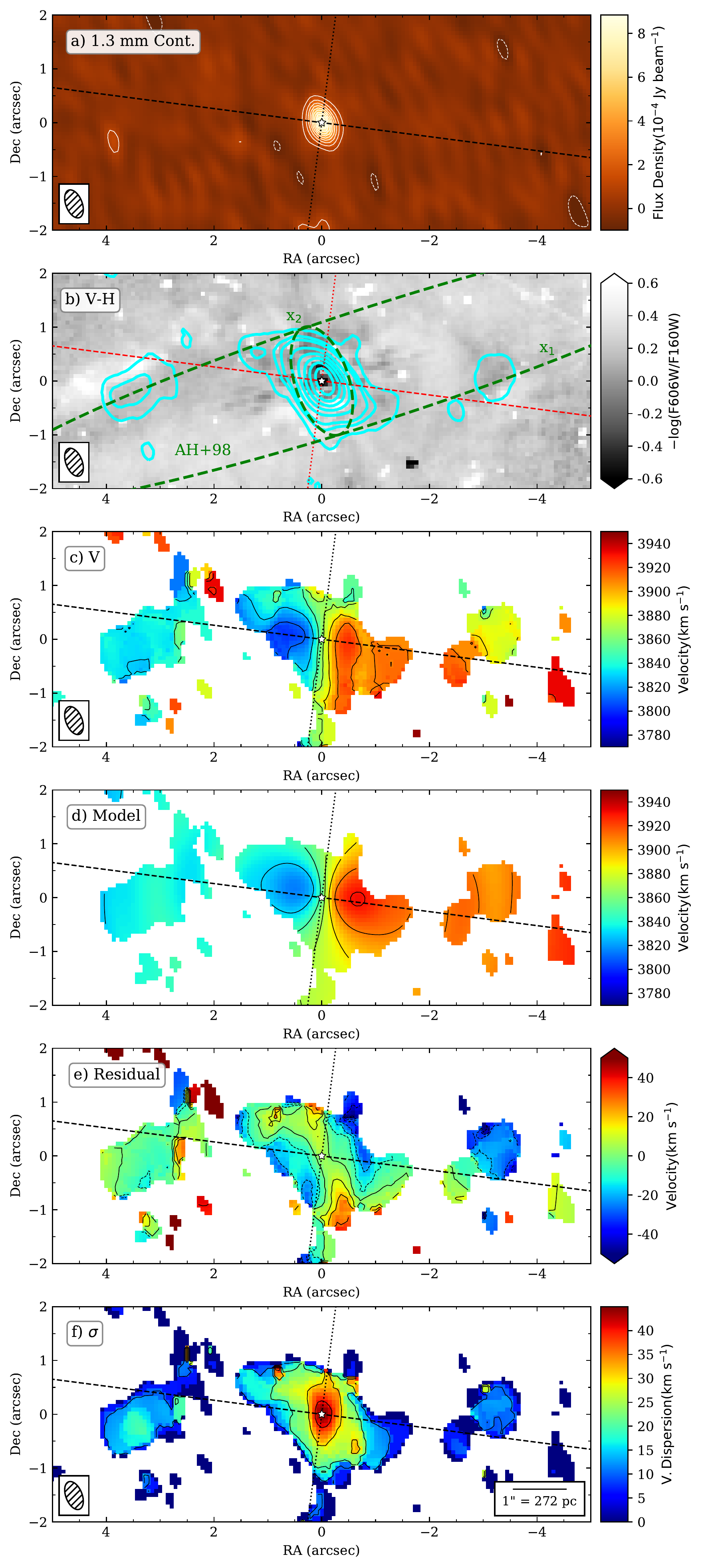}
	\caption{From top to bottom: the 1.3 mm continuum image, the CO(2-1) integrated intensity contours overlaid on the $V-H$ color map, the CO(2-1) observed mean velocity field, the $^{3D}$BAROLO velocity model, the residual velocity map (observed-model), and the CO(2-1) observed velocity dispersion map. The green dashed ellipses depicted in panel b are examples of x$_{1}$ and x$_{2}$ orbits induced by the main stellar bar assuming the parameters obtained by \cite{alonso1998near} and a bar width equals to that of the nuclear region (2".1). The integrated spectra extracted for this galaxy are shown in Fig. \ref{fig:spectra_4253_all}. The rest is as in Fig. \ref{fig:six_1066}. The continuum contours are -3$\sigma$ (white dashed line) and 3$\sigma$ to 33$\sigma$, in steps of 3$\sigma$ (white solid line), with $\sigma$=28 $\mu$Jy beam$^{-1}$. The velocity contours (b and c panels) are from 3770 to 3950 km s$^{-1}$ in steps of 20 km s$^{-1}$. The residual velocity contours are from $-40$ to 40 km s$^{-1}$ in steps of 10 km s$^{-1}$. The velocity dispersion  contours are in steps of 10 km s$^{-1}$, starting at 10 km s$^{-1}$.}
	\label{fig:six_4253}
\end{figure*}

\begin{figure}
\centering
\begin{subfigure}[b]{0.46\textwidth}
    \centering
    \includegraphics[width=\textwidth]{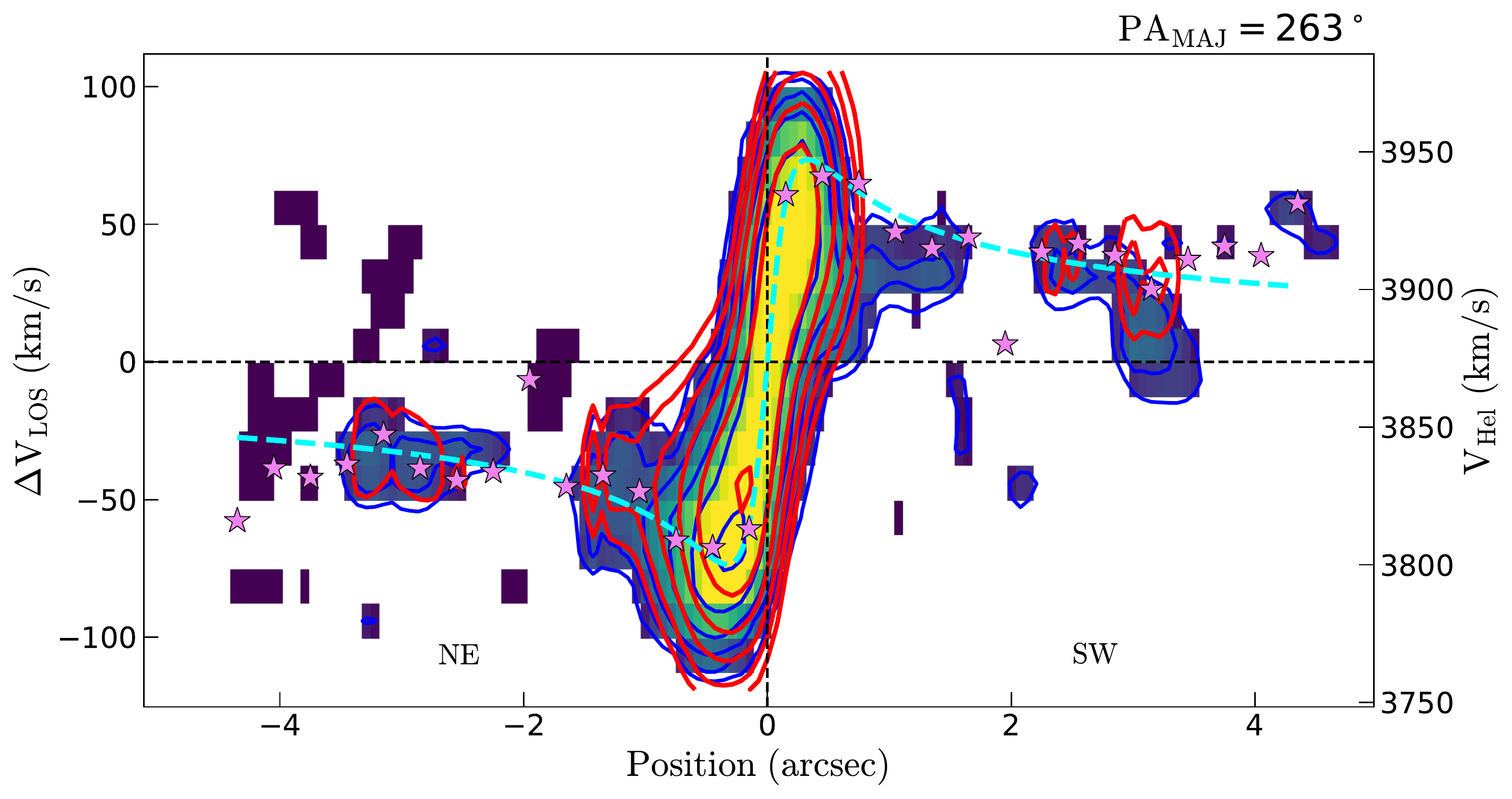}
\end{subfigure}\hfill
\begin{subfigure}[b]{0.46\textwidth}
    \centering
    \includegraphics[width=\textwidth]{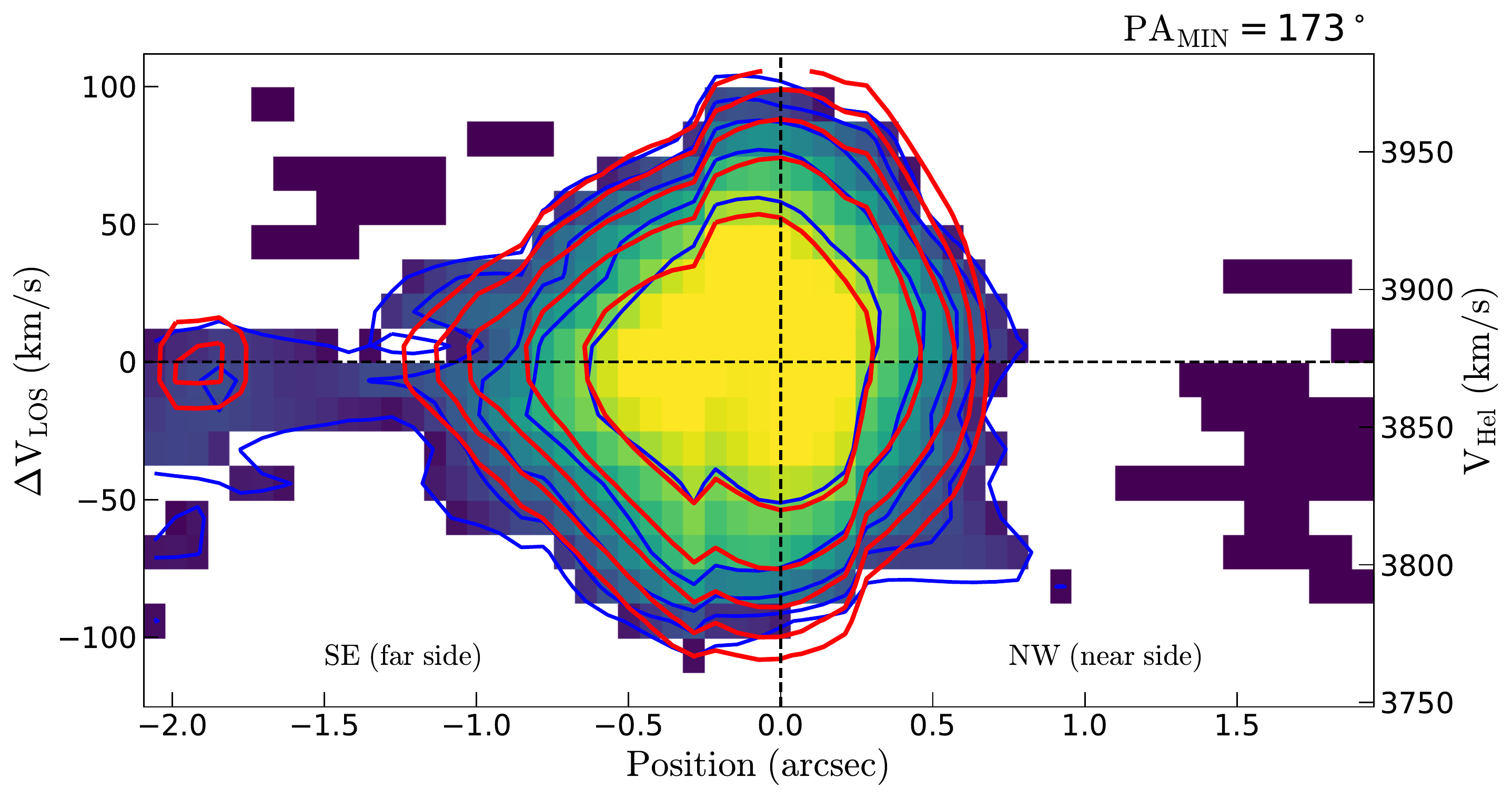}
\end{subfigure}\hfill

\caption{Position-Velocity diagrams taken along the kinematic major (PA$\sim$263$^{\circ}$) and minor (PA$\sim 173^{\circ}$) axes of NGC 4253 extracted with ${}^{\rm 3D}$BAROLO. Contours are 1 (3$\sigma$, see Table \ref{tab:obs}), 2, 4, 8, 16 and 32 times 3.0$\times$ 10$^{-3}$ Jy beam$^{-1}$. Colors and lines as in Fig.~\ref{fig:pv_1066}.} 
\label{fig:pv_4253}
\end{figure}

The CO(2-1) velocity-integrated intensity map (see Fig.~\ref{fig:ngc2273_op_in}) of NGC~2273 shows  emission along the major axis of the galaxy at PA$\sim$54$^{\circ}$ and extending over approximately 6" (0.7\,kpc). In the innermost regions there are two bright CO(2-1) peaks which do not coincide with the AGN position but are located approximately 0".5 ($\sim$60\,pc) to the north and south of the AGN. These peaks are nearly perpendicular to the main stellar bar axis. Such features, often called 'twin peaks' in the literature, have been found in other galaxies (e.g., NGC 3351, NGC 6951) and might be caused by the confluence of the inward gas transported by the bar and the gas in roughly circular orbits inside the Inner Linblad Resonance (ILR) of the bar \citep{kenney1994observations,van2011molecular}. \cite{petitpas2002molecular} proposed the existence of a nuclear bar based on their CO(1-0) observations. The CO(2-1) emission shows a fairly similar morphology. The two peaks seen in CO(2-1) are spatially coincident with the two arms seen in the V-H color map \citep[see Fig. \ref{fig:six_2273}a, and also][]{erwin2003imaging}. Then, if there is a nuclear bar, the peaks can be interpreted as molecular gas in the leading edges of this secondary bar. However, if this interpretation is correct, the motion of the gas inside the second bar will be trailing instead of leading as suggested by \cite{petitpas2002molecular}.

In the inner 4" ($\sim$0.5\,kpc) there are other fainter CO(2-1) emitting regions, especially to the NE, where there is a star-forming region seen in the H$\alpha$+[NII] map of \cite{ferruit2000hubble}, and to the SW, coinciding with the end of the nuclear mini-spiral arms. A similar morphology is observed at $8.7\,\mu$m \citep{alonso2014nuclear}, again suggesting that the CO(2-1) emitting regions are (dusty) star forming regions. By contrast, \cite{sani2012physical} found that the HCN(1-0) and HCO$^{+}$(1-0) high density gas tracers peak close to the 3\,mm continuum. This probably indicates that the density and/or the excitation conditions of molecular gas are different between the AGN and its immediately surrounding regions ($\sim 0".5$) where there might be some nuclear star formation activity  \cite[e.g., detection on nuclear PAH emission, see][and next section]{alonso2014nuclear}. 
 
The cold molecular gas is well correlated with the dust lanes traced by the $V-H$ color map (see Fig.~\ref{fig:six_2273}a). The majority of the dust is in the NW side, suggesting that it is the near side of the galaxy \citep[see also][]{barbosa2009gemini}. As for Mrk~1066, the arms seem to trail according to the dust distribution, and, therefore, the gas motion is anticlockwise. 

Excluding the emission to the NE, which is likely related to a star-forming region, the [OIII] emission inside the inner ring is seen mainly to east of the nucleus \citep[See Fig. 7 of][]{ferruit2000hubble}. This indicates that the approaching cone is projected on the eastern side and the receding one is obscured by the galaxy disk (see Fig. \ref{fig:ngc2273_op_in}).

\subsubsection{Kinematics}

The CO(2-1) observed mean-velocity field of NGC~2273 is shown in Fig. \ref{fig:six_2273}b. The isovelocity pattern is quite similar to a spider diagram, suggesting that the deviations from pure rotation of the gas motion are quite small. Interestingly, the channel maps of HCN(1-0) and HCO$^{+}$(1-0) of \cite[]{sani2012physical} (see their Fig. 11, upper panels) suggest that the molecular gas close to the nucleus has a velocity gradient almost opposite those of the CO(2-1) and the stellar velocity fields (see below). As our spatial and spectral resolutions are better than those of \cite{sani2012physical}, the divergence does not come from the different resolutions. We hypothesize that they could have observed a warp in the galaxy disk in denser molecular gas tracers. The existence of such warp has been suggested by \cite{barbosa2006gemini}. However, we note that the misalignment between the PA of the major axis of  CO(2-1) and stellar kinematics is significantly lower than that inferred from the H$_{2}$O maser distribution \citep[see Fig. 3 in][]{kuo2010megamaser}.

The corresponding observed velocity dispersion  map (Fig. \ref{fig:six_2273}f) shows the highest values ($\sim$ 50 km s$^{-1}$) near the 1.3 mm continuum peak. Apart from that, it shows moderately high values towards  the two peaks detected in the CO(2-1) integrated-intensity map.

The ${}^{\rm 3D}$BAROLO model reproduces well the observed CO(2-1) velocity field (panels b and c of Fig.~\ref{fig:six_2273}) as well as the position-velocity diagrams along the kinematic major and minor axes (see Fig.~\ref{fig:pv_2273}). Our derived disk parameters (see Table \ref{tab:Baroloparameters}) are in remarkable good agreement with those obtained from the stellar kinematics by \cite{barbosa2006gemini} and from HI gas by \cite{van1991study}. They are also similar to those reported by \cite{moiseev2004structure} from the NIR and optical surface brightness distributions. All this strongly suggests that the cold molecular gas follows the same kinematics as the stellar and neutral gas components. 

\cite{barbosa2009gemini} using optical IFU observations inferred the presence of ionized gas outflows mostly in the E-W direction launched at a small angle ($<40^{\circ}$) with respect to the host galaxy. Under this geometry we would expect some interaction between the AGN wind and the molecular gas in the host galaxy. We find, however, that the typical CO(2-1) velocity residuals (panel e of Fig.~\ref{fig:six_2273}) are only  $\sim 10-20$ km s$^{-1}$ in absolute value. This also supports that the bulk of the cold molecular gas is rotating in the galaxy disk. The most significant velocity residuals are close to the edge of the modeled FoV  near the minor axis of the galaxy. However, the extracted spectra in these regions did not show clear evidence for several kinematic components, in part due to the lower signal-to-noise ratios in these regions,  thus preventing us from any further analysis.

\subsection{NGC 4253 (Mrk 766)}\label{subsec:ngc4253}

NGC 4253 is a Seyfert 1.5 \citep{osterbrock1985spectra,osterbrock1993spectroscopic}. 
It shows a large scale bar at a PA$\sim 108^\circ$, which is asymmetric (7"+9" diameter) with respect to the nucleus \citep[][see Fig.~\ref{fig:ngc4253_op_in}]{alonso1998near, marquez1999near}, and possibly a nuclear bar with a radius of 1" at a position angle 5$^{\circ}$ \citep{marquez1999near}. \cite{mulchaey1996emission} obtained ground-based narrow-band images of the [OIII] and H$\alpha$ ionized gas emission of this galaxy. Their [OIII]/H$\alpha$ excitation map shows a biconical morphology, which is more apparent to the SE, with a PA=160$^\circ$ and a large opening angle of 110$^\circ$. \cite{nagar1999radio} derived a similar PA for the 
radio emission at 3.6 and 20\,cm.

\subsubsection{Morphology}

According to our fit of the 1.3 mm continuum (see Fig.~\ref{fig:six_4253} a and Table~\ref{tab:uvfit}), this source is only marginally resolved. The AGN coordinates obtained from 1.3 mm continuum with the {\tt UV$\_$FIT} task are in excellent agreement (differences of less than 0".1) with those derived from radio observations \citep{kukula1995high,thean2001merlin}. 

\begin{figure*}[h]
    \centering
    \includegraphics[width=\textwidth]{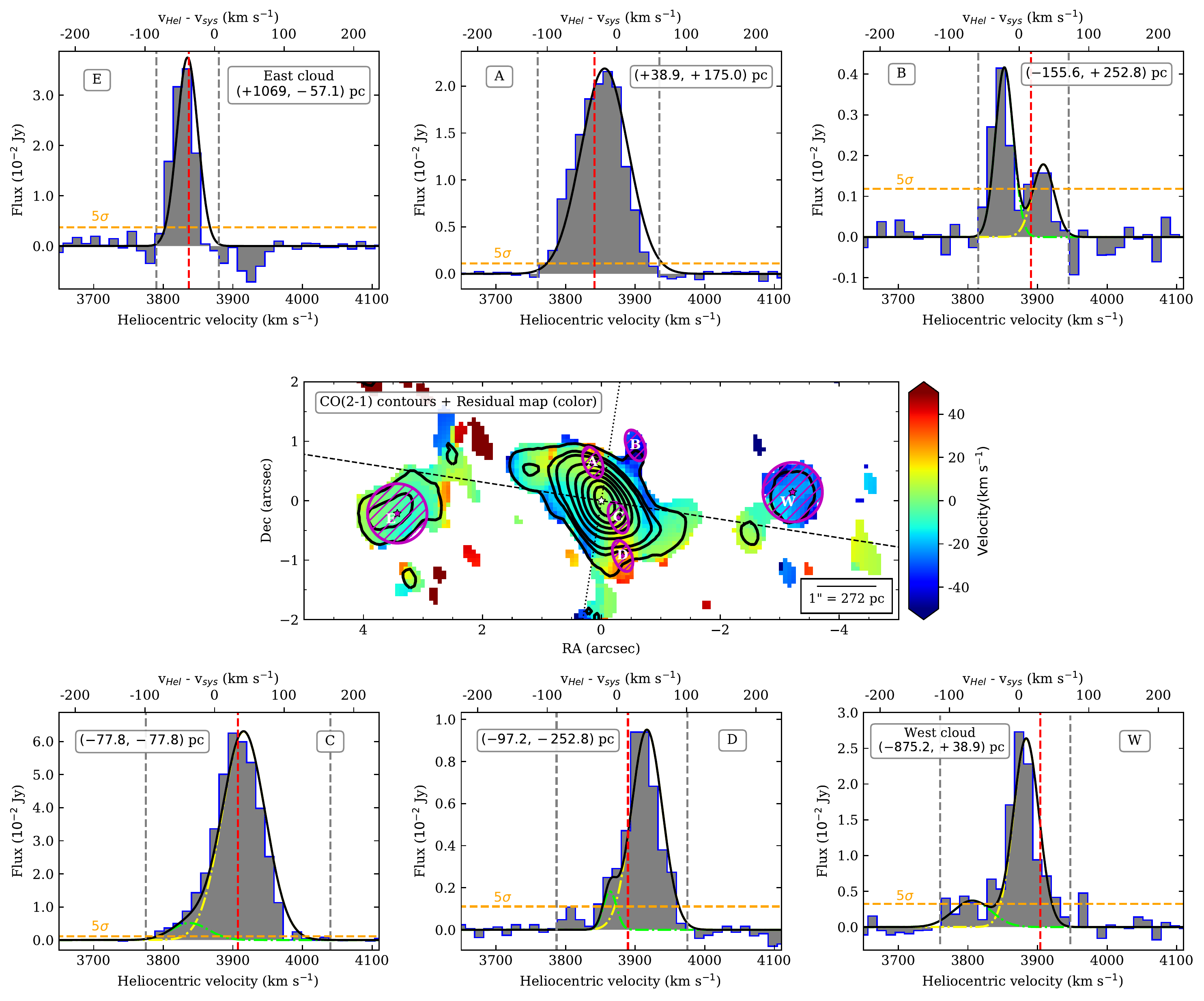}
\caption{Top and bottom panels: CO(2-1) extracted spectra of NGC~4253 (distances from the AGN are given in the top left/right boxes of each panel). We extracted the spectra with an aperture equal to the beam size except for the first and last spectra which are integrated on a circular aperture of radius 0".5. Middle: CO(2-1) integrated intensity contours overlaid on the mean-velocity residual map. Lines of the Spectra are as in Fig.~\ref{fig:spectra_1066} and CO(2-1) integrated intensity contours are as in Fig.\ref{fig:ngc4253_op_in}.} 
\label{fig:spectra_4253_all}
\end{figure*}

The only previous CO detection in this galaxy is that of the CO(1-0) single-dish observations \citep[][see also \citealt{taniguchi1990circumnuclear,vila1998co} ]{Maiolino1997molecular}. Thus, we present in Fig.~\ref{fig:ngc4253_op_in}, right panel, the first high-resolution observations of CO(2-1) emission in NGC 4253. The CO(2-1) emission extends for almost 8"$\simeq$2.2\,kpc but most of the emission is concentrated in the inner  2"$\simeq$0.5\,kpc and it peaks at the AGN position. This centrally peaked emission has a disk-like morphology oriented at an approximate PA of 30$^\circ$ and is similar to that traced by the NIR H$_2$ $2.12\,\mu$m line \citep{schonell2014feeding}. The fainter extended CO(2-1) emission is observed as two distinct regions (hereafter east and west clouds, respectively) located almost symmetrically $\sim 3".2$ ($\sim$0.8\,kpc) to the east and west of the AGN and are coincident with dust lanes observed in the $V-H$ color map presented by \cite{martini2003circumnuclear} (see also Fig. \ref{fig:six_4253}b). 

Unfortunately, the HST images for this galaxy are saturated at the AGN position and thus it is difficult to interpret them near the nucleus of the galaxy \citep[see  Fig. \ref{fig:six_4253}b, and also][]{malkan1998hubble,martini2003circumnuclear}. The large scale $V-H$ map shows redder colors to the north, suggesting that this is the near side of the galaxy. The concave sides of the leading edges of the main bar (see below) are redder than the convex ones, suggesting that the gas motion is clockwise. 

As mentioned above, \cite{alonso1998near} modeled the main bar of NGC 4253 (see Table \ref{tab:Bars}). We show am example of an x$_{1}$ orbit of their model in Fig.~\ref{fig:six_4253}b with a green dashed line. To do that, we  assumed that the width of the bar is that of the nuclear region (2".1). Assuming that the motion of the gas is clockwise in the plane of the galaxy, this figure clearly shows that the east and west clouds are located right at the leading edges of the main bar. This behavior has been found in many other galaxies \citep[e.g.,][]{sheth2002molecular}. The assumption that the cold molecular gas distribution is mainly the result of the main stellar bar is also supported by the location of star-forming regions, traced by brigth knots in the near-UV \citep{munoz2009nature}, in the leading side of the east cloud \citep[e.g.,][]{sheth2002molecular,heller1994fueling}. This cloud is also spatially coincident with H$_{\alpha}$ emission \citep{gonzalez1996spectrophotometric}. On the contrary, no significant H${\alpha}$ nor near-UV emission is seen in the location of the west cloud.

The main bar of NGC~4253 is relatively strong since the dust lanes \citep[][see also Fig. \ref{fig:six_4253}b]{martini2003circumnuclear} along it are nearly straight \citep{athanassoula1992existence}. This kind of bars is supposed to be more efficient in transporting gas towards nuclear regions \citep[e.g.,][]{kuno2008nobeyama}. In fact, the majority of the CO(2-1) emission concentrated in the nuclear region (r$\sim1"$). Additionally, it shows strong H$_{\alpha}$ emission and its the brightest region in the near-UV \citep{munoz2009nature,gonzalez1996spectrophotometric}, undoubtedly tracing nuclear star formation.

We also show an example of an x2 orbit (PA=18$^{\circ}$) in Fig.~\ref{fig:six_4253}b, which clearly suggests that the CO concentration in this region is, at least in part, the result of the gas motion in such orbits in the main-bar potential. If it exists, the secondary, nested bar \citep[][]{marquez1999near} could also have favoured the concentration of the gas in the nuclear region.

The [OIII] distribution in NGC 4253 is quite concentrated around the nucleus, showing a nearly circular shape of 1".9 diameter in both ground-based and HST observations \citep{mulchaey1996emission,schmitt2003hubble}. However, the excitation map presented by \cite{mulchaey1996emission} suggests that this galaxy has a ionization bicone with a large projected opening angle (see Table \ref{tab:Bars}). The cone to the SE of the nucleus is the most prominent, suggesting that is the approaching one.

\subsubsection{Kinematics}

The observed CO(2-1) mean-velocity field  (Fig.~\ref{fig:six_4253}, panel c) shows a general rotating pattern but it is quite flat due to the low inclination of the galaxy. The shapes of the isovels are different in the receding side, where they have a rigid-body like shape, and in the approaching one, where they resemble to that of a 'spider diagram'.   

As with Mrk~1066 (Section~\ref{mrk1066}), the CO(2-1) observed mean-velocity dispersion map presents the highest values in an oval structure centered at the 1.3 mm continuum peak. It also shows relatively high values in two finger-like structures, which begin from the ends of the oval structure at $\sim$0".8 from the nucleus and extend $\sim$0".6 to the east and west, respectively. The two clouds to the east and west also show relatively high velocity dispersions, especially the east one. Such enhancements could be due to streaming motions in the leading edges of the main bar.

The fitted ${}^{\rm 3D}$BAROLO model reproduces the overall rotating pattern. Our systemic velocity is compatible with most of the values reported in the literature \citep[e.g.,][]{gonzalez1996spectrophotometric,schnorr2014gas}. The derived inclination from the cold molecular gas kinematics ($31^{\circ}$) is similar to that derived from the NIR Pa$\beta$ and H$_2$ 2.12\,$\mu$m emission lines  \citep{schonell2014feeding} and slightly higher than that derived from the modeling of the stellar kinematics \citep[18.2$^{\circ}$, ][]{riffel2017gemini}. Our derived PA is almost the same as those obtained by \cite{schonell2014feeding} but 13$^{\circ}$ larger than the stellar one. It should be noticed that the disk of NGC 4253 has a low inclination and, thus, the fitting is more sensitive to small variations, which can explain these deviations.

\begin{figure*}[!ht]
	\centering
	\includegraphics[width=\hsize]{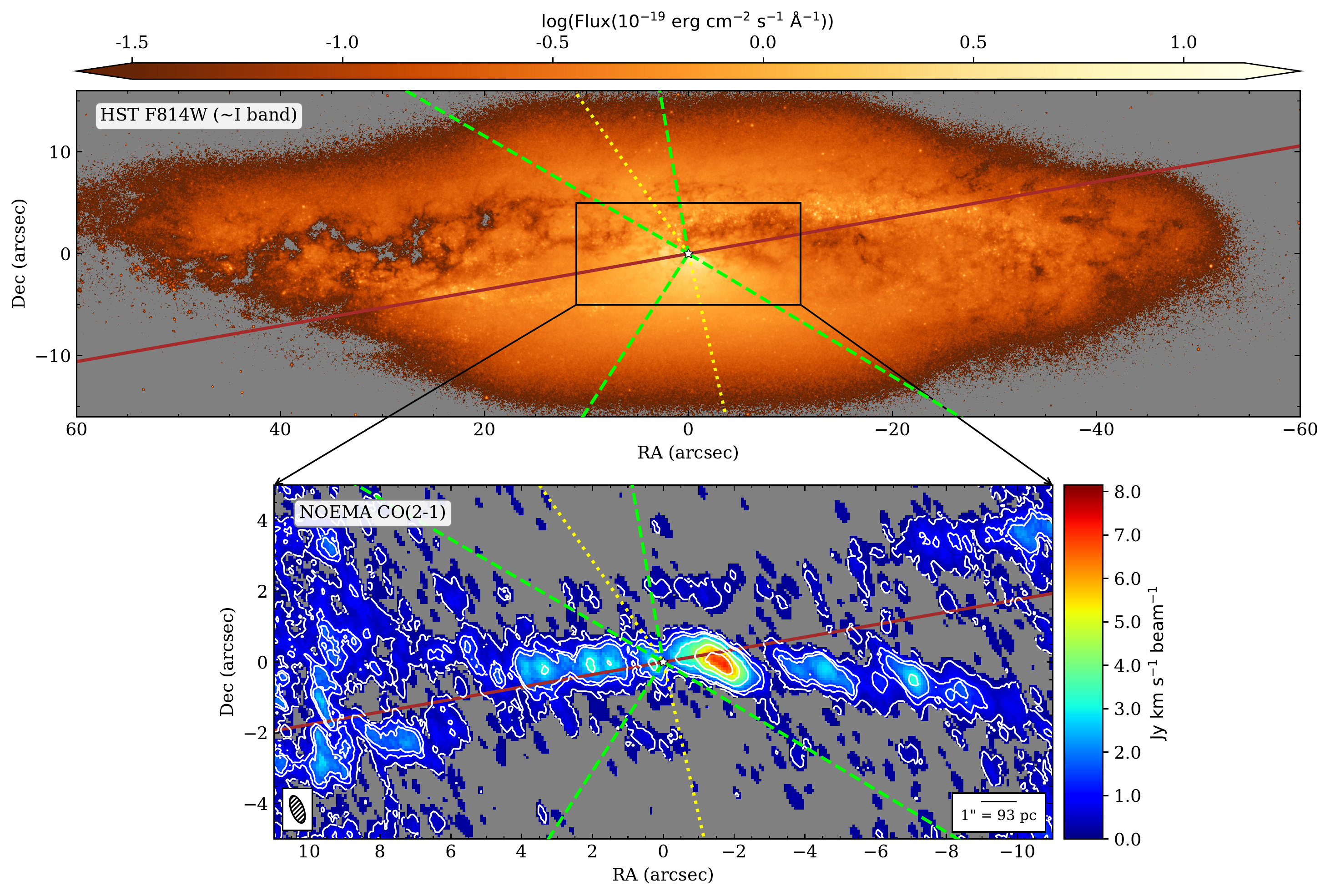}
	\caption{NGC 4388. Top: HST/WFC3/UVIS F814W ($\sim$ I band) image. Bottom: NOEMA CO(2-1) integrated intensity map (0th moment). CO(2-1) integrated intensity contours: $\sigma = 0.28$ Jy km s$^{-1}$ beam$^{-1}$, 1.0 ($\sim 3.6 \sigma$) and from 1.5 to 7.5 Jy km s$^{-1}$ beam$^{-1}$, in steps of 1.5 Jy km s$^{-1}$ beam$^{-1}$. The star marks the assumed location of the AGN, which in this case corresponds to the 21-cm continuum peak from VLBI observations \citep{giroletti2009faintest} since this is the only source in which we did not detect the 1.3 mm continuum (see text for futher details). Lines and other details as in Fig. \ref{fig:mrk1066_op_in}.}
	\label{fig:ngc4388_op_in}
\end{figure*}

\begin{figure*}
	\centering
	\includegraphics[width=0.33\hsize]{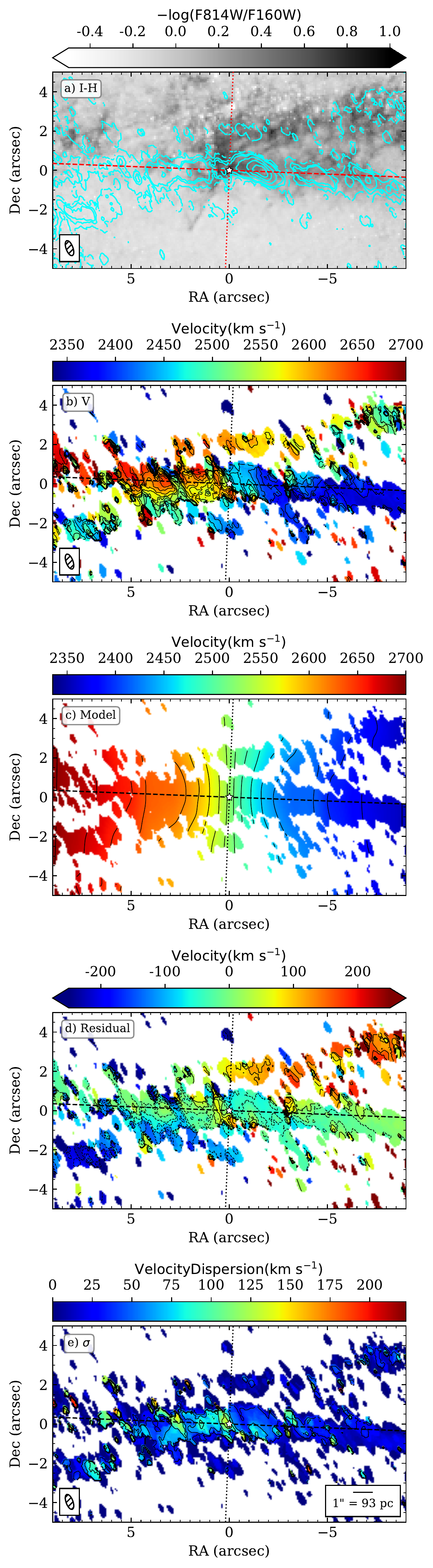}
	\caption{NGC 4388. From top to bottom: CO(2-1) integrated intensity contours overlaid on the $I-H$ color map, CO(2-1) observed mean velocity field, velocity model, residual velocity map (observed-model), and CO(2-1) observed velocity dispersion  map. The white star shows the adopted AGN position. The rest is as in Fig. \ref{fig:six_1066}. Contours: CO(2-1) integrated intensity contours are the same as in Fig. \ref{fig:mrk1066_op_in}. The velocity contours (b and c panels) are from 2335 to 2700 km s$^{-1}$ in steps of 20 km s$^{-1}$. The residual velocity contours are from $-200$ to 200 km s$^{-1}$ in steps of 30 km s$^{-1}$. The velocity dispersion contours are in steps of 30 km s$^{-1}$, starting at 30 km s$^{-1}$.}
	\label{fig:five_4388}
\end{figure*}

Although the velocity residuals (Fig.~\ref{fig:six_4253}{, see also the middle panel of Fig. \ref{fig:spectra_4253_all}}) are not strong ($\leq$ 40 km s$^{-1}$ in projection or [$(v_{obs}-v_{model})/sin(i)] \leq$78 km s$^{-1}$ deprojected, both in absolute value), they show an S-pattern that is not antisymmetric with respect to the nucleus. Such features might be related to a bad fit of the model \citep[e.g.,][]{van1978kinematics}. However, we allowed all our five parameters to vary in the first run with $^{3D}$BAROLO, we tested different input values and we obtained similar values. Moreover, this pattern is unlikely the result of an offset between our assumed AGN location and the real one \citep[see Fig.1 in][]{van1978kinematics} since it is in excellent agreement with that of the radio observations (see above). On the other hand, the S-pattern appears to be related to the leading edges of the main bar. Theoretical and observational arguments show that the isovels tend to warp along the major axis of the bar \citep[e.g.,][and references therein]{huntley1977gas,buta1987structure,mazzalay2014molecular}.

The position-velocity diagram along the minor axis (Fig.~\ref{fig:pv_4253}) shows some gradient in velocity in the inner $\sim 2"$, which is more evident in the blueshifted non-circular motions to the SE. In other words, the emission to the SE along the minor axis shows more negative velocities than those predicted by the ${}^{\rm 3D}$BAROLO model. Given the orientation of the galaxy and that these regions are within the nuclear bar corotation\footnote{We assumed, as in Mrk 1066, that the corotation radius of the secondary bar is similar to its radius.}, the non-circular motions are compatible with an inflow. 

To illustrate the complexity of the molecular gas kinematics in the nuclear region of NGC~4253, we extracted spectra of selected regions  close to the minor axis (regions A-D in Fig.~\ref{fig:spectra_4253_all}). The A spectrum is an example of the findings in the majority of these regions. Although this region displays a relatively high velocity dispersion ($\sim$ 30 km s$^{-1}$), and  the velocity residuals ($\sim$15 km s$^{-1}$) are close to our resolution ($\sim$13 km s$^{-1}$),  we only see one velocity component. However, the line profile is slightly asymmetric with an excess towards blue velocities. On the other hand, the B-D regions (see Fig.~\ref{fig:spectra_4253_all}) clearly show two velocity components. Assuming that the bright component is associated with the rotating disk, the second velocity component is redshifted ($+60\,{\rm km\,s}^{-1}$) to the NW (B area) and blueshifted (on average $-60\,{\rm km\,s}^{-1}$) to the SW (C and D areas) of the AGN. We note that these components are shifted in the opposite sense suggested by the residual map, which points out to the importance of interpreting the residual map along with extracted spectra (see Section \ref{characterization}).

\cite{schonell2014feeding} modeled the Pa$\beta$ and warm H$_{2}$ nuclear kinematics and found clear redshifted non-circular motions to the south and SE of the AGN in their residual maps, which are coincident with the position of the radio jet. In their channel maps, they found both blue and redshifted velocities in [Fe II], Pa$\beta$ and warm H$_{2}$, reaching $\pm$250 km s$^{-1}$ in the [Fe II] line at approximately 0.5" SE of the AGN. They interpreted the high velocity ionized gas to the SE (PA$\sim$135$^{\circ}$) of the AGN as evidence of an AGN-driven ionized outflow at an orientation close to the plane of the galaxy. However, as we saw in the position-velocity diagram along the kinematic minor axis, in this region we only found evidence of inflowing motions of the molecular gas.

\begin{figure}
\centering
\begin{subfigure}[b]{0.46\textwidth}
    \centering
    \includegraphics[width=\textwidth]{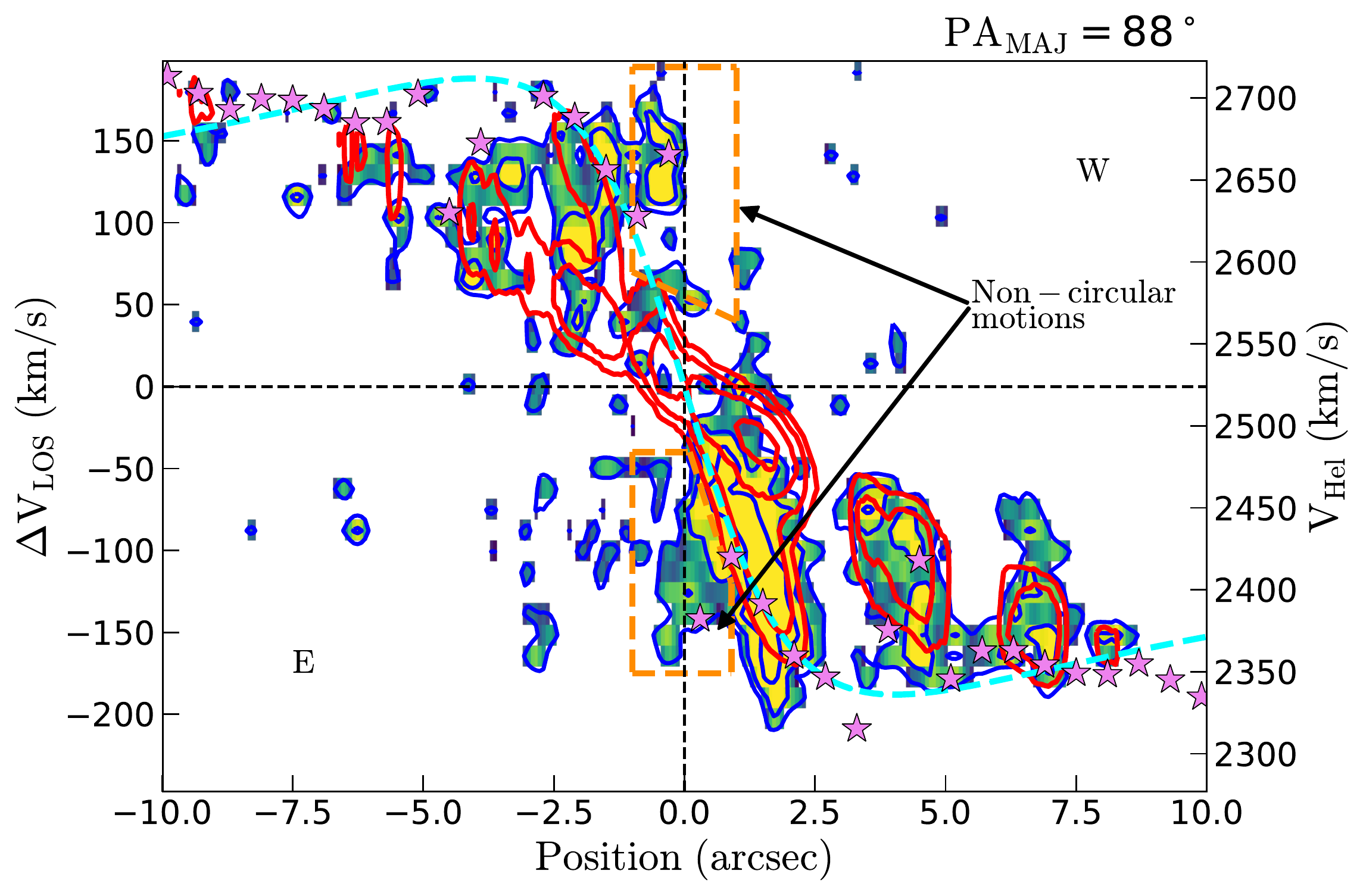}
\end{subfigure}\hfill
\begin{subfigure}[b]{0.46\textwidth}
    \centering
    \includegraphics[width=\textwidth]{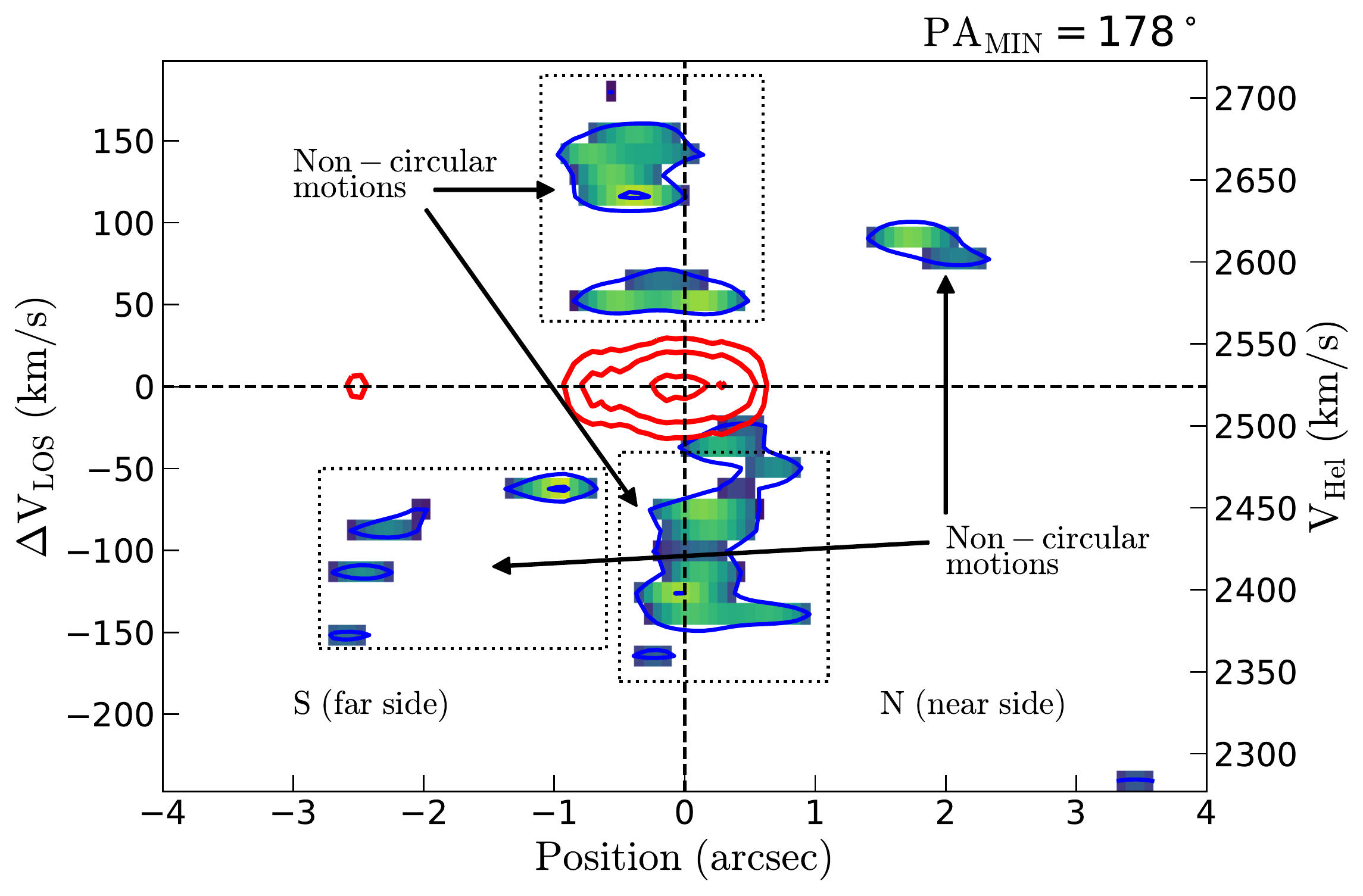}
\end{subfigure}\hfill

\caption{Position-Velocity diagrams taken along the kinematic major (PA$\sim$88$^{\circ}$) and minor (PA$\sim 178^{\circ}$) axes of NGC 4388 extracted with ${}^{\rm 3D}$BAROLO. Contours are 1 ($\sim 3 \sigma$, see Table \ref{tab:obs}), 2, 4 and 8 times 1.0 $\times$ 10$^{-2}$ Jy beam$^{-1}$. The orange dashed lines (top panel) enclose the emission inside 1" from the assumed nucleus which show non-circular motions. The rest is as in Fig.~\ref{fig:pv_1066}.}
\label{fig:pv_4388}
\end{figure}

Finally, we extracted CO(2-1) spectra from the east and west clouds (E and W panels in Fig. \ref{fig:spectra_4253_all}). The former spectrum only shows one component with a centroid almost coincident with the velocity predicted for the central spaxel of the region by our ${}^{\rm 3D}$BAROLO model. In the west cloud (located  at 3".2 (0.9\,kpc) from the AGN) there is evidence for an additional kinematic component. The brightest peak has a velocity centroid close to the modeled velocity for the central spaxel of the region but it is blueshifted with respect to it. The other component is  fainter and its centroid is blueshifted (approximately 90\,km s$^{-1}$). If we assume that both components are due to motions in the plane of the galaxy, it is reasonable to ascribe the first one to the gas following roughly circular orbits. The extra component, as the motions are detected near the kinematic  major axis, is likely related to the tangential motions associated to shocks along the leading edges of the main stellar bar \citep[see][and the discussion above]{downes1996co}.

\subsection{NGC 4388}\label{subsec:ngc4388}

This highly-inclined, Seyfert 1.9 \citep[][see Fig.\ref{fig:ngc4388_op_in}]{mason2015nuclear} has been classified as a 'changing look' AGN because it shows rapid (timescales of hours) changes in the X-ray absorption state \citep{elvis2004unveiling}. When in a  normal state, it is classified as Compton thin (e.g., N$_{H}= 3.5 \substack{+0.4 \\ -0.3} \times 10^{23}$ cm$^{-2}$, \citealp{iwasawa2003x}; N$_{H}= 2.67 \substack{+0.02 \\ -0.03} \times 10^{23}$ cm$^{-2}$, \citealt{miller2019nicer}). NGC~4388 belongs to the Virgo cluster and is located close to its center. Although it has a recession velocity of $\sim$ 2500 km s$^{-1}$, around 1500 km s$^{-1}$ are due to its falling towards the center of the cluster \citep{yoshida2002discovery}.  On large scales, it shows a long tail detected in HI, starting on the east side of the galaxy and extending northeastward up to $\sim$ 130 kpc \citep{yoshida2002discovery,vollmer2003atomic,oosterloo2005large}. This tail has been interpreted as the consequence of a long-lived ram pressure stripping process that also explains why NGC 4388 is deficient in HI gas \citep{vollmer2003atomic,oosterloo2005large,chung2009vla,pappalardo2012herschel}. Closer to the nucleus but still on large scales, there is a plume (detected in H${\alpha}$ and [OIII]) with PA $\sim35^{\circ}$ and reaching $\sim$ 4 kpc from the nucleus.

The [OIII] line emission on large scales shows a V-shape pattern, characteristic of a biconical outflow. This emission is considerably more prominent towards the SE. \cite{mulchaey1996emissionii} modeled the emission and inferred that the NE cone has a different axis and opening angle than the SW cone (see Table~\ref{tab:Bars}). In the central few arseconds, there is a good match between the [OIII] spatial distribution and that of the [Si VI], which benefits from the lower extintion in the NIR \citep{greene2014circumnuclear}.  The nuclear radio jet of this galaxy has a PA$=24^\circ$ \citep{greene2013using}.

\subsubsection{Morphology}

This galaxy is the only one in our sample without a 1.3 mm continuum detection. The  3$\sigma$ upper limit is 276 $\mu$Jy beam$^{-1}$. Note that the $\sigma$ in this case is at least 2 times higher than in the other four cases (see Table \ref{tab:obs}), which is likely the main reason for the non-detection. Owing to that, we assume the AGN location as that of the 21-cm radio continuum peak ($ {\rm RA \ (J2000)}= \ 12^{h}25^{m}46.75^{s}$, $\rm Dec \ (J2000)=12^{\circ}39'43.51"$) from VLBI observations \citep{giroletti2009faintest}. This position produces a symmetric rotation curve while that from VLA observations \citep{carral199015}, the most accurate one previously, does not. The VLBI coordinates are also in better agreement with those of the point source seen in hard X-ray \citep{iwasawa2003x} and those of the 33 GHz (9.1 mm) continuum peak \citep{kamali2017radio}. 
	
The large scale CO(2-1) line emission of NGC~4388 (Fig.~\ref{fig:ngc4388_op_in}, bottom panel) shows a disk-like morphology, extending out to the edges of our FoV in an approximate east-west direction for at least 20" ($\sim$2\,kpc). This extended CO(2-1) morphology towards the SE and NW roughly resembles the spiral arms seen at other wavelengths \citep{veilleux1999galactic,greene2014circumnuclear,vargas2019changesxvii}. Overall the CO(2-1) emission follows reasonably well the dust lanes seen in the $I-H$ color map (see Fig. \ref{fig:five_4388}a). Since these dust lanes are mainly located on the north side  we assume this is the near side of the galaxy. The CO(2-1) emission does not trace the H${\alpha}$ emission in the central few arcseconds, which is mostly associated with the southern ionization cone, or the Br$\gamma$ or [SiVI] emissions which trace the emission on both sides of the ionization cone \citep{greene2014circumnuclear}. 

The most striking feature in the central few arcseconds in the CO(2-1) intensity map (Fig. \ref{fig:ngc4388_op_in}, bottom panel) is that the emission peak is located 1".7 ($\sim 150\,$pc) from the assumed AGN position in an elongated structure. The morphology and kinematics (see below) seem to suggest that this structure is part of a circumnuclear ring or a nuclear disk with an approximate radius of 2".5, which would be partially co-spatial with the stellar nuclear disk (r$\sim$1") reported by \citet{greene2013using,greene2014circumnuclear}. We note that the NUGA survey showed that strongly lopsided rings or disks are not uncommon in the circumnuclear regions of Seyfert galaxies \citep[e.g.,][]{garcia2003molecular,garcia2009molecular,combes2009molecular}. However, the warm H$_{2}$ gas is brighter east of the nucleus, coinciding with a $\sigma$-drop found in the stellar kinematics \citep{greene2014circumnuclear}. However, this component has a secondary peak which is approximately spatially coincident with that of the CO(2-1). 

The ring/disk accounts for a large fraction of the CO(2-1) line emission in the central regions of NGC~4388. A similar situation was observed in NGC~2273 (Section~\ref{ngc2273}) and other Seyfert galaxies \citep[e.g.,][]{alonso2019nuclear,rosario2019accreting}. Remarkably, the morphology of the central 5" CO(2-1) emission is similar to that of the H$_2$/Br$\gamma$ map which traces roughly the excitation conditions of the warm molecular gas \citep{greene2014circumnuclear}. At the AGN location this ratio is low, as found in other Seyfert galaxies, and thus consistent with the CO(2-1) deficit observed there. As we move outwards from the AGN position, the H$_2$/Br$\gamma$ ratio increases and takes values consistent with those observed in the spiral arms of other galaxies while at the same time the CO(2-1) emission becomes brighter. 

\begin{figure*}[!ht]
	\centering
	\includegraphics[width=\hsize]{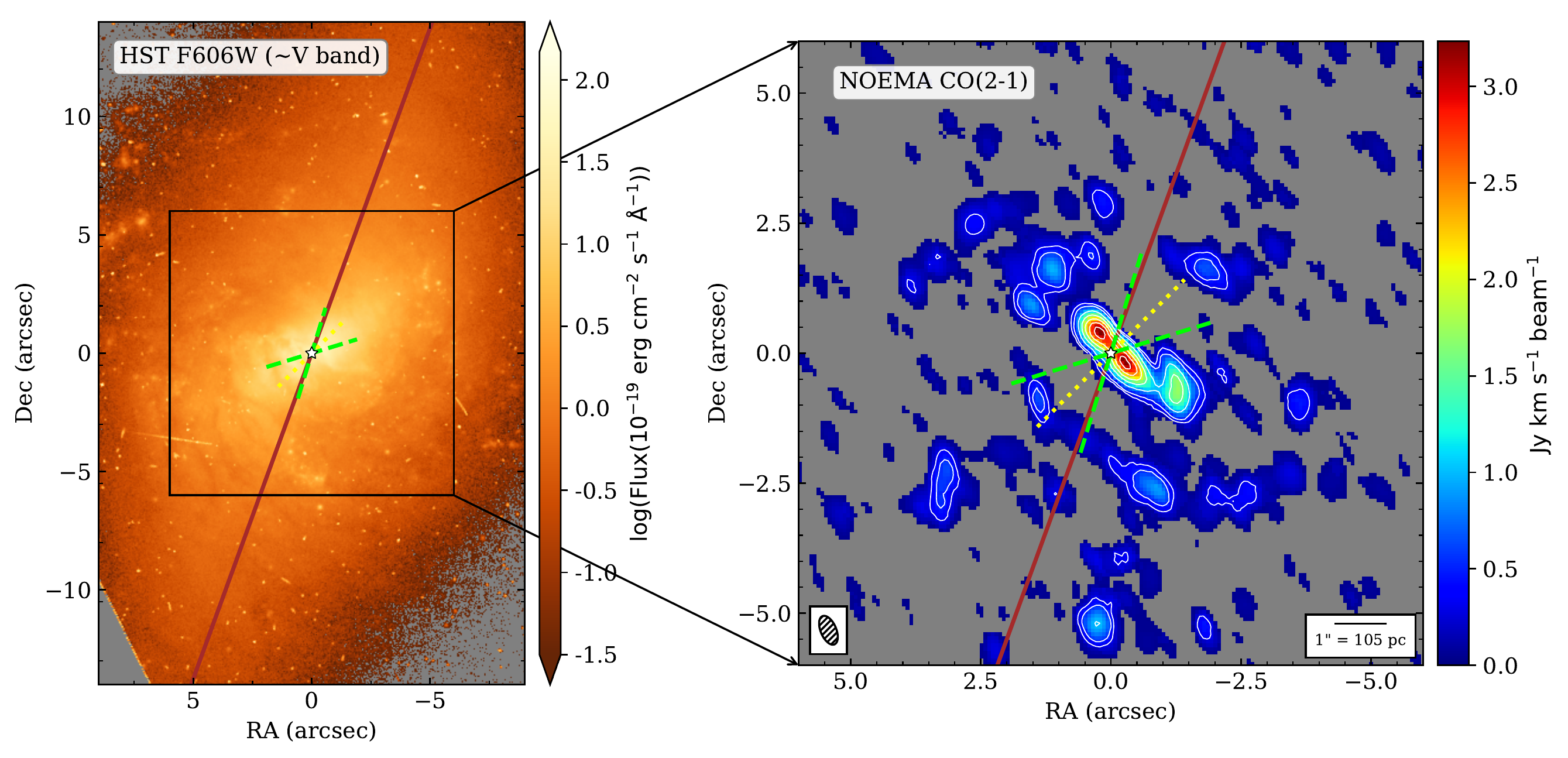}
	\caption{Same as Fig. \ref{fig:mrk1066_op_in} but for NGC~7465. CO(2-1) integrated-intensity contours: 0.3 ($\sim 4 \sigma$, $\sigma = 0.076$ Jy km s$^{-1}$ beam$^{-1}$) and from 0.5 to 3.0 Jy km s$^{-1}$ beam$^{-1}$, in steps of 0.5 Jy km s$^{-1}$ beam$^{-1}$.}
	\label{fig:ngc7465_op_in}
\end{figure*}

\begin{figure*}
	\centering
	\includegraphics[width=\hsize]{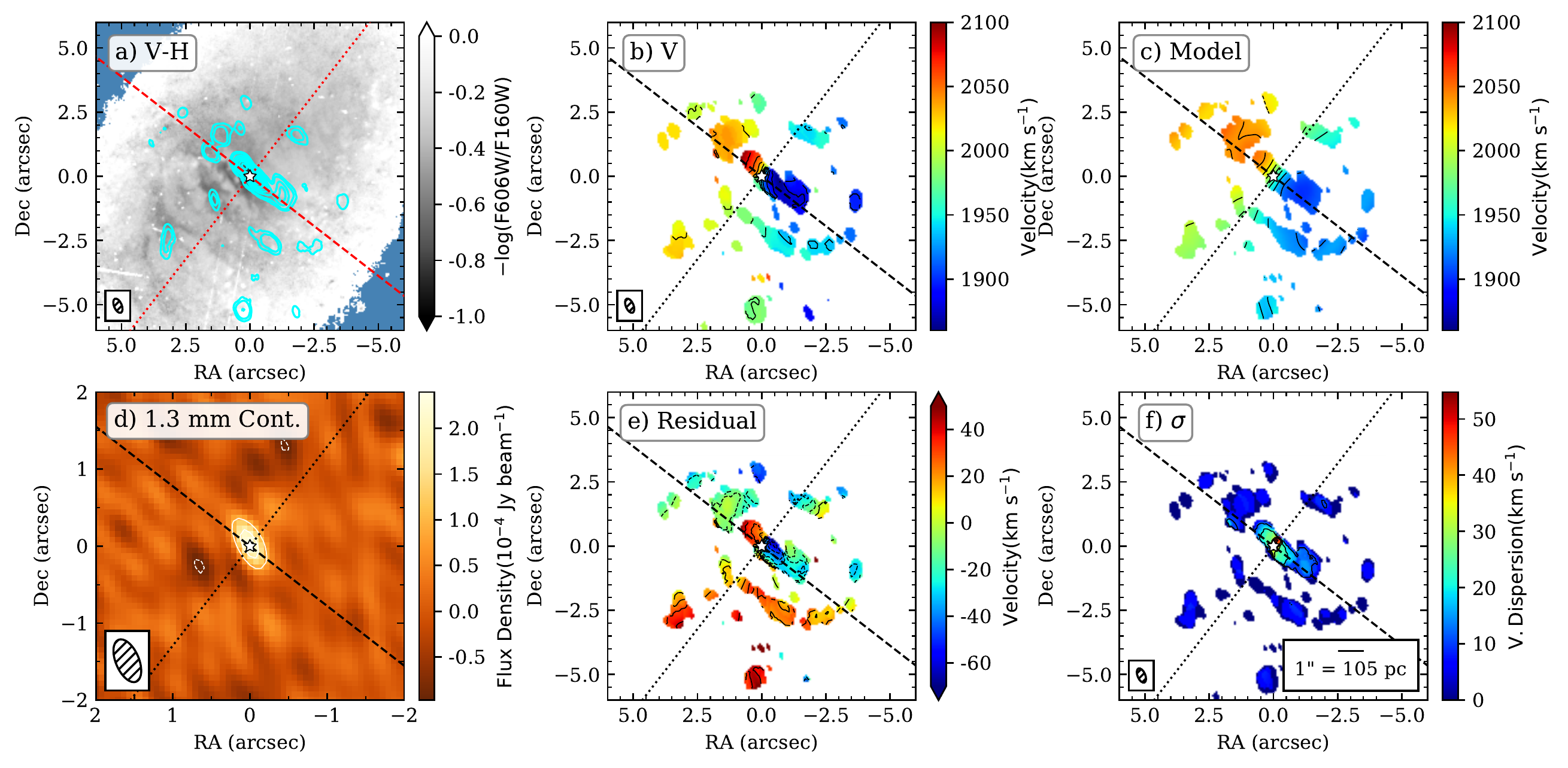}
	\caption{Same as Fig. \ref{fig:six_1066}, but for NGC~7465. The velocity contours in b and c panels are from 1860 to 2100 km s$^{-1}$ in steps of 20 km s$^{-1}$. The continuum contours are -3$\sigma$ (white dashed line), 3$\sigma$ and 6$\sigma$ (white solid line) with $\sigma$=30 $\mu$Jy beam$^{-1}$. The residual velocity contours are from $-50$ to 50 km s$^{-1}$ in steps of 10 km s$^{-1}$. The velocity dispersion contours are in steps of 10 km s$^{-1}$, starting at 10 km s$^{-1}$.}
	\label{fig:six_7465}
\end{figure*}

\begin{figure}
\centering
\begin{subfigure}[b]{0.46\textwidth}
    \centering
    \includegraphics[width=\textwidth]{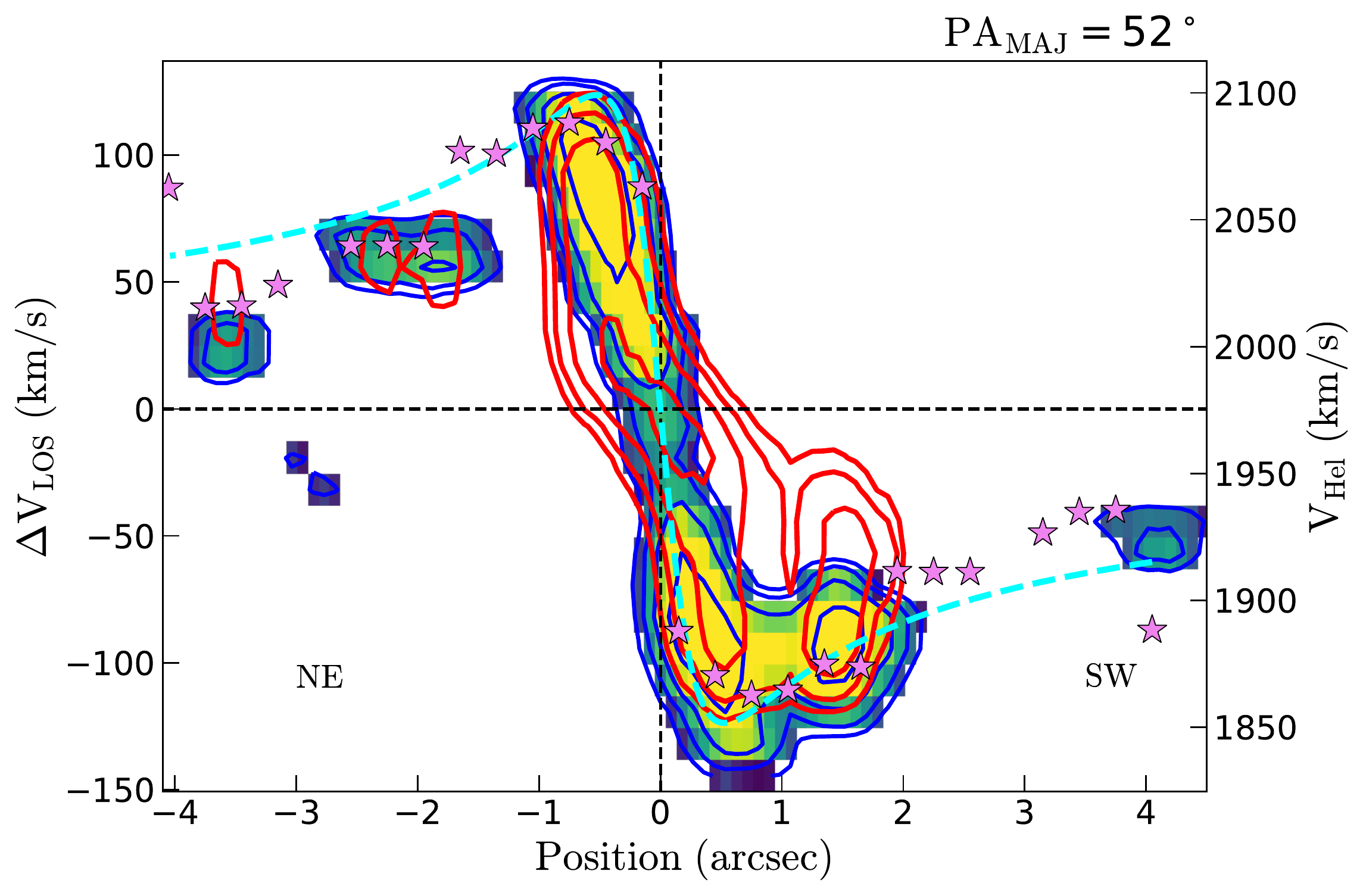}
\end{subfigure}\hfill
\begin{subfigure}[b]{0.46\textwidth}
    \centering
    \includegraphics[width=\textwidth]{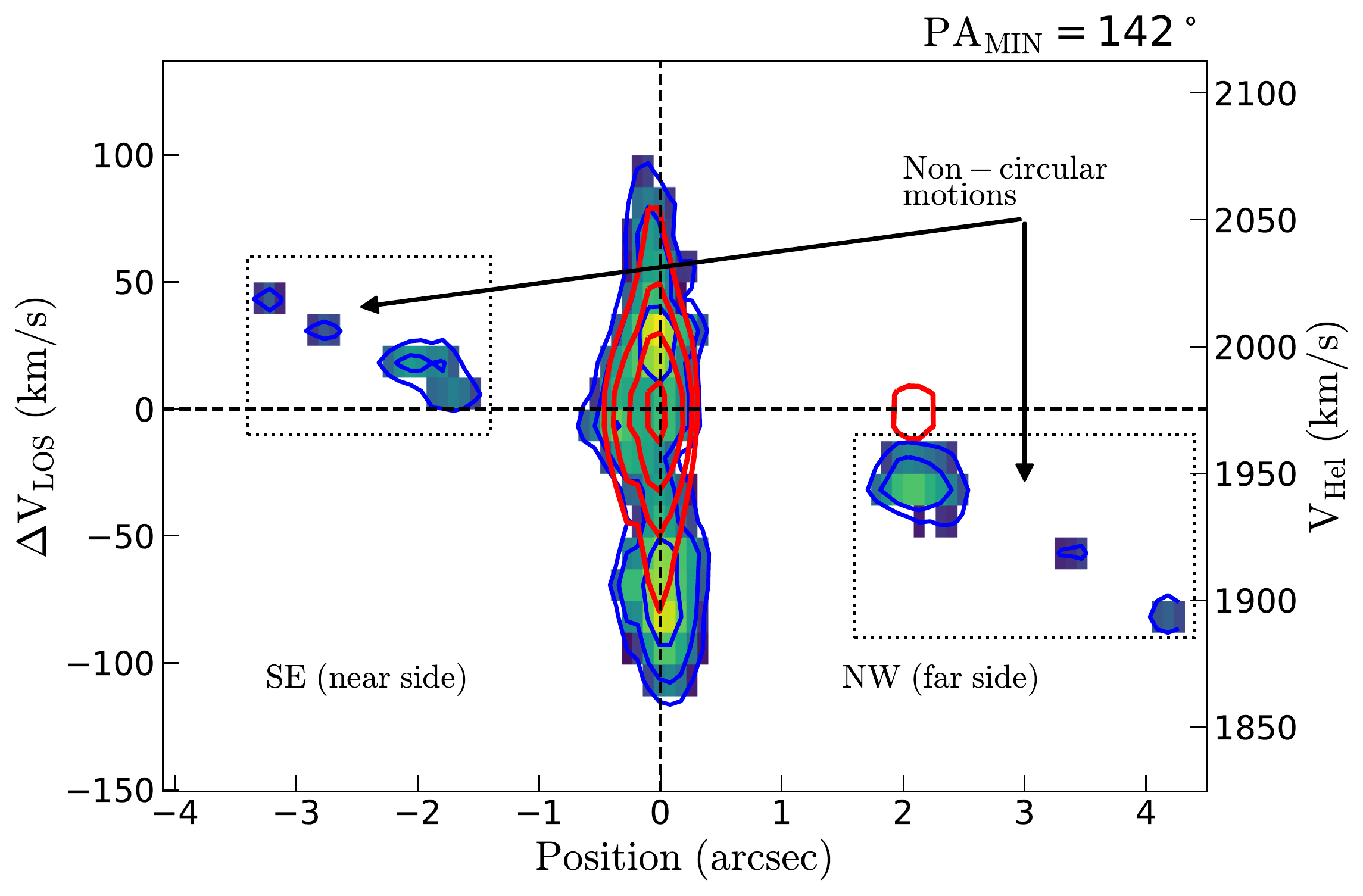}
\end{subfigure}\hfill

\caption{Position-Velocity diagrams taken along the kinematic major (PA$\sim$52$^{\circ}$) and minor (PA$\sim 142^{\circ}$) axes of NGC 7465 extracted with ${}^{\rm 3D}$BAROLO. Contours are 1 ($3 \sigma$, see Table \ref{tab:obs}), 2, 4 and 8 times 3.6 $\times$ 10$^{-3}$ Jy beam$^{-1}$. The dotted lines enclose the emission outside the inner 1". Colors and lines as in Fig.~\ref{fig:pv_1066}.} 
\label{fig:pv_7465}
\end{figure}

\subsubsection{Kinematics}

The observed CO(2-1) mean velocity field of NGC~4388 is complicated but it shows clear signs of rotation outside the nuclear region (Fig. \ref{fig:five_4388}, panel b), as already observed for the warm H$_2$ molecular gas traced by the NIR $2.12\,\mu$m line \citep{greene2014circumnuclear}. The deviations from a pure rotational pattern are mostly located in regions a few arcseconds to the SE and the NW of the AGN. These may correspond with the leading edges of the main bar, which was needed to reproduce the much less distorted H${\alpha}$+[NII] velocity field \citep{veilleux1999kinematic}. Moreover, other factors as the presence of an ionization cone, a radio jet and a ram-pressure stripped process due to the infalling of the galaxy in the Virgo cluster environment \citep[see][and references therein for a dicussion]{yoshida2002discovery} might also be at least in part responsible for such deviations. We will come to this issue later.  

The ${}^{\rm 3D}$BAROLO model for NGC~4388 (Fig.~\ref{fig:five_4388}, panel c and the parameters are listed in Table~\ref{tab:Baroloparameters})  reproduces the overall observed CO(2-1) mean velocity field over the approximate modeled $18" \times 10"$ (1.7\,kpc$\times$0.9\,kpc)  FoV.  The systemic velocity roughly agrees with that derived from the stellar \citep{greene2014circumnuclear}, H${\alpha}$+[NII] \citep{veilleux1999kinematic} and HI \citep{lu1993hi,vollmer2003atomic} kinematics and radio components \citep{kukula1995high}. Furthermore, the inclination and the PA are also consistent with those obtained from the H${\alpha}$+[NII] kinematics \citep{veilleux1999kinematic}. 

The CO(2-1) mean-velocity residual map of NGC~4388 (Fig.~\ref{fig:five_4388}, panel d) shows that along the kinematic major axis the residuals are relatively small and below $50\,{\rm km\,s}^{-1}$ (see also the p-v diagram along the major axis in Fig.~\ref{fig:pv_4388}). The largest velocity residuals ($\sim 250\,{\rm km\,s}^{-1}$ projected, in absolute value) are to the northwest (redshifted) and to the southeast (blueshited) extending east-west over the entire FoV modeled with $^{\rm 3D}$BAROLO. These residuals are similar to those obtained by \cite{veilleux1999kinematic} when modeling the H$\alpha$ velocity field with a rotating disk model over a larger FoV than ours. \cite{veilleux1999kinematic} were able to reproduce the strong non-circular motions as streaming motions, revealing a local 'inflow', due to the presence of a large scale stellar bar with a deprojected radius of 1.5\,kpc with a projected PA=100$^\circ$.

The position-velocity diagram taken along the major axis of NGC~4388 (Fig.~\ref{fig:pv_4388}, upper panel) shows that the CO(2-1) emission is asymmetric and more intense to the west. The projected rotation amplitude is around 150 km s$^{-1}$. The velocity gradient at the nucleus is $\sim$300 km s$^{-1}$, which is a signature of deviation from pure rotation (see the areas enclosed by the orange dashed lines in the upper panel of Fig. \ref{fig:pv_4388}). As the bar proposed by \cite{veilleux1999galactic} for this galaxy is almost edge-on (the major axis of the bar is almost perpendicular to the line-of-sight), these central deviations can be interpreted as molecular gas following the x$_{2}$ orbits induced by the bar in the central regions, probably forming a disk or ring \citep[see Fig.4 and Fig. 5 of][]{bureau1999bar}. In the position-velocity diagram taken along the minor axis (PA$\sim178^{\circ}$) a gradient is also detected within 1" of nucleus. The velocities of this gradient are preferentially redshifted to the south and blueshifted to the north. Assuming that the north is the near side (see above), it is consistent with an outflow in the plane of the galaxy. This cold molecular outflow is cospatial with base of the radio lobe \citep{hummel1991anomalous,irwin2019changesxviii}. We note that the PA of the 100-pc scale radio jet (PA$\sim30^{\circ}$) is quite different from the direction of the molecular outflow (178-358$^{\circ}$). We also note the projection to the north of the molecular outflow is not cospatial with the corresponding projected ionization cone (see Fig. \ref{fig:ngc4388_op_in}).

The position-velocity diagram along the minor axis also shows deviations from pure rotation on larger scales. Between around 1" and 3", there are faint emission with observed velocities of up to 150 km s$^{-1}$ in absolute value. The CO(2-1) emission shows blueshifted velocities to the S and redshifted to the N. If we assume that these motions are in the galaxy plane, then they are consistent with a local inflow. In this case, these streaming motions are naturally associated to the presence of the large-scale stellar bar.

\subsection{NGC 7465 (Mrk 313)} \label{subsec:ngc7465}

NGC~7465 (Fig.~\ref{fig:ngc7465_op_in}, left panel) belongs to the NGC~7448 group. Inside this group, there is a compact subgroup with NGC~7463, NGC~7464, UGC~12313 and UGC~12321 \citep{van1992study}. It is classified as a barred galaxy in the RC3 catalog. However, there are only a few discussions in the literature about the possible presence of a bar in this galaxy and no properties (PA, length) are reported \citep[see e.g., Khachikian \& Petrosyan 1983, ][]{van1992study}. The optical continuum images of this galaxy \citep{van1992study,ferruit2000hubble,merkulova2012study} show isophotes roughly elliptical beyond r>10" elongated at PA=160-165$^{\circ}$. In Fig. \ref{fig:ngc7465_op_in} we have assumed that these isophotes trace the main bar. The molecular and ionized gas components as well as the stellar component of this galaxy at scales of a few arcseconds appear to show different kinematic axes. All this is a clear signature of the recent/on-going interaction between NGC~7465 and one or more galaxies of its surroundings \citep{van1992study,li1994neutral,thomas2002extended,merkulova2012study,alatalo2013atlas3d}. The interaction with NGC~7464 is likely taking place since there is an HI spiral arm that connects the SE of NGC~7465 with the SW of NGC~7464, and the latter galaxy exhibits photometric distortions in the direction towards the former. Moreover, the existence of misaligned gas, the presence of low stellar metallicities, and a high $\alpha$-element to iron ratio in the central region of NGC~7465 have been interpreted as a sign that this galaxy has accreted gas from an external source \citep{young2014atlas3d}. 

NGC~7465 is the least luminous galaxy in X-rays of our sample (see Table \ref{tab:properties}). Its optical classification is between Seyfert 2 and a low-ionization nuclear emission-line region \cite[LINER, e.g.,][]{ferruit2000hubble}. However, \cite{almeida2009near} classified it as a type-1 LINER based on the presence of broad lines in their NIR long-slit spectra. \cite{guainazzi2005x} classified its nucleus as a Compton-thin but highly absorbed (N$_{H} \sim 5 \times 10^{23}$ cm$^{-2}$) source. \cite{muller2018keck} suggested the presence of a double nucleus in this galaxy based on the non-point-like shape of their 2.12 $\mu m$  continuum. 

\subsubsection{Morphology}

The nuclear source in the 1.3 mm continuum image (Fig. \ref{fig:six_7465}d) is  detected at a 8$\sigma$ level. We had to fix the size of the Gaussian function to be able to fit it in the uv-plane. The AGN coordinates obtained from 1.3 mm continuum are in excellent agreement with those derived from the 5 GHz (6 cm) continuum \citep{nyland2016atlas3d}. The resulting parameters indicate that the source is resolved along PA$\sim113^{\circ}$ (i.e., the minor axis of the beam of the observations) but not in the perpendicular direction, where previous radio works suggested extended emission \citep{nagar1999radio,mundell2009radio}. The [OIII] emission in the inner 2" has a PA=$135^{\circ}$ \citep{ferruit2000hubble}. We point out that the region surrounding the nucleus but not the nucleus itself displays the highest values in the $V-H$ color map suggesting strong reddening due to dust.

\cite{alatalo2013atlas3d} classified the cold molecular gas morphology of NGC~7465 as mildly disrupted, based on CARMA CO(1-0) $\sim$5.5" resolution observations. Our NOEMA CO(2-1) line emission map (Fig.~\ref{fig:ngc7465_op_in}, right panel) clearly resolves the central emission in  a set of spiral arms on scales of up to 5" ($\sim 500\,$pc) from the AGN. The spiral arms appear to be spatially coincident with regions of higher extinction in the $V-H$ color map (Fig,~\ref{fig:six_7465}, panel a) as well as some of the HII regions detected in the H$\alpha$+[NII] map of \cite{ferruit2000hubble}. The nuclear region shows a double CO(2-1) peak in an elongated bar-like structure. None of the nuclear CO(2-1) peaks coincides with the 1.3 mm continuum peak believed to mark the location of the AGN. Finally, the SE of the galaxy presents redder colors suggesting that it is the near side. 

\begin{table*}
	\caption{CO(2-1) integrated fluxes and derived molecular gas masses in several apertures. }            
	\label{tab:H2mass}     
	\centering           
	\begin{tabular}{c c c c c c c c c c}   
		\hline\hline        
		Galaxy &  \multicolumn{3}{c}{Nuclear (beam)} & 
		\multicolumn{3}{c}{Circumnuclear (4")} & \multicolumn{2}{c}{ 11-12" aperture}\\
		
        & 	ap.	& S$_{CO(2-1)}$  & M$_{H_{2}}$ &
        ap.	& S$_{CO(2-1)}$   & M$_{H_{2}}$ &
        ap. & S$_{CO(2-1)}$ 	\\
		& (pc$\times$pc) & (Jy km s$^{-1}$) & ($10^{6}$ M$_{\odot}$) & (kpc) & (Jy km s$^{-1}$) &  ($10^{8}$ M$_{\odot}$) & (kpc) & (Jy km s$^{-1}$)  \\
		\hline                    
		Mrk 1066 & $99\times 90$  & $7.24 \pm 0.03$ & $36\pm 2$ & 0.9 & $47 \pm 1$   & $ 2.3 \pm 0.2 $ & 2.7 & {$65 \pm 3 $}  \\
		NGC 2273 & $89\times 72$  & $5.84 \pm 0.07$ & $8.6\pm0.1$  & 0.5 & $86 \pm 3$   &  $1.26 \pm 0.09 $ & 1.5 & {$120 \pm 4 $}\\
		NGC 4253 & $150\times 82$ & $7.6 \pm 0.2$   & $55\pm 4$ & 1.1 & $46.7\pm 0.8$& $ 3.4 \pm 0.2 $ & 3.3 & $56 \pm 1 $\\
		NGC 4388 & $78 \times 33$ & $ 2.3 \pm 0.1$  & $1.9\pm 0.2$ & 0.4 & $69 \pm 9 $  & $0.57 \pm 0.08$ & 1.0 & $70  \pm 6 $ \\
		NGC 7465 & $62\times 32$  & $ 1.89\pm0.04 $ & $2.0\pm 0.1$ & 0.4 & $19.1 \pm 0.5$  &$ 0.20 \pm 0.02 $ & 1.1 & $36.6 \pm 0.8 $ \\
		\hline                  
	\end{tabular}
\end{table*}

\subsubsection{Kinematics}

The NOEMA observed CO(2-1) velocity field exhibits an overall rotating pattern  (Fig.~\ref{fig:six_7465}, panel b). However, the PA of  kinematic major axis clearly increases as we move away from the nuclear region. For example, this can be seen to the SE along the minor axis where the velocities near the nucleus are around 1975 km s$^{-1}$ while at a radius of $\sim$ 3".5 the velocities are $\sim$ 2025 km s$^{-1}$ (see Fig.~\ref{fig:six_7465}, panel b, and also Fig. \ref{fig:pv_7465}, bottom panel). Most of the molecular gas in the NE side (as a whole) of the central 12" of NGC 7465 recedes from us, while that in the SW side approaches. This orientation is rather different from that observed with the  CO(1-0) transition at $\sim$5.5" resolution \citep{alatalo2013atlas3d}. On larger scales the cold molecular gas is oriented in an approximate east-west direction ($\sim 106^\circ$) which is also
similar to that of the H$\alpha$ emitting gas \citep{merkulova2012study}. At radial distances greater than 10" the photometric major axis of the galaxy is PA$=160^\circ$ \citep{merkulova2012study}. {Regarding the correspondence with the high-ionized gas emission, the [OIII] emission in the nuclear region (r<2") is perpendicular to that of the CO(2-1). Further away, the former is much more diffuse and seems to be mostly related to HII regions \citep{ferruit2000hubble}.} The observed mean-velocity dispersions are overall below $\sim 30$ km s$^{-1}$ except at the AGN position. 

Despite the complex kinematics in the inner 12" of NGC~7465, we still attempted a ${}^{\rm  3D}$BAROLO fit as explained for the other galaxies (see Section~\ref{methodology}) by keeping the PA and inclination fixed in the second run. The model reproduces the general behavior of the observed velocity (Fig.~\ref{fig:six_7465} panels b and c, and the model parameters are given in Table~\ref{tab:Baroloparameters}) but it fails to account for the molecular gas motions with strong signs of perturbation. The systemic velocity obtained from the ${}^{\rm 3D}$BAROLO fit is almost in the middle of the values derived from the stellar \citep[1963 km s$^{-1}$,][]{merkulova2012study} and HI \citep[1989 km s$^{-1}$,][]{li1994neutral} components. The aforementioned moderate resolution CO(1-0) observation suggests a systemic velocity near the stellar one \citep[1960\,km\,s$^{-1}$,][]{alatalo2013atlas3d}. Regarding the inclination, the average value derived with ${}^{\rm 3D}$BAROLO is similar to that obtained from a tilted-ring method for the main stellar disk \citep{merkulova2012study} and quite different from the inclination derived for molecular gas on larger scales \citep[$\sim$70$^{\circ}$, see][]{alatalo2013atlas3d}. The average PA fitted from our CO(2-1) data is far from the values of both the stellar and CO(1-0) components (PA$_{\rm star}$=160$^{\circ}$, PA$_{\rm CO(1-0)}$=106$^{\circ}$). At face value, we could interpret this result as a kinematical decoupling of the molecular gas in the inner 12" with respect to the stellar and gas kinematics on larger scales. However, we cannot rule out that this is due to differences in the angular resolution, radial coverage, and/or modelling strategies of the different tracers.

The larger  CO(2-1) mean-velocity residuals ($\leq$ 70 km s$^{-1}$ in projection, i.e,  [$(v_{obs}-v_{model})/sin(i)]$ $\leq$ 88 km s$^{-1}$, see Fig.~\ref{fig:six_7465}, panel e) are found in the nuclear region of NGC~7465. This region appears to be further decoupled from that on larger scales of 12" of our NOEMA FoV,  and in the two clouds 5" to the south and 3" to SE. The velocity residuals of these two clouds can be explained by the approximately constant, anti-clockwise change of the PA of the kinematic major axis since they are the most distant structures from the nucleus. This is understood because the average PA in the $^{\rm 3D}$BAROLO model is strongly biased to that of the brighter regions, that is, the inner ones. In other words, these two clouds belong to regions rotating close to the orientation seen on larger scales with CO(1-0). 

The residual pattern outside the nuclear region at radial distances greater than approximately 2" shows positive/negative residuals to the SE/NW. These deviations from pure rotation are clearly seen in the position-velocity diagram along the minor axis (bottom panel of Fig.~\ref{fig:pv_7465}) and reach approximately projected velocities of +50$\,{\rm km\,s}^{-1}$ to the SE and $-80\,{\rm km \,s}^{-1}$ to the NW. 
If these motions are taking place in the galaxy disk, they are compatible with a local  inflow. In such case, the streaming motions might be due to the suspected main bar (see above).

\section{Molecular Gas Masses} \label{molecularmass}

We measured the CO(2-1) integrated intensity by fitting Gaussians (similar to the fits done in Figs.~\ref{fig:spectra_1066} and \ref{fig:spectra_4253_all}) to spectra extracted with three apertures. The first has a size equal to the beam and covers typically the central $\sim100$ pc (see Table~\ref{tab:H2mass} for the actual sizes) and is referred to as nuclear aperture. The second is a circular aperture with a 4" diameter covering physical scales of 0.4-1.1\,kpc (circumnuclear apertures). These will allow us to compare with properties measured from Spitzer spectroscopy in forthcoming works. The last one is also a circular aperture but with a 11-12" diameter, which is suitable to compare our recovered fluxes with those of the single-dish observations, and corresponds to 1.0-3.3\,kpc. The uncertainties of the CO(2-1) integrated intensity listed in Table \ref{tab:H2mass} for all apertures come from the Gaussian fitting.

Since interferometric observations may miss some of the total flux, we searched the literature for single-dish observations. We estimated the missing flux  by integrating the density flux (without applying any threshold) of the interferometic observations over an aperture equal to that of the single-dish observations and over a similar spectral width \cite[e.g.,][]{ueda2014cold}. We estimated the conversion factor between Jy and K through the formula: S$_{\nu}$/T$_{mb}$=$8.16\times10^{-7}(\nu (GHz))^{2}(\theta_{beam} (arcsec))^{2}$ \citep[e.g.,][]{grossi2016starforming}. For Mrk~1066, \cite{krugel1990co} obtained an integrated flux of 297 Jy km s$^{-1}$ in a 12" aperture while we measured 65 Jy km s$^{-1}$, that is, we recovered $\sim22 \%$ of the flux measured by the single-dish. For NGC~2273, our 11-12" measurement recovers most of the single-dish flux observed by \cite{albrecht2007dust} and approximately 30\% of the 21" JCMT flux derived by \citep{petitpas2003molecular}. However, we note that the \cite{krugel1990co} 11" single-dish measurement is approximately half that obtained by \cite{albrecht2007dust}. For NGC~7465, we recovered $\sim 34 \%$ of the flux measured by \cite{young2011atlas3d} for a 11" aperture. Finally in the case of NGC~4388  we recovered  $\sim 67 \%$ of the flux for a 11" aperture measured by \cite{verdugo2015ram}. As we mentioned before, there are no previous observations in CO(2-1) for NGC 4253, so we cannot estimate the missed flux for this source.

To derive the molecular gas masses, we used Eq. \ref{eq.mass} from \cite{sakamoto1999co}:

\begin{dmath}
	\Bigg(\frac{M_{H_{2}}}{M_{\odot}}\Bigg) = {1.18 \times 10^{4}} \times \Bigg(\frac{D^{2}_{L}}{Mpc}\Bigg) \times \Bigg(\frac{S_{CO(1-0)}}{Jy \ km \ s^{-1}}\Bigg) \times \Bigg[\frac{X_{CO}}{3.0 \times 10^{20} cm^{-2} \ (K \ km \ s^{-1})^{-1}}\Bigg] \label{eq.mass}
\end{dmath}

\noindent taking $X_{CO} = 2 \times 10^{20}$ cm$^{-2}$ (K km s$^{-1}$)$^{-1}$ \citep{bolatto2013co} and assuming a brightness temperature ratio, $r_{\rm 21}$=CO(J=2-1)/CO(J=1-0), of $0.89 \pm 0.06$ \citep{braine1992co}. $X_{\rm CO}$ is the typical Galactic conversion factor and is consistent with the resolution of our observations ($\sim$100 pc) according to numerical simulations \citep{gong2018x}. $r_{\rm 21}$ was obtained from measurements in the inner kpc of  nearby galaxies and could vary depending on excitation mechanisms in different regions of the galaxy \citep[e.g., between arm and inter-arm regions][or between nuclear and further regions \citealt{garcia2009molecular}]{hasegawa1997co}. It could also be different in isolated and perturbed galaxies \citep{braine1992co}. Estimations of r$_{21}$ have been reported for NGC~2273 and NGC~7465, $0.88 \pm 0.26$ \citep{petitpas2003molecular} and $0.95 \pm 0.04$ \citep{crocker2012atlas3d}, respectively, which are consistent with our assumed ratio. 

The derived molecular gas masses (see Table~\ref{tab:H2mass}) range from $2 \times 10^{6}$ to $5 \times 10^{7}$ M$_{\odot}$ and from $2 \times 10^{7}$ to $3 \times 10^{8}$ M$_{\odot}$ in the nuclear and circumnuclear apertures, respectively.  As suggested by their morphologies, in Mrk~1066 and NGC~4253 the CO(2-1) emission is more centrally peaked than in NGC~2273 and NGC~4388. Thus, the former galaxies have higher nuclear to circumnuclear CO(2-1) ratios (0.15) and corresponding molecular gas mass ratios than the latter (0.03-0.07). 

The circumnuclear cold molecular gas masses are typical values for Seyfert galaxies \citep[e.g.,][]{schinnerer2000distribution,hicks2009role,sani2012physical,garcia2014molecular}. The NOEMA nuclear apertures cover physical sizes smaller than 100\,pc, except for the most distant galaxy in our sample (NGC 4253). All the nuclear cold molecular gas masses for our sample are comparable to those measured in nuclear disks/tori of local low luminosity AGN on scales covering a few tens of parsecs with ALMA 
\citep{alonso2018resolving,combes2019alma}.

\section{Discussion}  \label{discussion}
\subsection{Kinematics of the cold molecular gas and ionized gas}

\begin{table*}
	\caption{Summary of the kinematics}          
	\label{tab:resume_kin}     
	\centering           
	\begin{tabular}{c c c c c c c c c c}   
		\hline\hline
		&  \multicolumn{3}{c}{Ionized gas outflows} & & 
		\multicolumn{5}{c}{Cold molecular gas flows}\\ \cline{2-4} \cline{6-10}
		Galaxy & Line & v$_{max}$ & Ref. & & Rotation & Streaming & v$_{max}$ & Outflows & v$_{max}$ \\
		& & (km s$^{-1}$) & & & & motions & (km s$^{-1}$) & &(km s$^{-1}$) \\
		\hline                    
		Mrk 1066 & [OIII] & 900 & 1 & & \cmark & \xmark & - & \cmark & $-$30, +90\\
		& [FeII] & 50 & 2 & &  & &  & & \\
		NGC 2273 & [S III] & [$-$400, 160] & 3 & & \cmark & \xmark & - & \xmark & - \\
		NGC 4253 & [FeII] & 277 & 4 & & \cmark & \cmark & $-$90 & \xmark & -\\
		NGC 4388 & [Si VI] & [$-$300, 300] & 5 & & \cmark & \cmark & [$-250$, +250] & \cmark & $\pm$150\\
		NGC 7465 & - & - & - & & \cmark & \cmark& [$-$80, +50] & \xmark & -\\
		\hline                  
	\end{tabular}
    \tablefoot{Column (1) Galaxy Name,  Columns (2)-(4) properties of ionized gas outflows: line (2), observed maximum velocity of the outflowing gas along the bicone axis (3) and reference (4). Columns (5)-(9) properties of the non-circular motions identified in the CO(2-1) kinematics: Is the general motion rotation? (5), there are signatures of streaming motions along bars/spiral arms? (6), projected maximum velocity of the streaming motions (7), there are signatures of outflows? (8) and projected maximum velocity of the outflows (9). 
    } 
	\tablebib{(1)~\citet{fischer2013determining}; (2)~\citet{riffel2011compact}; (3)~\citet{barbosa2009gemini}; (4)~\citet{schnorr2014gas}; (5)~\citet{greene2014circumnuclear}.
}
\end{table*}

As stated in the introduction, our main goal is to investigate the kinematics of the cold molecular gas in nearby Seyfert galaxies and look for evidence of non-circular motions. In this section we put together the results from the individual source analysis (Section~\ref{results}) and discuss how the non-circular motions relate to the presence of large scale bars and ionized gas outflows. In Table~\ref{tab:resume_kin}, we summarize the main results from the molecular gas kinematics and list some properties of the ionized gas outflows in our sample from the literature, including the maximum velocity and the ionized gas line used. 

The first result from this table is that the cold molecular gas in all five galaxies shows a clear rotational pattern on the scales probed by the NOEMA observations (a few hundred pc). Among them, three have signs of streaming motions along bars and/or spiral arms, two exhibit signs of outflowing material, and only one shows evidence of both. NGC~2273 is the only galaxy for which we did not find any evidence of streaming or outflowing motions. 

The observed maximum velocities of the streaming motions range from 50 to 250 km s$^{-1}$ in absolute value. These velocities correspond to deprojected velocities in the range $\sim$ 63- 252 km s$^{-1}$. We have calculated the deprojected values by using [$(v_{obs}-v_{model})/sin(i)$]. From those values, we conclude that the presence of bars affects the kinematics of the cold molecular gas on scales of hundreds of parsecs and the deviations from circular motions ascribed to them can be higher than those related to outflowing material for this phase (see below). It is therefore necessary to take them into account before looking for signs of molecular outflows.

Regarding the molecular outflows, the maximum observed velocities (in absolute value) are in the range from 30 to 150 km s$^{-1}$, which correspond to deprojected velocities in the range $\sim 40$-151 km s$^{-1}$ assuming that the outflows take place in the galaxy disk. These values are similar to those measured from hot molecular gas in the near-infrared for other Seyfert galaxies \citep{davies2014fueling}. Additionally, we find that the maximum velocities of the molecular outflows roughly follow the relation found by \cite{fiore2017agn}. In other words, for the AGN bolometric luminosities of our sample the expected maximum velocities for the molecular gas should 
be of the order of 100\,km s$^{-1}$. Moreover, the maximum velocity of the outflows in our galaxies (see Table~\ref{tab:resume_kin}) is significantly higher for the ionized phase than the molecular. This suggests that, at least for the AGN luminosities probed here, the cold molecular gas in the galaxy disk is swept up by the AGN wind rather than being part of the AGN wind. In other words, the bulk effect of the AGN wind is to entrain the cold molecular gas in the galaxy disk, when the geometry is favorable for this. This result is in agreement with previous observations \citep{davies2014fueling, garcia2014molecular, garcia2019, alonso2018resolving, alonso2019nuclear}. We also emphasize that the geometry and orientation of the ionization cone with respect to the host galaxy play a role on whether we should observe molecular outflows or not. Therefore, a detailed model of this emission \citep[see e.g.][]{fischer2013determining} is necessary for a correct interpretation of the non-circular motions of the molecular gas.

\subsection{Nuclear molecular outflow rate in Mrk~1066}

Mrk~1066 was the only galaxy where we were able to clearly resolve the molecular outflow components on opposite sides of nucleus and along the minor axis. In this section we estimate the molecular mass outflow rate. Following \cite{garcia2014molecular}, a lower limit of the mass rate of a multi-conical outflow uniformly filled by the outflowing clouds can be estimated through the following expression:
\begin{equation}
\dot{M}_{out}=3\times v_{max} \times \frac{M_{mol}}{R_{out}}\times tan(\boldsymbol{\gamma}), 
\end{equation}

\noindent where $v_{max}$ is the maximum observed velocity of the outflow, $M_{mol}$ is the characteristic mass of the outflow, $R_{out}$ the projected radius reached by the ouflowing material measured from the AGN position and $\boldsymbol{\gamma}$ is the angle between the ionization bicone axis and our line-of-sight. 

We first estimated the mass of the outflow by fitting a Gaussian function to the non-circular motion components seen in the integrated spectra of the two nuclear regions to the northeast and southwest of the AGN (see Fig.~\ref{fig:spectra_1066}) and using the expressions of Section~\ref{molecularmass} to convert the integrated fluxes into molecular gas masses. We obtained $M({\rm red-NE})=1.4\pm0.2\times 10^{6}$ and $M({\rm blue-SW})= 4.8\pm0.5\times 10^{6} \ M_{\odot}$, for the masses of the receding and approaching components of the outflow with respect to our line of sight, respectively. Comparing with the circumnuclear region (4\arcsec-aperture, see Table~\ref{tab:H2mass}), it is clear that the outflowing molecular gas mass is only a small fraction of the total mass in this region.

Taking the values of the maximum velocity of the outflow from Table~\ref{tab:resume_kin}, R$_{out}$=0".47$\times$224 pc/"=105 pc and assuming that the outflow takes place in the plane of the galaxy ($tan(\gamma)=1/tan(i_{disk}=50^{\circ})$, see Section~\ref{mrk1066}), we obtained $\dot{M}_{out}({\rm NE})\sim3.0$ and $\dot{M}_{out}({\rm SW})\sim3.5 \ M_{\odot}$ yr$^{-1}$. These values are similar to those found in the nuclear region of NGC~3227 \citep{alonso2019nuclear} but slightly lower than those found in other galaxies either using hot molecular gas H$_2$ $2.12\,\mu$m \citep{davies2014fueling} and ALMA CO(2-1) observations \citep{schnorrmuller2019outflows}. The molecular mass outflow rates in Mrk~1066 are about two orders of magnitude higher than the ionized gas mass outflow rate \citep[$6.2 \times 10^{-2}\,M_\odot$ yr$^{-1}$,][]{riffel2011compact}, which is in agreement with the previous findings in other moderate luminosity Seyfert galaxies \citep[e.g.,][]{fiore2017agn}.

\section{Summary and conclusions} \label{conclusions}

We presented new NOEMA observations of the CO(2-1) transition and adjacent continuum at $\sim 1.3$ mm of five nearby (D$_{L}$ between 19 and 58 Mpc) Seyfert galaxies. The achieved resolutions were  $0".3-0".8$, which provided physical resolutions $\sim 30-150$pc for the distances of our galaxies. The main conclusions are as follows.

\begin{enumerate}
    \item{{\bf 1.3 mm continuum:} All galaxies but one (NGC~4388) were detected at $1.3\,$mm. We found that the brightest continuum peak, which is mostly unresolved at our resolution, corresponds to the AGN position as determined by radio observations.}
    \item{{\bf CO(2-1) morphology:} The NOEMA CO(2-1) integrated line emission has been resolved in the five galaxies and extends over physical scales of 1-2\,kpc. The CO(2-1) morphologies resemble nuclear disks (Mrk~1066 and NGC~4253) with the emission peaking at the AGN position and circumnuclear rings/disks (NGC~2273, NGC~4388 and NGC~7465)  with several CO(2-1) emission regions not coincident with the AGN position. We also detected CO(2-1) spiral arm like features in the nuclear regions in four of them which were also well traced by dust extinction in $V-H$ or $I-H$ color maps.}
    \item{{\bf Molecular gas masses:} We measured significant amounts of molecular gas on both nuclear scales (30-150\,pc) ranging from $2 \times 10^{6}$ to $5 \times 10^{7}$ M$_{\odot}$ and circumnuclear scales (0.4-1.1\,kpc) ranging from $2 \times 10^{7}$ to $3 \times 10^{8}$ M$_{\odot}$. These values are similar to those measured in other Seyfert galaxies over similar physical regions.}
    \item{{\bf CO(2-1) kinematics:} All five galaxies in our sample show clear rotational patterns in the observed CO(2-1) mean-velocity field with varying degrees of non-circular motions. The galaxies in group/cluster environments (NGC~4388 and NGC~7465) show more complex CO(2-1) kinematics than those that appear to be isolated or relaxed (NGC~2273, Mrk~1066 and NGC~4253). Three out of the five galaxies show signs of streaming motions along bars and/or spiral arms. Mrk~1066 and NGC~4388 show signs of a molecular outflow (see Table~\ref{tab:resume_kin}). Only one galaxy (NGC~4388) shows both streaming motions and outflows. The typical velocity offsets of these motions (deprojected) with respect to the rotating component are in the range 175-252$\,{\rm km\,s}^{-1}$ and 116-151$\,{\rm km\,s}^{-1}$, respectively (see summary of the kinematics in Table 7).}
\end{enumerate}

Given the AGN luminosities of the galaxies in our sample, the expectation from the observed correlations is molecular outflows with maximum velocities of the order of $\sim 100\,{\rm km\,s}^{-1}$ \citep{fiore2017agn}. However, we only  found conclusive signs of molecular gas outflows in Mrk~1066 with an outflow rate of $3-3.5\,{\rm M}_\odot\,{\rm yr}^{-1}$
and tentative evidence in NGC~4388. Although our resolution may have an impact on these results, this clearly shows that it is only in cases of favorable geometry that we will be able to detect them. This includes the AGN wind impacting molecular gas in the host galaxy and non-circular motions close to the minor axis of the galaxy as in the cases of  NGC~1068 \citep{garcia2014molecular, garcia2019}, IC~5063 \citep{morganti2015radio}, and NGC~3227 \citep{alonso2019nuclear}. We also highlight the importance of assessing other non-circular motions (e.g., streaming motions due to the presence of a bar) which are of similar magnitude to those of AGN-driven outflows expected for moderate-luminosity AGN such as the typical local Seyfert galaxies.

\begin{acknowledgements} 

We thank the anonymous referee for her or his valuable comments that have improved significantly the quality of the paper. A.A.-H. acknowledges support from the Spanish Ministry of Science, Innovation and Universities through grant AYA2015-64346-C2-1-P, which was party funded by the FEDER program, and A.A.-H., S.G.-B. and A. U. through grant PGC2018-094671-B-I00 (MCIU, AEI, FEDER, EU).  S.G.-B. and C.R.A. acknowledge support from Spanish MINECO grant AYA2016-76682-C3-2-P. A.U acknowledges support from Spanish MINECO grants ESP2015-68964 and AYA2016-79006, partly funded by the FEDER program. A.L. acknowledges the support from Comunidad de Madrid through the Atracción de Talento grant 2017-T1/TIC-5213. A.A.-H. and A. L. work was done under project No. MDM-2017-0737 Unidad de Excelencia “María de Maeztu”- Centro de Astrobiología (INTA-CSIC). M.P.-S. acknowledges support from the Comunidad de Madrid, Spain, through Atracción de Talento Investigador Grant 2018-T1/TIC-11035 and STFC through grants ST/N000919/1 and ST/N002717/1. The authors wish to thank Erin K. S. Hicks for showing us the OSIRIS data prior to publication. We also wish to thank E. Di Teodoro for giving us the details of how position-velocity diagrams are produced in $^{\rm 3D}$BAROLO. This work is based on observations carried out under project numbers W14CB and W16BP with the IRAM NOEMA Interferometer. Data reduction was done using the GILDAS software package supported at IRAM (\url{http://www.iram.fr/IRAMFR/GILDAS}). Some of the data presented are observations made with the NASA/ESA Hubble Space Telescope, obtained from the data archive at the Space Telescope Science Institute and from the Hubble Legacy Archive. STScI is operated by the Association of Universities for Research in Astronomy, Inc. under NASA contract NAS 5-26555. HLA is a collaboration between the Space Telescope Science Institute (STScI/NASA), the Space Telescope European Coordinating Facility (ST-ECF/ESA) and the Canadian Astronomy Data Centre (CADC/NRC/CSA). This work has also made use of the NASA/IPAC Extragalactic Database (NED) which is operated by the Jet Propulsion Laboratory, California Institute of Technology, under contract with the National Aeronautics and Space Administration. In addition, it is acknowledged the usage of the HyperLeda database (\url{http://leda.univ-lyon1.fr}).

\end{acknowledgements}

\bibliographystyle{aa} 
\bibliography{bib.bib}

\begin{thebibliography}{149}
\expandafter\ifx\csname natexlab\endcsname\relax\def\natexlab#1{#1}\fi

\bibitem[{Afanasiev {et~al.}(1998)Afanasiev, Mikhailov, \&
  Shapovalova}]{afanasiev1998two}
Afanasiev, V., Mikhailov, V., \& Shapovalova, A. 1998, Astronomical and
  Astrophysical Transactions, 16, 257

\bibitem[{{Aguerri} {et~al.}(2015){Aguerri}, {M{\'e}ndez-Abreu},
  {Falc{\'o}n-Barroso}, {Amorin}, {Barrera-Ballesteros}, {Cid Fernandes},
  {Garc{\'\i}a-Benito}, {Garc{\'\i}a-Lorenzo}, {Gonz{\'a}lez Delgado},
  {Husemann}, {Kalinova}, {Lyubenova}, {Marino}, {M{\'a}rquez}, {Mast},
  {P{\'e}rez}, {S{\'a}nchez}, {van de Ven}, {Walcher}, {Backsmann},
  {Cortijo-Ferrero}, {Bland-Hawthorn}, {del Olmo}, {Iglesias-P{\'a}ramo},
  {P{\'e}rez}, {S{\'a}nchez-Bl{\'a}zquez}, {Wisotzki}, \&
  {Ziegler}}]{aguerri2015bar}
{Aguerri}, J.~A.~L., {M{\'e}ndez-Abreu}, J., {Falc{\'o}n-Barroso}, J., {et~al.}
  2015, \aap, 576, A102

\bibitem[{Alatalo {et~al.}(2013)Alatalo, Davis, Bureau, Young, Blitz, Crocker,
  Bayet, Bois, Bournaud, Cappellari, \& ...}]{alatalo2013atlas3d}
Alatalo, K., Davis, T.~A., Bureau, M., {et~al.} 2013, \mnras, 432, 1796

\bibitem[{{Albrecht} {et~al.}(2007){Albrecht}, {Kr{\"u}gel}, \&
  {Chini}}]{albrecht2007dust}
{Albrecht}, M., {Kr{\"u}gel}, E., \& {Chini}, R. 2007, \aap, 462, 575

\bibitem[{Alonso-Herrero {et~al.}(2016)Alonso-Herrero, Esquej, Roche,
  Ramos~Almeida, Gonz{\'a}lez-Mart{\'\i}n, Packham, Levenson, Mason,
  Hern{\'a}n-Caballero, Pereira-Santaella, \& ...}]{alonso2016mid}
Alonso-Herrero, A., Esquej, P., Roche, P., {et~al.} 2016, \mnras, 455, 563

\bibitem[{{Alonso-Herrero} {et~al.}(2019){Alonso-Herrero},
  {Garc{\'\i}a-Burillo}, {Pereira-Santaella}, {Davies}, {Combes},
  {Vestergaard}, {Raimundo}, {Bunker}, {D{\'\i}az-Santos}, {Gandhi},
  {Garc{\'\i}a-Bernete}, {Hicks}, {H{\"o}nig}, {Hunt}, {Imanishi}, {Izumi},
  {Levenson}, {Maciejewski}, {Packham}, {Ramos Almeida}, {Ricci}, {Rigopoulou},
  {Roche}, {Rosario}, {Schartmann}, {Usero}, \& {Ward}}]{alonso2019nuclear}
{Alonso-Herrero}, A., {Garc{\'\i}a-Burillo}, S., {Pereira-Santaella}, M.,
  {et~al.} 2019, \aap, 628, A65

\bibitem[{{Alonso-Herrero} {et~al.}(2018){Alonso-Herrero}, Pereira-Santaella,
  Garc{\'\i}a-Burillo, Davies, Combes, Asmus, Bunker, D{\'\i}az-Santos, Gandhi,
  Gonz{\'a}lez-Mart{\'\i}n, {et~al.}}]{alonso2018resolving}
{Alonso-Herrero}, A., Pereira-Santaella, M., Garc{\'\i}a-Burillo, S., {et~al.}
  2018, \apj, 859, 144

\bibitem[{Alonso-Herrero {et~al.}(2014)Alonso-Herrero, Ramos~Almeida, Esquej,
  Roche, Hern{\'a}n-Caballero, H{\"o}nig, Gonz{\'a}lez-Mart{\'\i}n, Aretxaga,
  Mason, Packham, \& ...}]{alonso2014nuclear}
Alonso-Herrero, A., Ramos~Almeida, C., Esquej, P., {et~al.} 2014, \mnras, 443,
  2766

\bibitem[{Alonso-Herrero {et~al.}(1998)Alonso-Herrero, Simpson, Ward, \&
  Wilson}]{alonso1998near}
Alonso-Herrero, A., Simpson, C., Ward, M.~J., \& Wilson, A.~S. 1998, \apj, 495,
  196

\bibitem[{{Athanassoula}(1992b)}]{athanassoula1992existence}
{Athanassoula}, E. 1992b, \mnras, 259, 345

\bibitem[{Awaki {et~al.}(2009)Awaki, Terashima, Higaki, \&
  Fukazawa}]{awaki2009detection}
Awaki, H., Terashima, Y., Higaki, Y., \& Fukazawa, Y. 2009, \pasj, 61, S317

\bibitem[{Barbosa {et~al.}(2006)Barbosa, Storchi-Bergmann, Fernandes, Winge, \&
  Schmitt}]{barbosa2006gemini}
Barbosa, F. K.~B., Storchi-Bergmann, T., Fernandes, R.~C., Winge, C., \&
  Schmitt, H. 2006, \mnras, 371, 170

\bibitem[{Barbosa {et~al.}(2009)Barbosa, Storchi-Bergmann, Fernandes, Winge, \&
  Schmitt}]{barbosa2009gemini}
Barbosa, F. K.~B., Storchi-Bergmann, T., Fernandes, R.~C., Winge, C., \&
  Schmitt, H. 2009, \mnras, 396, 2

\bibitem[{Bolatto {et~al.}(2013)Bolatto, Wolfire, \& Leroy}]{bolatto2013co}
Bolatto, A.~D., Wolfire, M., \& Leroy, A.~K. 2013, \araa, 51, 207

\bibitem[{Bower {et~al.}(1995)Bower, Wilson, Morse, Gelderman, Whittle, \&
  Mulchaey}]{bower1995radio}
Bower, G., Wilson, A., Morse, J.~A., {et~al.} 1995, \apj, 454, 106

\bibitem[{Braine \& Combes(1992)}]{braine1992co}
Braine, J. \& Combes, F. 1992, \aap, 264, 433

\bibitem[{{Bureau} \& {Athanassoula}(1999)}]{bureau1999bar}
{Bureau}, M. \& {Athanassoula}, E. 1999, \apj, 522, 686

\bibitem[{{Buta}(1987)}]{buta1987structure}
{Buta}, R. 1987, \apjs, 64, 383

\bibitem[{Carral {et~al.}(1990)Carral, Turner, \& Ho}]{carral199015}
Carral, P., Turner, J.~L., \& Ho, P.~T. 1990, \apj, 362, 434

\bibitem[{{Chen} {et~al.}(2019){Chen}, {Akiyama}, {Noda}, {Abdurro'uf},
  {Yamamura}, {Kawaguchi}, {Kokubo}, \& {Ichikawa}}]{chen2019discovery}
{Chen}, X., {Akiyama}, M., {Noda}, H., {et~al.} 2019, \pasj, 71, 29

\bibitem[{Chung {et~al.}(2009)Chung, Van~Gorkom, Kenney, Crowl, \&
  Vollmer}]{chung2009vla}
Chung, A., Van~Gorkom, J., Kenney, J.~D., Crowl, H., \& Vollmer, B. 2009, \apj,
  138, 1741

\bibitem[{Comastri(2004)}]{comastri2004compton}
Comastri, A. 2004, in Supermassive Black Holes in the Distant Universe
  (Springer), 245--272

\bibitem[{Combes {et~al.}(2009)Combes, Baker, Schinnerer, Garc{\'\i}a-Burillo,
  Hunt, Boone, Eckart, Neri, \& Tacconi}]{combes2009molecular}
Combes, F., Baker, A., Schinnerer, E., {et~al.} 2009, Astronomy \&
  Astrophysics, 503, 73

\bibitem[{Combes {et~al.}(2004)Combes, Boiss{\'e}, Mazure, \&
  Blanchard}]{combes2004galaxies}
Combes, F., Boiss{\'e}, P., Mazure, A., \& Blanchard, A. 2004, Galaxies and
  cosmology (Springer Science \& Business Media)

\bibitem[{Combes {et~al.}(2019)Combes, Garc{\'\i}a-Burillo, Audibert, Hunt,
  Eckart, Aalto, Casasola, Boone, Krips, Viti, {et~al.}}]{combes2019alma}
Combes, F., Garc{\'\i}a-Burillo, S., Audibert, A., {et~al.} 2019, \aap, 623,
  A79

\bibitem[{{Combes} {et~al.}(2013){Combes}, {Garc{\'{\i}}a-Burillo}, {Casasola},
  {Hunt}, {Krips}, {Baker}, {Boone}, {Eckart}, {Marquez}, {Neri}, {Schinnerer},
  \& {Tacconi}}]{combes2013alma}
{Combes}, F., {Garc{\'{\i}}a-Burillo}, S., {Casasola}, V., {et~al.} 2013, \aap,
  558, A124

\bibitem[{{Combes} {et~al.}(2014){Combes}, {Garc{\'{\i}}a-Burillo}, {Casasola},
  {Hunt}, {Krips}, {Baker}, {Boone}, {Eckart}, {Marquez}, {Neri}, {Schinnerer},
  \& {Tacconi}}]{combes2014alma}
{Combes}, F., {Garc{\'{\i}}a-Burillo}, S., {Casasola}, V., {et~al.} 2014, \aap,
  565, A97

\bibitem[{Contini {et~al.}(1998)Contini, Considere, \&
  Davoust}]{contini1998starbursts}
Contini, T., Considere, S., \& Davoust, E. 1998, \aaps, 130, 285

\bibitem[{Crocker {et~al.}(2012)Crocker, Krips, Bureau, Young, Davis, Bayet,
  Alatalo, Blitz, Bois, Bournaud, {et~al.}}]{crocker2012atlas3d}
Crocker, A., Krips, M., Bureau, M., {et~al.} 2012, \mnras, 421, 1298

\bibitem[{{Davies} {et~al.}(2011){Davies}, {F{\"o}rster Schreiber}, {Cresci},
  {Genzel}, {Bouch{\'e}}, {Burkert}, {Buschkamp}, {Genel}, {Hicks}, {Kurk},
  {Lutz}, {Newman}, {Shapiro}, {Sternberg}, {Tacconi}, \&
  {Wuyts}}]{davies2011how}
{Davies}, R., {F{\"o}rster Schreiber}, N.~M., {Cresci}, G., {et~al.} 2011,
  \apj, 741, 69

\bibitem[{Davies {et~al.}(2014)Davies, Maciejewski, Hicks, Emsellem, Erwin,
  Burtscher, Dumas, Lin, Malkan, M{\"u}ller-S{\'a}nchez,
  {et~al.}}]{davies2014fueling}
Davies, R., Maciejewski, W., Hicks, E., {et~al.} 2014, \apj, 792, 101

\bibitem[{Di~Teodoro \& Fraternali(2015)}]{teodoro20153d}
Di~Teodoro, E. \& Fraternali, F. 2015, \mnras, 451, 3021

\bibitem[{{Downes} {et~al.}(1996){Downes}, {Reynaud}, {Solomon}, \&
  {Radford}}]{downes1996co}
{Downes}, D., {Reynaud}, D., {Solomon}, P.~M., \& {Radford}, S.~J.~E. 1996,
  \apj, 461, 186

\bibitem[{Elvis {et~al.}(2004)Elvis, Risaliti, Nicastro, Miller, Fiore, \&
  Puccetti}]{elvis2004unveiling}
Elvis, M., Risaliti, G., Nicastro, F., {et~al.} 2004, \apjl, 615, L25

\bibitem[{{Erwin} \& {Sparke}(2002)}]{erwin2002double}
{Erwin}, P. \& {Sparke}, L.~S. 2002, \aj, 124, 65

\bibitem[{{Erwin} \& {Sparke}(2003)}]{erwin2003imaging}
{Erwin}, P. \& {Sparke}, L.~S. 2003, \apjs, 146, 299

\bibitem[{{Federrath} {et~al.}(2017){Federrath}, {Salim}, {Medling}, {Davies},
  {Yuan}, {Bian}, {Groves}, {Ho}, {Sharp}, {Kewley}, {Sweet}, {Richards},
  {Bryant}, {Brough}, {Croom}, {Scott}, {Lawrence}, {Konstantopoulos}, \&
  {Goodwin}}]{federrath2017sami}
{Federrath}, C., {Salim}, D.~M., {Medling}, A.~M., {et~al.} 2017, \mnras, 468,
  3965

\bibitem[{Ferruit {et~al.}(2000)Ferruit, Wilson, \&
  Mulchaey}]{ferruit2000hubble}
Ferruit, P., Wilson, A.~S., \& Mulchaey, J. 2000, \apjs, 128, 139

\bibitem[{Fiore {et~al.}(2017)Fiore, Feruglio, Shankar, Bischetti, Bongiorno,
  Brusa, Carniani, Cicone, Duras, Lamastra, {et~al.}}]{fiore2017agn}
Fiore, F., Feruglio, C., Shankar, F., {et~al.} 2017, \aap, 601, A143

\bibitem[{{Fischer} {et~al.}(2013){Fischer}, {Crenshaw}, {Kraemer}, \&
  {Schmitt}}]{fischer2013determining}
{Fischer}, T.~C., {Crenshaw}, D.~M., {Kraemer}, S.~B., \& {Schmitt}, H.~R.
  2013, \apjs, 209, 1

\bibitem[{{Gallimore} {et~al.}(2016){Gallimore}, {Elitzur}, {Maiolino},
  {Marconi}, {O'Dea}, {Lutz}, {Baum}, {Nikutta}, {Impellizzeri}, {Davies},
  {Kimball}, \& {Sani}}]{gallimore2016high}
{Gallimore}, J.~F., {Elitzur}, M., {Maiolino}, R., {et~al.} 2016, \apjl, 829,
  L7

\bibitem[{{Garc{\'{\i}}a-Burillo} \& {Combes}(2012)}]{garcia2012feeding}
{Garc{\'{\i}}a-Burillo}, S. \& {Combes}, F. 2012, in Journal of Physics
  Conference Series, Vol. 372, Journal of Physics Conference Series, 012050

\bibitem[{Garc{\'\i}a-Burillo {et~al.}(2003)Garc{\'\i}a-Burillo, Combes, Hunt,
  Boone, Baker, Tacconi, Eckart, Neri, Leon, Schinnerer, \&
  ...}]{garcia2003molecular}
Garc{\'\i}a-Burillo, S., Combes, F., Hunt, L., {et~al.} 2003, \aap, 407, 485

\bibitem[{{Garcia-Burillo} {et~al.}(2019){Garcia-Burillo}, {Combes}, {Ramos
  Almeida}, {Usero}, {Alonso-Herrero}, {Hunt}, {Rouan}, {Aalto}, {Querejeta},
  {Viti}, {van der Werf}, {Vives-Arias}, {Fuente}, {Colina}, {Martin-Pintado},
  {Henkel}, {Martin}, {Krips}, {Gratadour}, {Neri}, \& {Tacconi}}]{garcia2019}
{Garcia-Burillo}, S., {Combes}, F., {Ramos Almeida}, C., {et~al.} 2019, arXiv
  e-prints, arXiv:1909.00675

\bibitem[{Garc{\'\i}a-Burillo {et~al.}(2005)Garc{\'\i}a-Burillo, Combes,
  Schinnerer, Boone, \& Hunt}]{garcia2005molecular}
Garc{\'\i}a-Burillo, S., Combes, F., Schinnerer, E., Boone, F., \& Hunt, L.
  2005, \aap, 441, 1011

\bibitem[{{Garc{\'{\i}}a-Burillo} {et~al.}(2014){Garc{\'{\i}}a-Burillo},
  {Combes}, {Usero}, {Aalto}, {Krips}, {Viti}, {Alonso-Herrero}, {Hunt},
  {Schinnerer}, {Baker}, {Boone}, {Casasola}, {Colina}, {Costagliola},
  {Eckart}, {Fuente}, {Henkel}, {Labiano}, {Mart{\'{\i}}n}, {M{\'a}rquez},
  {Muller}, {Planesas}, {Ramos Almeida}, {Spaans}, {Tacconi}, \& {van der
  Werf}}]{garcia2014molecular}
{Garc{\'{\i}}a-Burillo}, S., {Combes}, F., {Usero}, A., {et~al.} 2014, \aap,
  567, A125

\bibitem[{Garc{\'\i}a-Burillo {et~al.}(2009)Garc{\'\i}a-Burillo,
  Fern{\'a}ndez-Garc{\'\i}a, Combes, Hunt, Haan, Schinnerer, Boone, Krips, \&
  M{\'a}rquez}]{garcia2009molecular}
Garc{\'\i}a-Burillo, S., Fern{\'a}ndez-Garc{\'\i}a, S., Combes, F., {et~al.}
  2009, \aap, 496, 85

\bibitem[{{Garcia-Burillo} {et~al.}(1994){Garcia-Burillo}, {Sempere}, \&
  {Combes}}]{garcia1994gas}
{Garcia-Burillo}, S., {Sempere}, M.~J., \& {Combes}, F. 1994, Astronomy and
  Astrophysics, 287, 419

\bibitem[{Gimeno {et~al.}(2004)Gimeno, D{\'\i}az, \&
  Carranza}]{gimeno2004catalog}
Gimeno, G.~N., D{\'\i}az, R.~J., \& Carranza, G.~J. 2004, \apj, 128, 62

\bibitem[{Giroletti \& Panessa(2009)}]{giroletti2009faintest}
Giroletti, M. \& Panessa, F. 2009, \apjl, 706, L260

\bibitem[{Gong {et~al.}(2018)Gong, Ostriker, \& Kim}]{gong2018x}
Gong, M., Ostriker, E.~C., \& Kim, C.-G. 2018, \apj, 858, 16

\bibitem[{Gonzalez~Delgado \& P{\'e}rez(1996)}]{gonzalez1996spectrophotometric}
Gonzalez~Delgado, R.~M. \& P{\'e}rez, E. 1996, \mnras, 278, 737

\bibitem[{Goodrich \& Osterbrock(1983)}]{goodrich1983mrk}
Goodrich, R. \& Osterbrock, D. 1983, \apj, 269, 416

\bibitem[{Greene {et~al.}(2013)Greene, Seth, Den~Brok, Braatz, Henkel, Sun,
  Peng, Kuo, Impellizzeri, \& Lo}]{greene2013using}
Greene, J.~E., Seth, A., Den~Brok, M., {et~al.} 2013, \apj, 771, 121

\bibitem[{Greene {et~al.}(2014)Greene, Seth, Lyubenova, Walsh, Van De~Ven, \&
  L{\"a}sker}]{greene2014circumnuclear}
Greene, J.~E., Seth, A., Lyubenova, M., {et~al.} 2014, \apj, 788, 145

\bibitem[{{Grossi} {et~al.}(2016){Grossi}, {Corbelli}, {Bizzocchi},
  {Giovanardi}, {Bomans}, {Coelho}, {De Looze}, {Gon{\c{c}}alves}, {Hunt},
  {Leonardo}, {Madden}, {Men{\'e}ndez-Delmestre}, {Pappalardo}, \&
  {Riguccini}}]{grossi2016starforming}
{Grossi}, M., {Corbelli}, E., {Bizzocchi}, L., {et~al.} 2016, \aap, 590, A27

\bibitem[{Guainazzi {et~al.}(2005)Guainazzi, Matt, \& Perola}]{guainazzi2005x}
Guainazzi, M., Matt, G., \& Perola, G.~C. 2005, \aap, 444, 119

\bibitem[{Haan {et~al.}(2009)Haan, Schinnerer, Emsellem, Garc{\'\i}a-Burillo,
  Combes, Mundell, \& Rix}]{haan2009dynamical}
Haan, S., Schinnerer, E., Emsellem, E., {et~al.} 2009, \apj, 692, 1623

\bibitem[{Hasegawa(1997)}]{hasegawa1997co}
Hasegawa, T. 1997, in CO: Twenty-Five Years of Millimeter-Wave Spectroscopy
  (Springer), 39--46

\bibitem[{{Heller} \& {Shlosman}(1994)}]{heller1994fueling}
{Heller}, C.~H. \& {Shlosman}, I. 1994, \apj, 424, 84

\bibitem[{Henkel {et~al.}(2005)Henkel, Peck, Tarchi, Nagar, Braatz, Castangia,
  \& Moscadelli}]{henkel2005new}
Henkel, C., Peck, A., Tarchi, A., {et~al.} 2005, \aap, 436, 75

\bibitem[{Hern{\'a}ndez-Garc{\'\i}a {et~al.}(2015)Hern{\'a}ndez-Garc{\'\i}a,
  Masegosa, Gonz{\'a}lez-Mart{\'\i}n, \& M{\'a}rquez}]{hernandez2015x}
Hern{\'a}ndez-Garc{\'\i}a, L., Masegosa, J., Gonz{\'a}lez-Mart{\'\i}n, O., \&
  M{\'a}rquez, I. 2015, \aap, 579, A90

\bibitem[{Hicks {et~al.}(2009)Hicks, Davies, Malkan, Genzel, Tacconi,
  S{\'a}nchez, \& Sternberg}]{hicks2009role}
Hicks, E., Davies, R., Malkan, M., {et~al.} 2009, \apj, 696, 448

\bibitem[{{H{\"o}nig} \& {Kishimoto}(2017)}]{honig2018dusty}
{H{\"o}nig}, S.~F. \& {Kishimoto}, M. 2017, \apjl, 838, L20

\bibitem[{{H{\"o}nig} {et~al.}(2010){H{\"o}nig}, {Kishimoto}, {Gandhi},
  {Smette}, {Asmus}, {Duschl}, {Polletta}, \& {Weigelt}}]{honig2010dusty}
{H{\"o}nig}, S.~F., {Kishimoto}, M., {Gandhi}, P., {et~al.} 2010, \aap, 515,
  A23

\bibitem[{Hummel \& Saikia(1991)}]{hummel1991anomalous}
Hummel, E. \& Saikia, D. 1991, \aap, 249, 43

\bibitem[{{Huntley}(1978)}]{huntley1977gas}
{Huntley}, J.~M. 1978, \apjl, 225, L101

\bibitem[{Iorio {et~al.}(2016)Iorio, Fraternali, Nipoti, Di~Teodoro, Read, \&
  Battaglia}]{iorio2016little}
Iorio, G., Fraternali, F., Nipoti, C., {et~al.} 2016, \mnras, 466, 4159

\bibitem[{{Irwin} {et~al.}(2019){Irwin}, {Damas-Segovia}, {Krause},
  {Miskolczi}, {Li}, {Stein}, {English}, {Henriksen}, {Beck}, {Wiegert}, \&
  {Dettmar}}]{irwin2019changesxviii}
{Irwin}, J., {Damas-Segovia}, A., {Krause}, M., {et~al.} 2019, Galaxies, 7, 42

\bibitem[{Iwasawa {et~al.}(2003)Iwasawa, Wilson, Fabian, \&
  Young}]{iwasawa2003x}
Iwasawa, K., Wilson, A., Fabian, A., \& Young, A. 2003, \mnras, 345, 369

\bibitem[{Izumi {et~al.}(2018)Izumi, Wada, Fukushige, Hamamura, \&
  Kohno}]{izumi2018circumnuclear}
Izumi, T., Wada, K., Fukushige, R., Hamamura, S., \& Kohno, K. 2018, \apj, 867,
  48

\bibitem[{Kamali {et~al.}(2017)Kamali, Henkel, Brunthaler, Impellizzeri,
  Menten, Braatz, Greene, Reid, Condon, Lo, {et~al.}}]{kamali2017radio}
Kamali, F., Henkel, C., Brunthaler, A., {et~al.} 2017, \aap, 605, A84

\bibitem[{{Kenney}(1994)}]{kenney1994observations}
{Kenney}, J.~D.~P. 1994, in Mass-Transfer Induced Activity in Galaxies, ed.
  I.~{Shlosman}, 78

\bibitem[{Krips {et~al.}(2007)Krips, Neri, Garc{\'\i}a-Burillo, Combes,
  Schinnerer, Baker, Eckart, Boone, Hunt, Leon, {et~al.}}]{krips2007molecular}
Krips, M., Neri, R., Garc{\'\i}a-Burillo, S., {et~al.} 2007, \aap, 468, L63

\bibitem[{Kr{\"u}gel {et~al.}(1990)Kr{\"u}gel, Chini, \& Steppe}]{krugel1990co}
Kr{\"u}gel, E., Chini, R., \& Steppe, H. 1990, \aap, 229, 17

\bibitem[{Kukula {et~al.}(1995)Kukula, Pedlar, Baum, \&
  O’Dea}]{kukula1995high}
Kukula, M.~J., Pedlar, A., Baum, S.~A., \& O’Dea, C.~P. 1995, \mnras, 276,
  1262

\bibitem[{{Kuno} {et~al.}(2008){Kuno}, {Sato}, {Nakanishi}, {Hirota}, {Tosaki},
  {Shioya}, {Sorai}, {Nakai}, {Nishiyama}, \&
  {Vila-Vilar{\'o}}}]{kuno2008nobeyama}
{Kuno}, N., {Sato}, N., {Nakanishi}, H., {et~al.} 2008, Astrophysics and Space
  Science Proceedings, 4, 170

\bibitem[{Kuo {et~al.}(2010)Kuo, Braatz, Condon, Impellizzeri, Lo, Zaw,
  Schenker, Henkel, Reid, \& Greene}]{kuo2010megamaser}
Kuo, C., Braatz, J., Condon, J., {et~al.} 2010, \apj, 727, 20

\bibitem[{Levenson {et~al.}(2001)Levenson, Weaver, \&
  Heckman}]{levenson2001seyfert}
Levenson, N., Weaver, K., \& Heckman, T. 2001, \apj, 550, 230

\bibitem[{Li \& Seaquist(1994)}]{li1994neutral}
Li, J.~G. \& Seaquist, E. 1994, \apj, 107, 1953

\bibitem[{{Lu} {et~al.}(1993){Lu}, {Hoffman}, {Groff}, {Roos}, \&
  {Lamphier}}]{lu1993hi}
{Lu}, N.~Y., {Hoffman}, G.~L., {Groff}, T., {Roos}, T., \& {Lamphier}, C. 1993,
  \apjs, 88, 383

\bibitem[{{Lynden-Bell} \& {Pringle}(1974)}]{lynden1974evolution}
{Lynden-Bell}, D. \& {Pringle}, J.~E. 1974, \mnras, 168, 603

\bibitem[{{Maiolino} {et~al.}(1997){Maiolino}, {Ruiz}, {Rieke}, \&
  {Papadopoulos}}]{Maiolino1997molecular}
{Maiolino}, R., {Ruiz}, M., {Rieke}, G.~H., \& {Papadopoulos}, P. 1997, \apj,
  485, 552

\bibitem[{Malkan {et~al.}(1998)Malkan, Gorjian, \& Tam}]{malkan1998hubble}
Malkan, M.~A., Gorjian, V., \& Tam, R. 1998, \apjs, 117, 25

\bibitem[{{Marinucci} {et~al.}(2012){Marinucci}, {Bianchi}, {Nicastro}, {Matt},
  \& {Goulding}}]{marinucci2012link}
{Marinucci}, A., {Bianchi}, S., {Nicastro}, F., {Matt}, G., \& {Goulding},
  A.~D. 2012, \apj, 748, 130

\bibitem[{{M{\'a}rquez} {et~al.}(1999){M{\'a}rquez}, {Durret}, {Gonz{\'a}lez
  Delgado}, {Marrero}, {Masegosa}, {Maza}, {Moles}, {P{\'e}rez}, \&
  {Roth}}]{marquez1999near}
{M{\'a}rquez}, I., {Durret}, F., {Gonz{\'a}lez Delgado}, R.~M., {et~al.} 1999,
  \aaps, 140, 1

\bibitem[{Martini {et~al.}(2003)Martini, Regan, Mulchaey, \&
  Pogge}]{martini2003circumnuclear}
Martini, P., Regan, M.~W., Mulchaey, J.~S., \& Pogge, R.~W. 2003, \apjs, 146,
  353

\bibitem[{Mason {et~al.}(2015)Mason, Rodriguez-Ardila, Martins, Riffel,
  Mart{\'\i}n, Almeida, Dutra, Ho, Thanjavur, Flohic,
  {et~al.}}]{mason2015nuclear}
Mason, R.~E., Rodriguez-Ardila, A., Martins, L., {et~al.} 2015, \apjs, 217, 13

\bibitem[{{Mazzalay} {et~al.}(2014){Mazzalay}, {Maciejewski}, {Erwin},
  {Saglia}, {Bender}, {Fabricius}, {Nowak}, {Rusli}, \&
  {Thomas}}]{mazzalay2014molecular}
{Mazzalay}, X., {Maciejewski}, W., {Erwin}, P., {et~al.} 2014, \mnras, 438,
  2036

\bibitem[{Mazzarella \& Boroson(1993)}]{mazzarella1993optical}
Mazzarella, J.~M. \& Boroson, T.~A. 1993, \apjs, 85, 27

\bibitem[{Merkulova {et~al.}(2012)Merkulova, Karataeva, Yakovleva, \&
  Burenkov}]{merkulova2012study}
Merkulova, O., Karataeva, G., Yakovleva, V., \& Burenkov, A. 2012, Astronomy
  letters, 38, 290

\bibitem[{{Miller} {et~al.}(2019){Miller}, {Kammoun}, {Ludlam}, {Gendreau},
  {Arzoumanian}, {Cackett}, \& {Tombesi}}]{miller2019nicer}
{Miller}, J.~M., {Kammoun}, E., {Ludlam}, R.~M., {et~al.} 2019, arXiv e-prints,
  arXiv:1908.08023

\bibitem[{Mirabel \& Wilson(1984)}]{mirabel1984neutral}
Mirabel, I. \& Wilson, A. 1984, \apj, 277, 92

\bibitem[{Moiseev {et~al.}(2004)Moiseev, Vald{\'e}s, \&
  Chavushyan}]{moiseev2004structure}
Moiseev, A.~V., Vald{\'e}s, J., \& Chavushyan, V. 2004, \aap, 421, 433

\bibitem[{{Morganti} {et~al.}(2015){Morganti}, {Oosterloo}, {Oonk},
  {Frieswijk}, \& {Tadhunter}}]{morganti2015radio}
{Morganti}, R., {Oosterloo}, T., {Oonk}, J.~B.~R., {Frieswijk}, W., \&
  {Tadhunter}, C. 2015, \aap, 580, A1

\bibitem[{Mulchaey {et~al.}(1996{\natexlab{a}})Mulchaey, Wilson, \&
  Tsvetanov}]{mulchaey1996emission}
Mulchaey, J.~S., Wilson, A.~S., \& Tsvetanov, Z. 1996{\natexlab{a}}, \apjs,
  102, 309

\bibitem[{Mulchaey {et~al.}(1996{\natexlab{b}})Mulchaey, Wilson, \&
  Tsvetanov}]{mulchaey1996emissionii}
Mulchaey, J.~S., Wilson, A.~S., \& Tsvetanov, Z. 1996{\natexlab{b}}, \apj, 467,
  197

\bibitem[{M{\"u}ller-S{\'a}nchez {et~al.}(2018)M{\"u}ller-S{\'a}nchez, Hicks,
  Malkan, Davies, Yu, Shaver, \& Davis}]{muller2018keck}
M{\"u}ller-S{\'a}nchez, F., Hicks, E., Malkan, M., {et~al.} 2018, \apj, 858, 48

\bibitem[{Mundell {et~al.}(2009)Mundell, Ferruit, Nagar, \&
  Wilson}]{mundell2009radio}
Mundell, C., Ferruit, P., Nagar, N., \& Wilson, A. 2009, \apj, 703, 802

\bibitem[{Mu{\~n}oz~Mar{\'\i}n {et~al.}(2009)Mu{\~n}oz~Mar{\'\i}n,
  Storchi-Bergmann, Delgado, Schmitt, Spinelli, P{\'e}rez, \&
  Cid~Fernandes}]{munoz2009nature}
Mu{\~n}oz~Mar{\'\i}n, V., Storchi-Bergmann, T., Delgado, R.~G., {et~al.} 2009,
  \mnras, 399, 842

\bibitem[{Nagar {et~al.}(1999)Nagar, Wilson, Mulchaey, \&
  Gallimore}]{nagar1999radio}
Nagar, N.~M., Wilson, A.~S., Mulchaey, J.~S., \& Gallimore, J.~F. 1999, \apjs,
  120, 209

\bibitem[{Nyland {et~al.}(2016)Nyland, Young, Wrobel, Sarzi, Morganti, Alatalo,
  Blitz, Bournaud, Bureau, Cappellari, {et~al.}}]{nyland2016atlas3d}
Nyland, K., Young, L.~M., Wrobel, J.~M., {et~al.} 2016, \mnras, 458, 2221

\bibitem[{Oosterloo \& van Gorkom(2005)}]{oosterloo2005large}
Oosterloo, T. \& van Gorkom, J. 2005, \aap, 437, L19

\bibitem[{Osterbrock \& Martel(1993)}]{osterbrock1993spectroscopic}
Osterbrock, D.~E. \& Martel, A. 1993, \apj, 414, 552

\bibitem[{Osterbrock \& Pogge(1985)}]{osterbrock1985spectra}
Osterbrock, D.~E. \& Pogge, R.~W. 1985, \apj, 297, 166

\bibitem[{Pappalardo {et~al.}(2012)Pappalardo, Bianchi, Corbelli, Giovanardi,
  Hunt, Bendo, Boselli, Cortese, Magrini, Zibetti,
  {et~al.}}]{pappalardo2012herschel}
Pappalardo, C., Bianchi, S., Corbelli, E., {et~al.} 2012, \aap, 545, A75

\bibitem[{Petitpas \& Wilson(2002)}]{petitpas2002molecular}
Petitpas, G.~R. \& Wilson, C.~D. 2002, \apj, 575, 814

\bibitem[{Petitpas \& Wilson(2003)}]{petitpas2003molecular}
Petitpas, G.~R. \& Wilson, C.~D. 2003, \apj, 587, 649

\bibitem[{{Pringle}(1981)}]{pringle1981accretion}
{Pringle}, J.~E. 1981, \araa, 19, 137

\bibitem[{Ramos~Almeida {et~al.}(2014)Ramos~Almeida, Alonso-Herrero, Esquej,
  Gonz{\'a}lez-Mart{\'\i}n, Riffel, Garc{\'\i}a-Bernete,
  Rodr{\'\i}guez~Espinosa, Packham, Levenson, Roche, \& ...}]{ramos2014mid}
Ramos~Almeida, C., Alonso-Herrero, A., Esquej, P., {et~al.} 2014, \mnras, 445,
  1130

\bibitem[{Ramos~Almeida {et~al.}(2009)Ramos~Almeida, Garc{\'\i}a, \&
  Acosta-Pulido}]{almeida2009near}
Ramos~Almeida, C., Garc{\'\i}a, A.~P., \& Acosta-Pulido, J.~A. 2009, \apj, 694,
  1379

\bibitem[{Ricci {et~al.}(2017)Ricci, Trakhtenbrot, Koss, Ueda, Del~Vecchio,
  Treister, Schawinski, Paltani, Oh, Lamperti, {et~al.}}]{ricci2017bat}
Ricci, C., Trakhtenbrot, B., Koss, M.~J., {et~al.} 2017, \apjs, 233, 17

\bibitem[{Riffel \& Storchi-Bergmann(2011)}]{riffel2011compact}
Riffel, R.~A. \& Storchi-Bergmann, T. 2011, \mnras, 411, 469

\bibitem[{Riffel {et~al.}(2017)Riffel, Storchi-Bergmann, Riffel, Dahmer-Hahn,
  Diniz, Sch{\"o}nell, \& Dametto}]{riffel2017gemini}
Riffel, R.~A., Storchi-Bergmann, T., Riffel, R., {et~al.} 2017, \mnras, 470,
  992

\bibitem[{{Rosario} {et~al.}(2019){Rosario}, {Togi}, {Burtscher}, {Davies},
  {Shimizu}, \& {Lutz}}]{rosario2019accreting}
{Rosario}, D.~J., {Togi}, A., {Burtscher}, L., {et~al.} 2019, \apjl, 875, L8

\bibitem[{Sakamoto {et~al.}(1999)Sakamoto, Okumura, Ishizuki, \&
  Scoville}]{sakamoto1999co}
Sakamoto, K., Okumura, S.~K., Ishizuki, S., \& Scoville, N. 1999, \apjs, 124,
  403

\bibitem[{{Salak} {et~al.}(2016){Salak}, {Nakai}, {Hatakeyama}, \&
  {Miyamoto}}]{salak2016gas}
{Salak}, D., {Nakai}, N., {Hatakeyama}, T., \& {Miyamoto}, Y. 2016, \apj, 823,
  68

\bibitem[{Sani {et~al.}(2012)Sani, Davies, Sternberg, Graci{\'a}-Carpio, Hicks,
  Krips, Tacconi, Genzel, Vollmer, Schinnerer, \& ...}]{sani2012physical}
Sani, E., Davies, R., Sternberg, A., {et~al.} 2012, \mnras, 424, 1963

\bibitem[{Schinnerer {et~al.}(2000{\natexlab{a}})Schinnerer, Eckart, Tacconi,
  Genzel, \& Downes}]{schinnerer2000bars}
Schinnerer, E., Eckart, A., Tacconi, L., Genzel, R., \& Downes, D.
  2000{\natexlab{a}}, \apj, 533, 850

\bibitem[{Schinnerer {et~al.}(2000{\natexlab{b}})Schinnerer, Eckart, \&
  Tacconi}]{schinnerer2000distribution}
Schinnerer, E., Eckart, A., \& Tacconi, L.~J. 2000{\natexlab{b}}, \apj, 533,
  826

\bibitem[{Schmitt {et~al.}(2003)Schmitt, Donley, Antonucci, Hutchings, \&
  Kinney}]{schmitt2003hubble}
Schmitt, H.~R., Donley, J.~L., Antonucci, R., Hutchings, J., \& Kinney, A.
  2003, \apjs, 148, 327

\bibitem[{Schneider(2007)}]{schneider2007extragalactic}
Schneider, P. 2007, Extragalactic astronomy and cosmology: an introduction
  (Springer Science \& Business Media)

\bibitem[{Schnorr-M{\"u}ller {et~al.}(2014)Schnorr-M{\"u}ller,
  Storchi-Bergmann, Nagar, \& Ferrari}]{schnorr2014gas}
Schnorr-M{\"u}ller, A., Storchi-Bergmann, T., Nagar, N.~M., \& Ferrari, F.
  2014, \mnras, 438, 3322

\bibitem[{Sch{\"o}nell {et~al.}(2014)Sch{\"o}nell, Riffel, Storchi-Bergmann, \&
  Winge}]{schonell2014feeding}
Sch{\"o}nell, A.~J., Riffel, R.~A., Storchi-Bergmann, T., \& Winge, C. 2014,
  \mnras, 445, 414

\bibitem[{{Sheth} {et~al.}(2002){Sheth}, {Vogel}, {Regan}, {Teuben}, {Harris},
  \& {Thornley}}]{sheth2002molecular}
{Sheth}, K., {Vogel}, S.~N., {Regan}, M.~W., {et~al.} 2002, \aj, 124, 2581

\bibitem[{{Shlosman} {et~al.}(1989){Shlosman}, {Frank}, \&
  {Begelman}}]{Shlosman1989accretion}
{Shlosman}, I., {Frank}, J., \& {Begelman}, M.~C. 1989, \nat, 338, 45

\bibitem[{{Slater} {et~al.}(2019){Slater}, {Nagar}, {Schnorr-M{\"u}ller},
  {Storchi-Bergmann}, {Finlez}, {Lena}, {Ramakrishnan}, {Mundell}, {Riffel},
  {Peterson}, {Robinson}, \& {Orellana}}]{schnorrmuller2019outflows}
{Slater}, R., {Nagar}, N.~M., {Schnorr-M{\"u}ller}, A., {et~al.} 2019, \aap,
  621, A83

\bibitem[{Smirnova {et~al.}(2010)Smirnova, Moiseev, \&
  Afanasiev}]{smirnova2010seyfert}
Smirnova, A., Moiseev, A., \& Afanasiev, V. 2010, \mnras, 408, 400

\bibitem[{{Tacconi} {et~al.}(2013){Tacconi}, {Neri}, {Genzel}, {Combes},
  {Bolatto}, {Cooper}, {Wuyts}, {Bournaud}, {Burkert}, {Comerford}, {Cox},
  {Davis}, {F{\"o}rster Schreiber}, {Garc{\'\i}a-Burillo}, {Gracia-Carpio},
  {Lutz}, {Naab}, {Newman}, {Omont}, {Saintonge}, {Shapiro Griffin}, {Shapley},
  {Sternberg}, \& {Weiner}}]{tacconi2013phibss}
{Tacconi}, L.~J., {Neri}, R., {Genzel}, R., {et~al.} 2013, \apj, 768, 74

\bibitem[{Taniguchi {et~al.}(1990)Taniguchi, Kameya, Nakai, \&
  Kawara}]{taniguchi1990circumnuclear}
Taniguchi, Y., Kameya, O., Nakai, N., \& Kawara, K. 1990, \apj, 358, 132

\bibitem[{Thean {et~al.}(2001)Thean, Gillibrand, Pedlar, \&
  Kukula}]{thean2001merlin}
Thean, A., Gillibrand, T., Pedlar, A., \& Kukula, M. 2001, \mnras, 327, 369

\bibitem[{Thomas {et~al.}(2002)Thomas, Dunne, Clemens, Alexander, Eales, Green,
  \& James}]{thomas2002extended}
Thomas, H., Dunne, L., Clemens, M., {et~al.} 2002, \mnras, 331, 853

\bibitem[{Ueda {et~al.}(2014)Ueda, Iono, Yun, Crocker, Narayanan, Komugi,
  Espada, Hatsukade, Kaneko, Matsuda, {et~al.}}]{ueda2014cold}
Ueda, J., Iono, D., Yun, M.~S., {et~al.} 2014, The Astrophysical Journal
  Supplement Series, 214, 1

\bibitem[{Ulvestad \& Wilson(1989)}]{ulvestad1989radio}
Ulvestad, J.~S. \& Wilson, A.~S. 1989, \apj, 343, 659

\bibitem[{{van der Kruit} \& {Allen}(1978)}]{van1978kinematics}
{van der Kruit}, P.~C. \& {Allen}, R.~J. 1978, \araa, 16, 103

\bibitem[{{van der Laan} {et~al.}(2011){van der Laan}, {Schinnerer}, {Boone},
  {Garc{\'{\i}}a-Burillo}, {Combes}, {Haan}, {Leon}, {Hunt}, \&
  {Baker}}]{van2011molecular}
{van der Laan}, T.~P.~R., {Schinnerer}, E., {Boone}, F., {et~al.} 2011, \aap,
  529, A45

\bibitem[{Van~Driel {et~al.}(1992)Van~Driel, Augarde, Bottinelli, Gouguenheim,
  Hamabe, Maehara, Baan, Goudfrooij, \& Groenewegen}]{van1992study}
Van~Driel, W., Augarde, R., Bottinelli, L., {et~al.} 1992, \aap, 259, 71

\bibitem[{Van~Driel \& Buta(1991)}]{van1991study}
Van~Driel, W. \& Buta, R. 1991, \aap, 245, 7

\bibitem[{{Vargas} {et~al.}(2019){Vargas}, {Walterbos}, {Rand }, {Stil},
  {Krause}, {Li}, {Irwin}, \& {Dettmar}}]{vargas2019changesxvii}
{Vargas}, C.~J., {Walterbos}, R. A.~M., {Rand }, R.~J., {et~al.} 2019, \apj,
  881, 26

\bibitem[{Veilleux {et~al.}(1999{\natexlab{a}})Veilleux, Bland-Hawthorn, \&
  Cecil}]{veilleux1999kinematic}
Veilleux, S., Bland-Hawthorn, J., \& Cecil, G. 1999{\natexlab{a}}, \apj, 118,
  2108

\bibitem[{Veilleux {et~al.}(1999{\natexlab{b}})Veilleux, Bland-Hawthorn, Cecil,
  Tully, \& Miller}]{veilleux1999galactic}
Veilleux, S., Bland-Hawthorn, J., Cecil, G., Tully, R.~B., \& Miller, S.~T.
  1999{\natexlab{b}}, \apj, 520, 111

\bibitem[{Verdugo {et~al.}(2015)Verdugo, Combes, Dasyra, Salom{\'e}, \&
  Braine}]{verdugo2015ram}
Verdugo, C., Combes, F., Dasyra, K., Salom{\'e}, P., \& Braine, J. 2015, \aap,
  582, A6

\bibitem[{Vila-Vilar{\'o} {et~al.}(1998)Vila-Vilar{\'o}, Taniguchi, \&
  Nakai}]{vila1998co}
Vila-Vilar{\'o}, B., Taniguchi, Y., \& Nakai, N. 1998, \apj, 116, 1553

\bibitem[{Vollmer \& Huchtmeier(2003)}]{vollmer2003atomic}
Vollmer, B. \& Huchtmeier, W. 2003, \aap, 406, 427

\bibitem[{{Wada}(2012)}]{wada2012radiation}
{Wada}, K. 2012, \apj, 758, 66

\bibitem[{{Williamson} {et~al.}(2018){Williamson}, {Venanzi}, \&
  {H{\"o}nig}}]{williamson2018radiation}
{Williamson}, D., {Venanzi}, M., \& {H{\"o}nig}, S. 2018, arXiv e-prints
  [\eprint[arXiv]{1812.07448}]

\bibitem[{Yoshida {et~al.}(2002)Yoshida, Yagi, Okamura, Aoki, Ohyama, Komiyama,
  Yasuda, Iye, Kashikawa, Doi, \& ...}]{yoshida2002discovery}
Yoshida, M., Yagi, M., Okamura, S., {et~al.} 2002, \apj, 567, 118

\bibitem[{Young {et~al.}(2011)Young, Bureau, Davis, Combes, McDermid, Alatalo,
  Blitz, Bois, Bournaud, Cappellari, \& ...}]{young2011atlas3d}
Young, L.~M., Bureau, M., Davis, T.~A., {et~al.} 2011, \mnras, 414, 940

\bibitem[{Young {et~al.}(2014)Young, Scott, Serra, Alatalo, Bayet, Blitz, Bois,
  Bournaud, Bureau, Crocker, {et~al.}}]{young2014atlas3d}
Young, L.~M., Scott, N., Serra, P., {et~al.} 2014, \mnras, 444, 3408

\end{thebibliography}
\end{document}